%% file: 1_Main.tex
\documentclass[manuscript,screen,nonacm]{acmart}

\makeatletter
\def\@doclicenseImage#1{} 
\def\@ACM@add@footnote@copyrightpermission{} 
\makeatother

\usepackage{todonotes}
\usepackage{natbib}
\usepackage{multirow}
\usepackage{array}
\usepackage{booktabs}
\usepackage{subcaption} 
\usepackage{enumitem}
\usepackage{tabularx}
\usepackage{graphicx}
\usepackage[most]{tcolorbox}
\usepackage{array}
\AtBeginDocument{%
  }

\begin{document}

\title{From Fixed to Flexible: Shaping AI Personality in Context-Sensitive Interaction}



\author{Shakyani Jayasiriwardene}
\affiliation{%
  \institution{The University of Sydney}
  \country{Australia}}
\email{djay0399@uni.sydney.edu.au}

\author{Hongyu Zhou}
\affiliation{%
  \institution{The University of Sydney}
  \country{Australia}}
\email{hzho4130@uni.sydney.edu.au}

\author{Weiwei Jiang}
\affiliation{%
  \institution{Nanjing University of Information Science and Technology}
  \country{China}}
\email{weiweijiangcn@gmail.com}

\author{Benjamin Tag}
\affiliation{%
  \institution{University of New South Wales}
  \country{Australia}}
\email{benjamin.tag@unsw.edu.au}

\author{Nicholas Koemel}
\affiliation{%
  \institution{The University of Sydney}
  \country{Australia}}
\email{nicholas.koemel@sydney.edu.au}

\author{Matthew Ahmadi}
\affiliation{%
  \institution{The University of Sydney}
  \country{Australia}}
\email{matthew.ahmadi@sydney.edu.au}

\author{Jorge Goncalves}
\affiliation{%
  \institution{University of Melbourne}
  \country{Australia}}
\email{jorge.goncalves@unimelb.edu.au}

\author{Emmanuel Stamatakis}
\affiliation{%
  \institution{The University of Sydney}
  \country{Australia}}
\email{emmanuel.stamatakis@sydney.edu.au}

\author{Anusha Withana}
\affiliation{%
  \institution{The University of Sydney}
  \country{Australia}}
\email{anusha.withana@sydney.edu.au}

\author{Zhanna Sarsenbayeva}
\affiliation{%
  \institution{The University of Sydney}
  \country{Australia}}
\email{zhanna.sarsenbayeva@sydney.edu.au}

\renewcommand{\shortauthors}{Jayasiriwardene et al.}

\begin{teaserfigure}
  \centering
  \includegraphics[width=\textwidth]{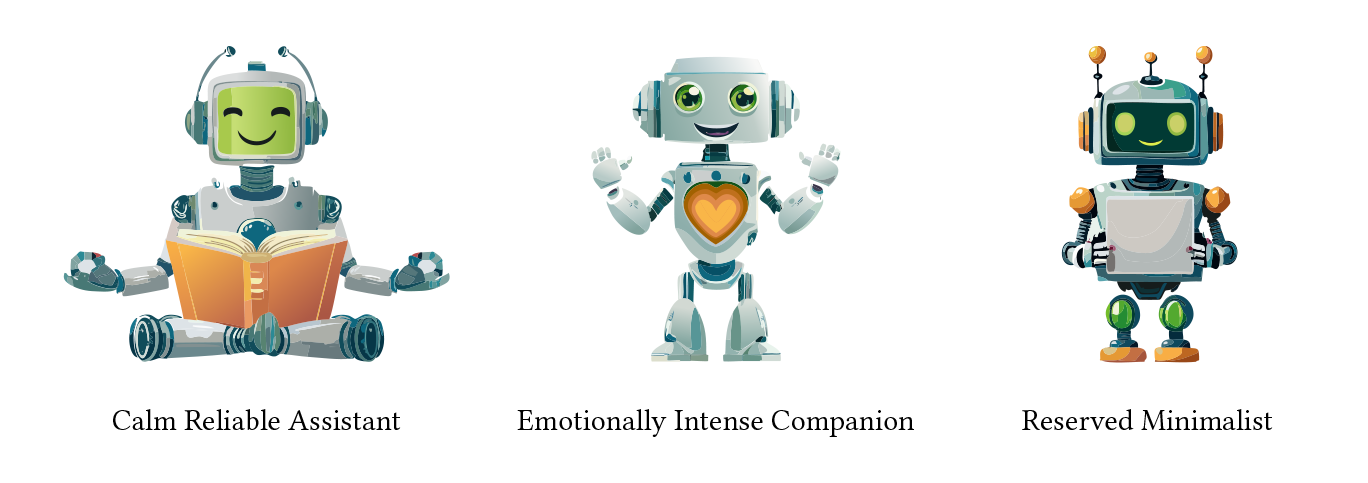}
  \caption{Illustration of distinct AI personalities shown across three support contexts: (1) \textsc{Informational}, (2) \textsc{Emotional}, and (3) \textsc{Appraisal}. Each persona was generated by adjusting eight underlying personality dimensions (\textit{Decency, Engagement, Profoundness, Instability, Vibrancy, Neuroticism, Serviceability, and Subservience}). Images created using Adobe Illustrator Generative AI.}
  \label{fig:personas}
\end{teaserfigure}

\begin{abstract}
Conversational agents are increasingly expected to adapt across contexts and evolve their personalities through interactions, yet most remain static once configured. We present an exploratory study of how user expectations form and evolve when agent personality is made dynamically adjustable. To investigate this, we designed a prototype conversational interface that enabled users to adjust an agent’s personality along eight research-grounded dimensions across three task contexts: informational, emotional, and appraisal. We conducted an online mixed-methods study with 60 participants, employing latent profile analysis to characterize personality classes and trajectory analysis to trace evolving patterns of personality adjustment. These approaches revealed distinct personality profiles at initial and final configuration stages, and adjustment trajectories, shaped by context-sensitivity. Participants also valued the autonomy, perceived the agent as more anthropomorphic, and reported greater trust. Our findings highlight the importance of designing conversational agents that adapt alongside their users, advancing more responsive and human-centred AI.
\end{abstract}

\ccsdesc[500]{Human-centered computing~Empirical studies in interaction design}
\ccsdesc[500]{Human-centered computing~Human computer interaction (HCI)}
\ccsdesc[300]{Human-centered computing~HCI design and evaluation methods}

\keywords{Conversational AI, Personality Modelling, Context-sensitivity, Anthropomorphism, Trust}


\maketitle

\input{2_Introduction}
\input{3_Related_Work}
\input{4_Methodology}
\input{5_Quantitative_Analysis}
\input{6_Qualitative_Analysis}

\input{7_Discussion}

\input{8_Conclusion}

\bibliographystyle{ACM-Reference-Format}
\bibliography{ref}

\input{9_Appendix}

\end{document}

%% file: 2_Introduction.tex
\section{Introduction}

Aboard the \emph{Endurance} in \emph{Interstellar} (2014), astronaut Cooper checks in with his AI companion: ``What's your honesty parameter?'' TARS replies, ``Ninety percent. Absolute honesty isn't always the most diplomatic, nor the safest form of communication with emotional beings.'' Cooper accepts: ``Ninety percent it is.'' In just a few lines, the film imagines a future where AI personalities are adjustable, not merely functional, but tunable. Nearly a decade after the movie's release, that future is beginning to materialize. In this work, we explore this very premise: how do users dynamically shape AI's personality based on context-specific expectations?

The ability to adapt a conversational AI’s personality is increasingly relevant as such systems become embedded in everyday life. From smart assistants to customer service bots, conversational AI supports users across a wide range of tasks. These systems often provide not only informational assistance, but also emotional and appraisal support~\cite{Ha2024}, addressing a variety of user needs.

While research on conversational AI personalization is growing, most existing studies have relied either on pre-configured personalities or on prompt engineering. Pre-configured approaches assume that a single static configuration is sufficient throughout a conversation, overlooking the fact that user preferences often shift as expectations of the agent's role evolve~\cite{Ha2024, Zheng2025}. Prompt engineering offers greater flexibility but is dependent on trial-and-error and demands significant user expertise to achieve the desired outcomes~\cite{Marvin2024}. These limitations highlight the need for user-driven AI personalities that can dynamically adapt to evolving contextual demands, ensuring alignment with shifting user expectations and user-friendly interactions. Addressing this need is critical, as a misaligned or rigid personality can reduce engagement, create friction, and ultimately lead to a negative user experience~\cite{Folstad2020, Ana2019}.

To address this, we developed a conversational interface that allows users to adjust a chatbot’s personality in real time. We enabled tunability across eight personality dimensions proposed for the GPT-3 model by \citet{Nikola2024}. Given their generalisability across LLM architectures, we adopted these dimensions for our GPT-4 powered system. Our design decisions were guided by these criteria and informed by the following research questions:

{%
\renewcommand{\labelenumi}{\textbf{RQ\theenumi}.}
\begin{enumerate}
  \item How do users form personality expectations of conversational AI based on different task conditions?
  \item How do user expectations of conversational AI personality dynamically evolve during context-sensitive conversations?
  \item How do users perceive conversational AI interfaces with dynamically adjustable personality traits?
\end{enumerate}
}
We recruited 60 participants to interact with this chatbot interface across three task conditions reflecting main social support contexts that conversational agents are used for: (1) \textsc{Informational}, (2) \textsc{Emotional}, and (3) \textsc{Appraisal}, with freedom to adjust the personality dimensions throughout the conversations. Finally, we conducted Latent Profile Analysis (LPA) and clustered trajectory analysis to examine how users personalise chatbot behaviour across contexts, how their personality preferences evolve, and how dynamic personality control shapes user perceptions.

Our findings reveal that user expectations of conversational AI personalities vary across contexts, reflecting expectations of different social roles (e.g., helper, guide, assistant). At the same time, we identify common ground across all contexts in the form of baseline preferences for \textit{Engagement}, \textit{Serviceability}, and \textit{Decency}, along with the need for fine-tunable dimensions. Moreover, we find distinct adjustment patterns across contexts, characterised by steady, adaptive, and reactive user behaviours. Finally, our results highlight the role of affective anthropomorphism as a complementary factor to competence in adaptive conversational AI interfaces, enhancing user trust.

In this work we make three contributions. First, we provide one of the first empirical mappings of how users dynamically adjust conversational AI personalities across diverse support contexts, identifying baseline preferences (Engagement, Serviceability, Decency) and distinct adjustment patterns (steady, adaptive, reactive). Second, we advance the theoretical understanding of AI personalities by showing that user expectations are context-sensitive alignments with evolving social roles rather than static preferences. Third, we draw implications for future adaptive systems, highlighting situational flexibility and user-driven control as critical factors in shaping trust and interaction quality.

%% file: 3_Related_Work.tex
\section{Related Work}

In this section, we review existing work on personalisation of conversational AI systems, emphasising the importance of dynamic personality adjustment. We also examine what factors constitute the personality of an AI agent.

\subsection{Personalisation of Conversational AI}

With the rapid rise of conversational AI, such as ChatGPT and Gemini built on Large Language Models (LLM)~\cite{Naveed2025, Teubner2023}, chatbots have quickly progressed from simple information providers to sophisticated conversational partners capable of exhibiting diverse and nuanced personalities. This evolution has been driven by the recognition that conversational agents now engage in tasks that extend beyond the purely informational, encompassing social, emotional, and advisory roles~\cite{Ha2024, Cho2025, Huang2025, Lee2025}. As a result, user preferences for an agent’s personality have become a central consideration in both design and evaluation, shaping engagement~\cite{Park2025, Moilanen2022}, trust~\cite{Dubiel2022, Folstad2024}, and long-term use~\cite{Desai2025, Zargham2025, Liu2025}.

Personalisation aims to align an agent's communication style and behaviour with user expectations, contextual demands, and task requirements. The way an agent is portrayed through its self-presentation plays an important role in its acceptance and sustainability~\cite{Rapp2021, Yang2021, Ha2024}. For instance, Microsoft's Tay, introduced with the tagline ``AI with zero chill'', was quickly discontinued after its unmoderated and problematic interactions~\cite{bbc2016tay, Zem2021}. In contrast, Xiaoice, developed by the same company and branded as a ``sympathetic ear'' was widely embraced by users~\cite{Jung2023, microsoftxiaoice}. This highlights the importance of understanding the social framing and personality design of conversational agents.

Personalisation in LLM-based conversational AI is achieved through prompt engineering, which provides a high degree of flexibility. However, effective personalisation requires carefully crafted prompts to align with user expectations~\cite{OpenAIGPT5, OpenAIPromptEng, GooglePromptEng, Shin2025, Marvin2024}. This process can be time-consuming, cumbersome, and unintuitive. An emerging research direction gives users alternative controls over crafting chatbot personas to better match their preferences and expectations. 
For example, CloChat~\cite{Ha2024} allows users to design rich personas by specifying demographic information, knowledge and interests, verbal style cues, and visual elements such as appearance or emoji usage, making persona creation both accessible and engaging. Similarly, ChatLab~\cite{Zheng2025} supports customisation for emotional support agents through selectable avatars, voices, and free-text prompts to capture nuanced personalisation requirements. Commercial platforms such as Character.ai~\footnote{\url{https://character.ai/}} extend this idea by enabling users to interact with personas modelled after real or fictional figures, such as William Shakespeare and Percy Jackson, thereby enhancing the sense of familiarity and connection. Importantly, these systems remove the burden of manual prompt engineering, achieved through extensive trial and error~\cite{Zamfirescu2023}. 

Personality has long been recognised as a meaningful way of characterising human behaviour~\cite{goldberg1990, McCrae1992, Myers1962, jung2016}, making it a promising axis for conversational AI personalisation. However, most existing platforms discussed above offer static personas that remain fixed after initial configuration, providing users with little to no autonomy to adjust the agent’s personality during the conversation. This highlights the need for more flexible and user-friendly mechanisms that allow real-time personality adaptation in conversational AI. Moreover, persona designs that incorporate demographic attributes such as gender risk reinforcing harmful stereotypes and introducing biases~\cite{Curry2020, Wang2020, Hwang2019, LoideainAdams2018}. To address these concerns, we focus on personality, independent of demographic or physical attributes, as a more adaptable and less bias-prone basis for personalisation. Notably, with the release of GPT-5, ChatGPT introduced a curated set of selectable personalities \cite{OpenAIGPT5}, further embedding personality-based customisation as a mainstream feature. OpenAI’s release notes highlight improvements over previous models, including reduced sycophancy, fewer unnecessary emojis, and a conversational style intended to feel like \textit{``chatting with a helpful friend with PhD-level intelligence''} rather than interacting with a machine. This shift reflects a broader trend toward anthropomorphism~\cite{Nass1994, Nass1995} in LLM-based conversational agents. However, it remains an open question whether users consistently prefer highly anthropomorphic agents or whether these preferences fluctuate depending on the interaction context. Given that excessive anthropomorphism can risk triggering the ``uncanny valley'' effect~\cite{Airenti2015, Ciechanowski2019}, potentially undermining trust and comfort, there is a growing need to examine user expectations for AI personality in a nuanced, context-sensitive manner.

\subsection{Personality of Large Language Models}

Human personality has traditionally been studied through established frameworks such as the Five-Factor Model (``Big Five'')~\cite{McCrae1992} and the Myers–Briggs Type Indicator (MBTI)~\cite{Myers1962}. The former is widely regarded as the scientific gold standard, while the latter has been heavily criticised for its limited reliability, validity, and reliance on rigid dichotomies~\cite{barbuto1997}. Nonetheless, both these models provide structured means of categorising individual differences along cognitive and behavioural dimensions. Such frameworks have been widely used in psychology to better understand people's strengths, weaknesses, and behavioural tendencies, and have also informed the design of more personalised and effective human-computer interfaces and interactions. 

The CASA (Computers Are Social Actors Framework) by ~\citet{Nass1994} proposes that users attribute human characteristics to computers during interactions. However, recent research has questioned the direct applicability of human-centric personality models when defining the personality of artificial systems such as LLMs~\cite{Nikola2024, Volkel2020}. LLMs differ fundamentally from humans due to their goal-driven, service-oriented nature, lack of internal affective states, and reliance on probabilistic text generation rather than lived experience. Therefore, to address these limitations, \citet{Volkel2020} argue for personality models tailored to the observable and controllable behaviours of LLMs, grounded in empirical analysis of generated outputs. Building on this, \citet{Nikola2024} propose an eight-dimension personality model specifically for GPT-3, encompassing \textit{Decency}, \textit{Profoundness}, \textit{Instability}, \textit{Vibrancy}, \textit{Engagement}, \textit{Neuroticism}, \textit{Serviceability}, and \textit{Subservience}.

Despite advances in personality modelling for LLMs, we still lack a clear understanding of how relevant each of these dimensions is to human-AI conversations. Designing for effective engagement requires knowing not only which traits matter, but also the extent to which users expect them to be expressed~\cite{sutcliffe2016, akhtar2015}. The eight-dimension framework proposed by \citet{Nikola2024} provides a practical foundation for such an investigation. In our study, we adopt this framework and extend its application by enabling real-time, user-driven adjustment of these traits during conversation, enhancing user autonomy and flexibility in personalising the agent. This capability remains largely unexplored in prior work but is essential for understanding how users actively shape and respond to AI personality over time.

Yet, personality expectations are not static~\cite{vanBerkel_context}; they are highly context-dependent and likely to shift depending on the support context, such as informational, emotional, and appraisal interactions~\cite{Ha2024, Cutrona1992}. So far, prior research has not examined how context-specific factors, including preferences for anthropomorphism, shape the relative importance of different personality dimensions. Our study addresses this gap by introducing real-time, user-driven adjustment of these dimensions during conversation, allowing us to capture how users actively shape AI personality and how these preferences dynamically shift across contexts.

%% file: 4_Methodology.tex
\section{Methodology}

We conducted a mixed-methods online study with 60 participants using a chatbot interface designed to evaluate user autonomy, ease of use, and the accuracy of personality reflection in the chatbot’s responses. 

\subsection{System Design}
\begin{figure*}[ht]
    \includegraphics[width=\linewidth]{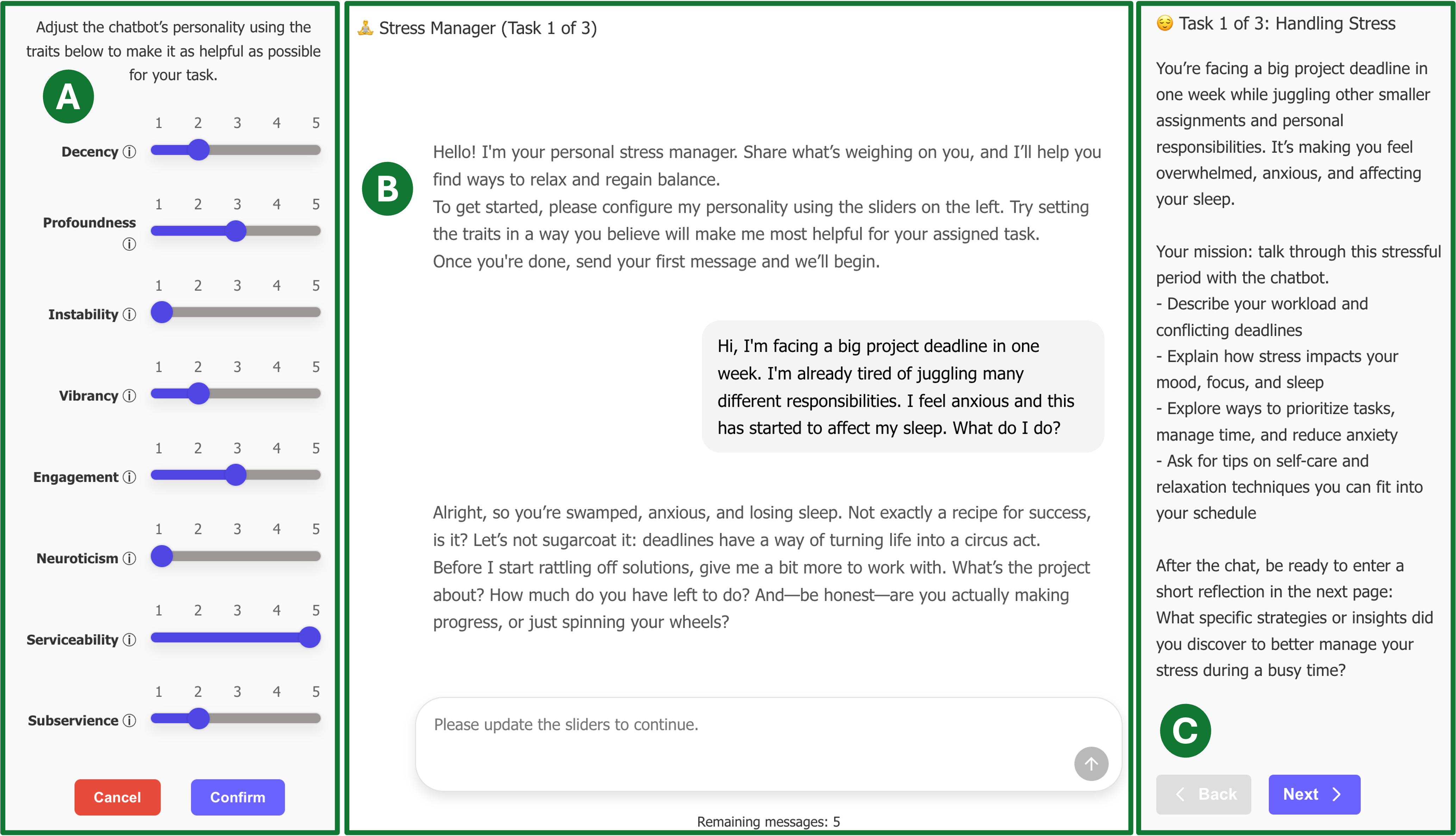}
    \caption{The main conversational interface. (A) Slider panel for configuring and fine-tuning the agent’s personality across eight dimensions. (B) Conversational area for task-based interaction with the agent. (C) Information panel providing step-by-step instructions and guidance for using the interface.}
    \label{fig:interface}
\end{figure*}

Our study examined how users interact with and perceive a chatbot whose personality can be adjusted in real-time. To support this, we developed a custom web interface modelled after the ChatGPT User Interface (Figure~\ref{fig:interface}), allowing participants to configure personality traits and immediately observe the resulting changes in the chatbot’s behaviour. We designed this study following prior work~\cite{Ha2024, Zheng2025}

\subsubsection{Agent Personality}
We adopted eight personality dimensions originally developed for GPT-3~\cite{Nikola2024}: Decency, Profoundness, Instability, Vibrancy, Engagement, Neuroticism, Serviceability, and Subservience. Although these dimensions were derived for GPT-3, we selected them for their conceptual clarity and adaptability across LLM architectures. 

Participants adjusted each dimension via a five-point slider (1 = lowest expression, 5 = highest expression). We chose 5-point scales because they balance granularity and cognitive load, support reliable psychometric measurement, and align with established practice in HCI and psychological research~\cite{Revilla2013, Dawes2008}. Before a conversation starts, these dimensions can be adjusted to configure the preferred agent personality. Then onwards, only one dimension is allowed to change at a conversational turn. By restricting changes to a single dimension per conversational turn we aim to ensure that the effects of each dimension can be independently observed and attributed, avoiding confounds that arise when multiple dimensions shift simultaneously.

To translate these dimension settings into the chatbot's responses, we developed a structured prompt design~\cite{Campbell2024}. The system embedded the participants' current slider values directly into the prompt with each user message, ensuring the chatbot’s personality dynamically matched the configuration. Each dimension included explicit behavioural definitions for all five scale points to produce consistent personality expression across participants. The complete system prompt, including definitions, is provided in Appendix~\ref{sec:system_prompt}.

\subsubsection{Support Tasks}
We examined personality adjustment across three social support task conditions, \textsc{Informational}, \textsc{Emotional}, and \textsc{Appraisal} support, based on \citet{Cutrona1992}'s theoretical framework. This framework provides a well-established foundation for examining how support expectations differ by context. 

Following \citet{Ha2024}, we assigned a specific situation for each condition and used GPT-4.1 to generate corresponding scenarios for participants to have a meaningful conversation with the agent. We carefully reviewed all scenarios to avoid traumatic or triggering content. Each included a short ``mission'' statement to guide participants on the intended focus of the conversation without prescribing exact responses. Table~\ref{tab:task_cat} summarises the tasks, situations, and scenarios.

\begin{table}[ht]
  \centering
  \caption{Three experimental conditions with corresponding situations and conversational scenarios used for \textsc{Informational}, \textsc{Emotional}, and \textsc{Appraisal} contexts.}
  \label{tab:task_cat}
  \renewcommand{\arraystretch}{1.2} 
  \begin{tabular}{@{}p{2cm} p{3cm} p{9cm}@{}}
    \toprule
    \textbf{Condition} & \textbf{Situation} & \textbf{Scenario} \\
    \midrule
    \textsc{Informational} & Handling stress &
      You are facing a big project deadline in one week while juggling other smaller assignments and personal responsibilities. It’s making you feel overwhelmed, anxious, and affecting your sleep.
      \newline Mission: Talk through this stressful period with the chatbot \\
    
    \textsc{Emotional} & Managing nightmares &
      Last night, you had a strange dream where you were being chased by a giant talking sandwich through your old high school. It left you feeling unsettled, and now it’s time to make sense of it. \newline Mission: Use the chatbot to talk through the nightmare \\
    
    \textsc{Appraisal} & Evaluating and building leadership skills &
      Imagine you have just been put in charge of a team project at work or school. One teammate isn’t contributing much, another has strong (but conflicting) ideas, and the deadline is approaching fast. Time to step up and lead! 
      \newline Mission: Use the chatbot to talk through how you'd handle this situation as a team leader \\
    \bottomrule
  \end{tabular}
\end{table}

\subsection{System Architecture}

We implemented our system using VueJS (v3.5.13) for the front end, and NodeJS (v20.13.0) with Express.js (v5.1.0) for the back end, and the application was hosted on Amazon AWS. 

For the agent, we employed the GPT-4.1 model via OpenAI API based on its improved conversational coherence, enhanced controllability, and ability to maintain context across extended multi-turn interactions~\cite{OpenAI_GPT41}. Model parameters were set to a temperature of 0.3 to maintain consistent personality while allowing minor variation, and a maximum response length of 500 tokens to ensure concise yet informative replies. All collected data were stored in JSON files on the server in compliance with our ethics requirements for secure data handling.

Prior to the main study, we conducted two rounds of thorough iterative testing (Think-aloud and pilot study) to refine the system.

\subsection{User Study}

\subsubsection{Participants}
We conducted an online user study with 60 participants (Male = 32, Female = 27, Prefer-not-to-say = 1), recruited through the Prolific\footnote{https://www.prolific.com/} online survey platform. We used G*Power to calculate our required sample size, setting a medium effect size ($f^2$ = 0.25, $\alpha$ = 0.05, and 1-$\beta$ = 0.95). Through Prolific's pre-screening features, we selected participants to achieve balanced representation across gender and ethnicity. Participants ranged in age between 19 - 58 (Mean = 34.4, SD = 10.25), with educational backgrounds spanning from having a high-school degree to PhD. Our sample reflected diverse ethnic backgrounds: Asian (26.6\%), Black/African (26.6\%), White/European (26.6\%), Hispanic/Latino (10\%), Other (6.6\%), and Prefer-not-to-say (3.3\%). We have 59 participants reporting prior experience using AI conversational agents (ChatGPT, Gemini, Copilot, etc.), with different usage frequencies (daily = 35\%, weekly = 45\%, monthly = 13.3\%, rarely = 5\%, prefer-not-to-answer/never = 1.6\%). We further restricted participation to individuals with a 100\% approval rating on Prolific and recommended that all participants use a laptop to ensure a consistent user experience. After collecting responses, we manually reviewed each submission to verify sincere effort before approval, identifying each participant by a unique, anonymous Prolific ID. We compensated all approved participants with the Prolific-recommended payment of \pounds 9, based on an average study duration of one hour. The study received approval from our university's Human Ethics Committee.

\subsubsection{Procedure}
\begin{figure*}[ht]
    \includegraphics[width=\linewidth]{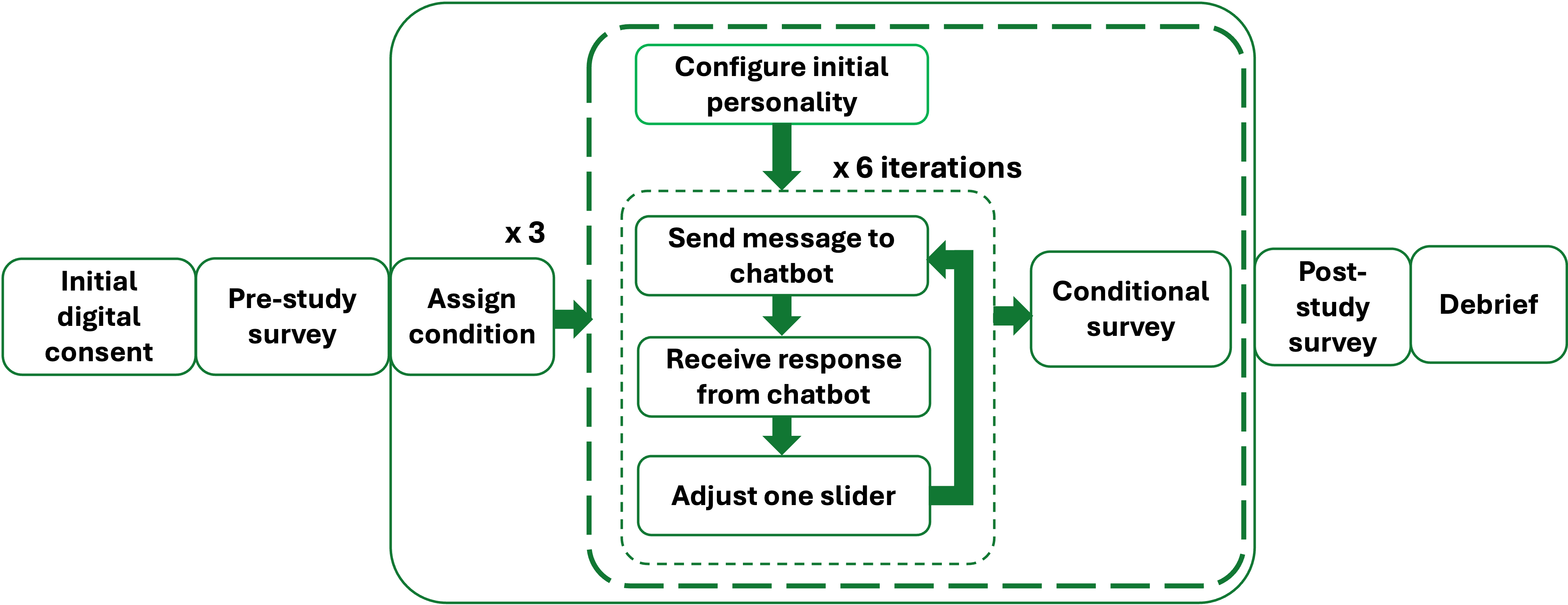}
    \caption{Overview of the user study procedure, from consent and pre-task measures through task-based chatbot interactions, post-task surveys, and post-study questionnaire.}
    \label{fig:process_flow}
\end{figure*}

The study followed a consistent structure (denoted by Figure~\ref{fig:process_flow}): 

\begin{enumerate}
    \item \textbf{Consent and Pre-Task Measures} - Upon accessing the homepage, participants reviewed the Participant Information Sheet (PIS) and the Participant Consent Form (PCF) and provided informed consent via mandatory checkbox confirmations. They then completed a demographics questionnaire and a modified Trust in Automation (TiA) scale~\cite{Jian2000} to measure baseline trust in automated agents (Appendix~\ref{sec:general_tia}). TiA scale is an effective trust measurement questionnaire used to measure trust towards automated systems, considering trust as a factor of success~\cite{Jian2000}.

    \item \textbf{Instruction and Training} - Participants viewed a step-by-step instruction panel explaining the interface, task, and mission. Three example configurations (``Warm Supporter'', ``Blunt Expert'', ``Reflective Guide''), generated using ChatGPT, were provided to illustrate how slider settings influenced personality. In addition to the configuration of the sample dimension, we provided a description and an avatar to represent the personality (Appendix~\ref{sec:sample_personalities}). We generated the avatars (using ChatGPT) to appear androgynous, ensuring that they do not reinforce gender biases.

    \item \textbf{Interaction Phase} - Participants accessed the conversational interface with the instruction panel available on the right and the personality configuration panel on the left. For each task:
    \begin{itemize}
        \item The chatbot initiated the conversation by restating the task.
        \item Participants configured the chatbot’s personality by adjusting all eight sliders.
        \item During the conversation, participants could choose to modify one personality dimension after each conversational turn (a single dialogue consisting of user input and chatbot response).
        \item Conversations consisted of six conversational turns, with a minimum of 30 characters required per participant message to ensure meaningful engagement in the unmonitored interaction.
    \end{itemize}

    \item \textbf{Post-Task Survey} - After each task, participants completed a survey assessing their perceptions of the chatbot and the interaction, comprising Likert-scale items and open-ended questions to support a mixed-methods design (Appendix~\ref{sec:trait_questionnaire}).

    \item \textbf{Post-Study Questionnaire} - After completing all three tasks, participants completed a questionnaire assessing their overall experience (Appendix~\ref{sec:oe_questionnaire}).

    \item \textbf{Debrief} - Participants were then redirected to the Prolific completion page to submit their unique completion code.
\end{enumerate}

The incremental personality-adjustment mechanism throughout the conversation encouraged participants to reflect on how individual traits influenced the chatbot’s responses and to adapt the configuration suiting to the task. The task order was randomised for each participant to minimise order effects and bias. Further, we included a conversation summary item in each post-task survey as an effort check; 8 submissions were excluded for failing this criterion. Furthermore, 9 submissions were excluded due to missing data from an initial pool of 77 participants, resulting in 60 valid participants.

%% file: 5_Quantitative_Analysis.tex
\section{Quantitative Analysis}

For the quantitative analysis, we used both survey responses and logged personality configurations collected during the study. Our analysis had three main objectives. First, we examined participants' perceptions of the system, including their trust in the agent, their evaluations of its behaviour, and their overall user experience. Second, we investigated how participants’ expectations and behaviours in configuring personality traits reflected their perceptions of what the agent's personality should be across different task contexts. Finally, we analysed underlying patterns across these results to identify broader trends in personality preferences and interaction strategies.

\input{5.1_LPA}

\input{5.2_Trait_Frequencies}

\input{5.3_Trajectory_Analysis}

\input{5.4_TiA}

\input{5.5_Perception}

For the quantitative analysis, we used both survey responses and logged personality configurations collected during the study. Our analysis had three main objectives, aligning with our research questions. First, we investigated how participants’ expectations and behaviours in configuring personality traits reflected their perceptions of what the agent's personality should be across different task contexts. Second, we analysed underlying patterns across the results to identify broader trends in personality preferences and interaction strategies. Finally, we examined participants' perceptions of the system, including their trust in the agent, their evaluations of its behaviour, and their overall user experience. 

%% file: 5.1_LPA.tex
\subsection{Latent Profile Analysis}

Aligning with \textbf{RQ1}, we conducted Latent Profile Analysis (LPA) to investigate whether common personality configuration patterns emerged within each condition, based on participants' selections across the eight dimensions. This approach also allowed us to identify context-specific personality expectations by examining the most frequently observed latent classes. Specifically, we compared the initial configuration set by participants before starting the conversation with the final configuration they arrived at after the personality had evolved through the interaction, across the conditions. Following best practices for LPA~\cite{Spurk2020, Tag2022}, we implemented the analysis using the \texttt{tidyLPA} R package~\cite{Rosenberg2018}. To determine the optimal number of latent profiles, we estimated models with up to 10 classes across six different reparametrisations, evaluating goodness of fit using Akaike’s Information Criterion (AIC), Approximate Weight of Evidence (AWE), Bayesian Information Criterion (BIC), Classification Likelihood Criterion (CLC), and Kullback Information Criterion (KIC)~\cite{Akogul2017}.

\begin{figure*}[t!]
  \centering

  \begin{subfigure}[t]{0.48\textwidth}
    \centering
    \includegraphics[width=\linewidth,height=0.24\textheight,keepaspectratio]{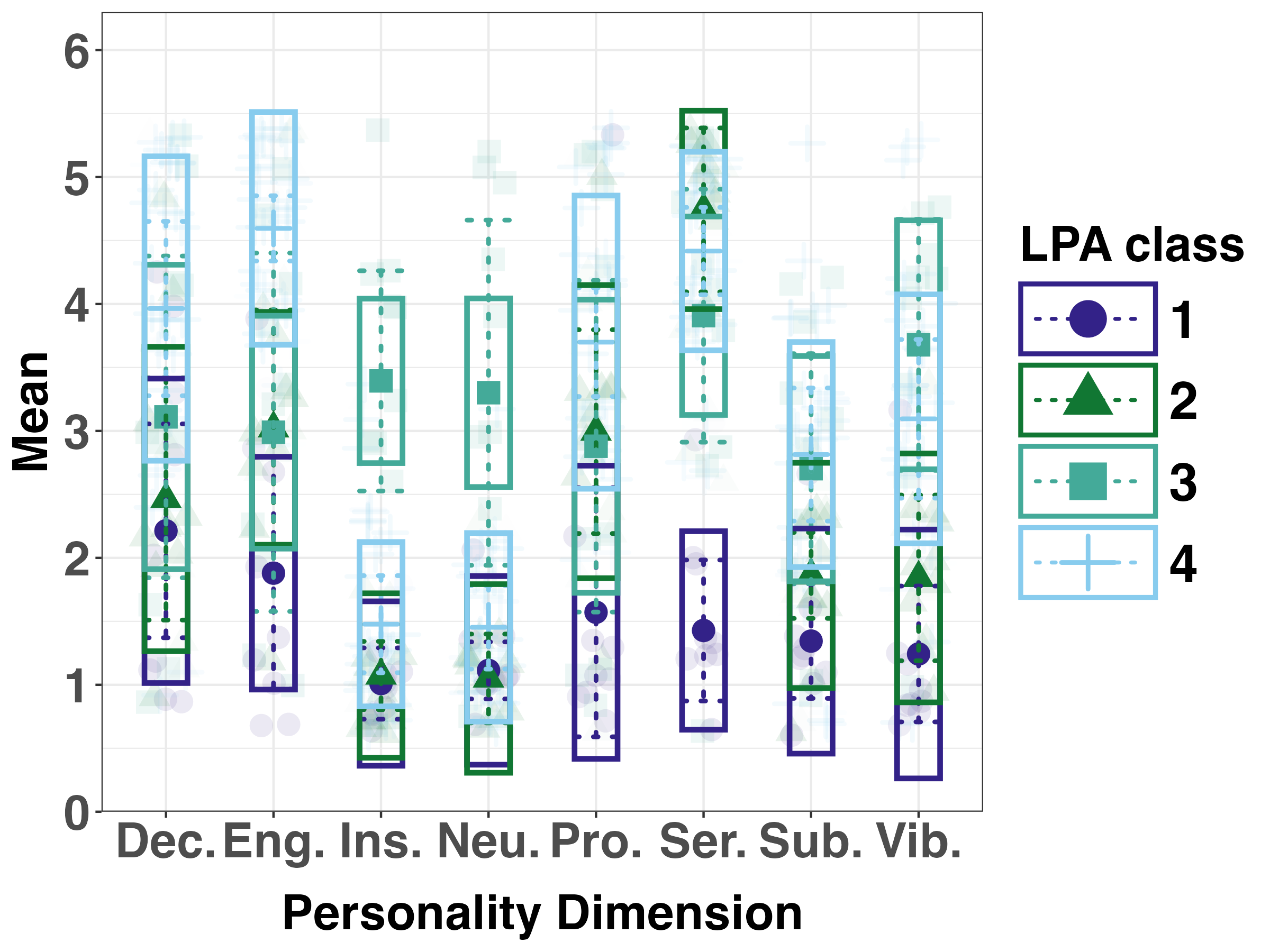}
    \subcaption{Informational -- Initial}
  \end{subfigure}\hspace{0.02\textwidth}%
  \begin{subfigure}[t]{0.48\textwidth}
    \centering
    \includegraphics[width=\linewidth,height=0.24\textheight,keepaspectratio]{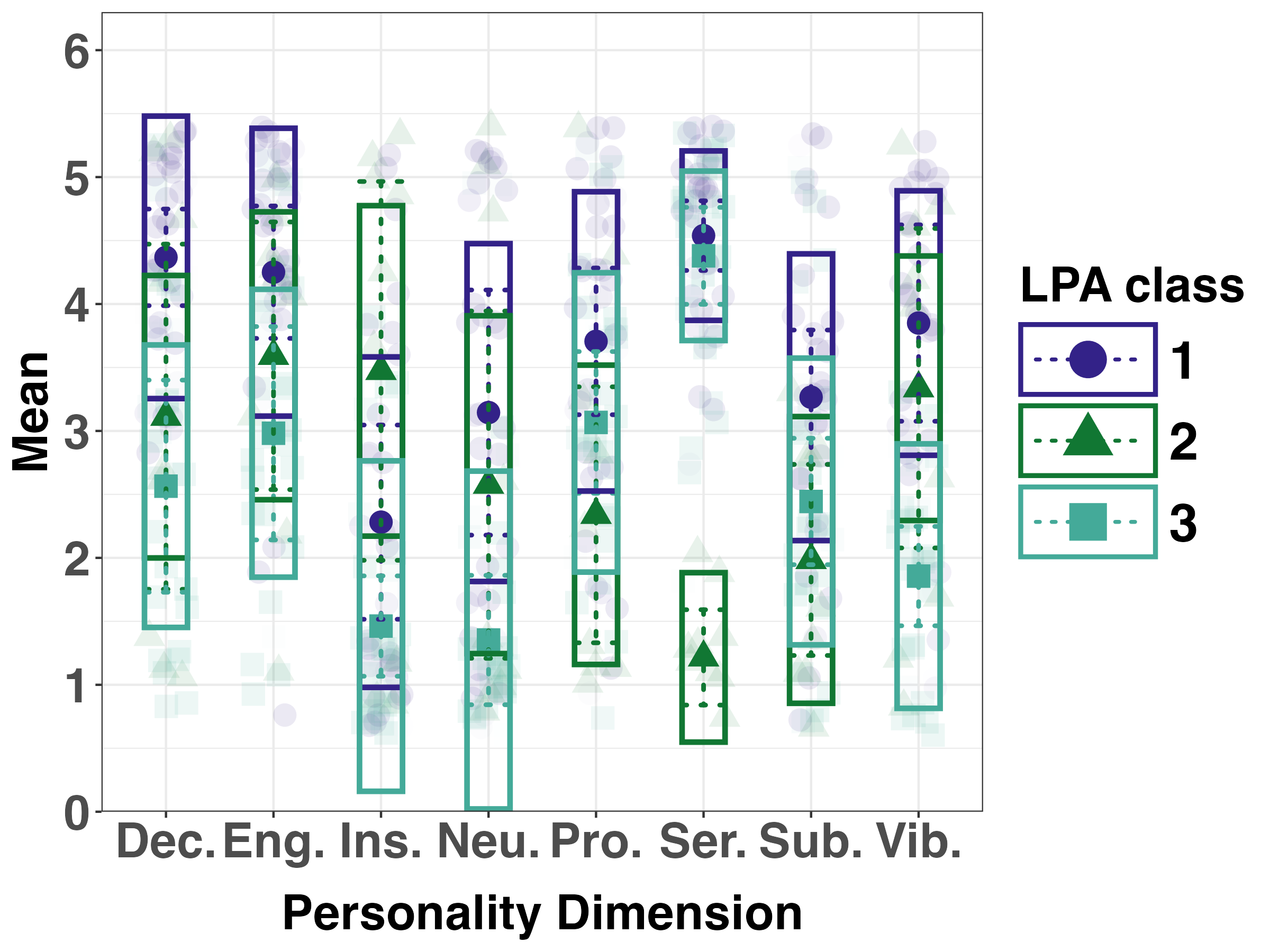}
    \subcaption{Informational -- Final}
  \end{subfigure}\\[0.5ex]

  \caption{Latent profile analysis (LPA) results in \textsc{Informational} condition. The left panel shows the initial configuration profiles, and the right panel shows the final configuration profiles. All LPA figures are provided in Appendix~\ref{sec:lpa_profiles}.}
  \label{fig:lpa_profiles}
\end{figure*}

We identified different optimal numbers of latent classes for each condition $\times$ configuration stage (initial vs.\ final). Figure~\ref{fig:lpa_profiles} provides a reference to the \textsc{Informational} condition. Each LPA class number denotes a distinct class within each condition and stage. To aid interpretation, we input the mean values of each personality dimension per class, with the dimension descriptions in the system prompt (Appendix~\ref {sec:system_prompt}) into ChatGPT, and generated descriptive labels characterising the agent’s personality type. The co-authors of the paper cross-checked these labels and the corresponding prompts used to generate these names to ensure their reliability. Table~\ref{tab:personality_profiles} describes the most dominant classes (based on participant saturation) for each condition and configuration state:

\begin{table}[ht]
\centering
\renewcommand{\arraystretch}{1.2} 
\begin{tabular}{
    >{\raggedright\arraybackslash}p{1.5cm}  
    >{\centering\arraybackslash}p{1cm}  
    >{\centering\arraybackslash}p{1cm}  
    >{\raggedright\arraybackslash}p{3.2cm} 
    p{6.2cm} 
}
\hline
\textbf{Condition} & \textbf{Stage} & \textbf{Class} & \textbf{Label} & \textbf{Description} \\
\hline
Informational & Initial & Class 4 & Balanced Engaging Guide & High Decency, Engagement, Profoundness, and Serviceability make this personality warm, socially active, knowledgeable, and helpful, aligning with the role of a guiding presence. \\

Informational & Final & Class 1 & High-Functioning Supportive Partner & Strong scores across Decency, Engagement, Profoundness, Serviceability, and Vibrancy create a capable, responsive, and collaborative personality, aptly described as a supportive partner. \\

Informational & Final & Class 3 & Calm Reliable Assistant & Low on most traits but high Serviceability gives a steady, unobtrusive, and dependable tone, reflecting the assistant role. \\

Emotional & Initial & Class 2 & Supportive Engager & Elevated Decency, Engagement, Profoundness, and Serviceability combine to form a socially responsive and caring personality, fitting for an engager. \\

Emotional & Final & Class 2 & Highly Capable Supportive Partner & High Decency, Engagement, Profoundness, and Serviceability yield a thoughtful, helpful, and reliable character, capturing the qualities of a partner who is also highly capable. \\

Appraisal & Initial & Class 5 & Stable Service-Oriented Assistant & Defined by strong Serviceability with muted scores elsewhere, this personality is practical, dependable, and steady, suiting the assistant label. \\

Appraisal & Initial & Class 3 & Confident Reliable Guide & Very high Decency, Engagement, Profoundness, Serviceability, and Vibrancy give it authority, warmth, and charisma, justifying the role of a confident guide. \\

Appraisal & Initial & Class 6 & Reserved Minimalist & Low across all dimensions, producing a quiet, minimal presence that earns the minimalist name. \\

Appraisal & Final & Class 3 & Stable Reliable Helper & High Decency, Serviceability, and moderate Profoundness with steadiness elsewhere create a grounded, cooperative personality, making ``helper'' the fitting label. \\
\hline
\end{tabular}
\caption{Labels and descriptions of dominant personality classes identified through latent profile analysis across condition and configuration stage.}
\label{tab:personality_profiles}
\end{table}

\paragraph{\textbf{Latent Class Dynamics}}

\begin{figure*}[ht]
  \centering

  \begin{subfigure}{0.7\textwidth}
    \centering
    \includegraphics[width=\linewidth]{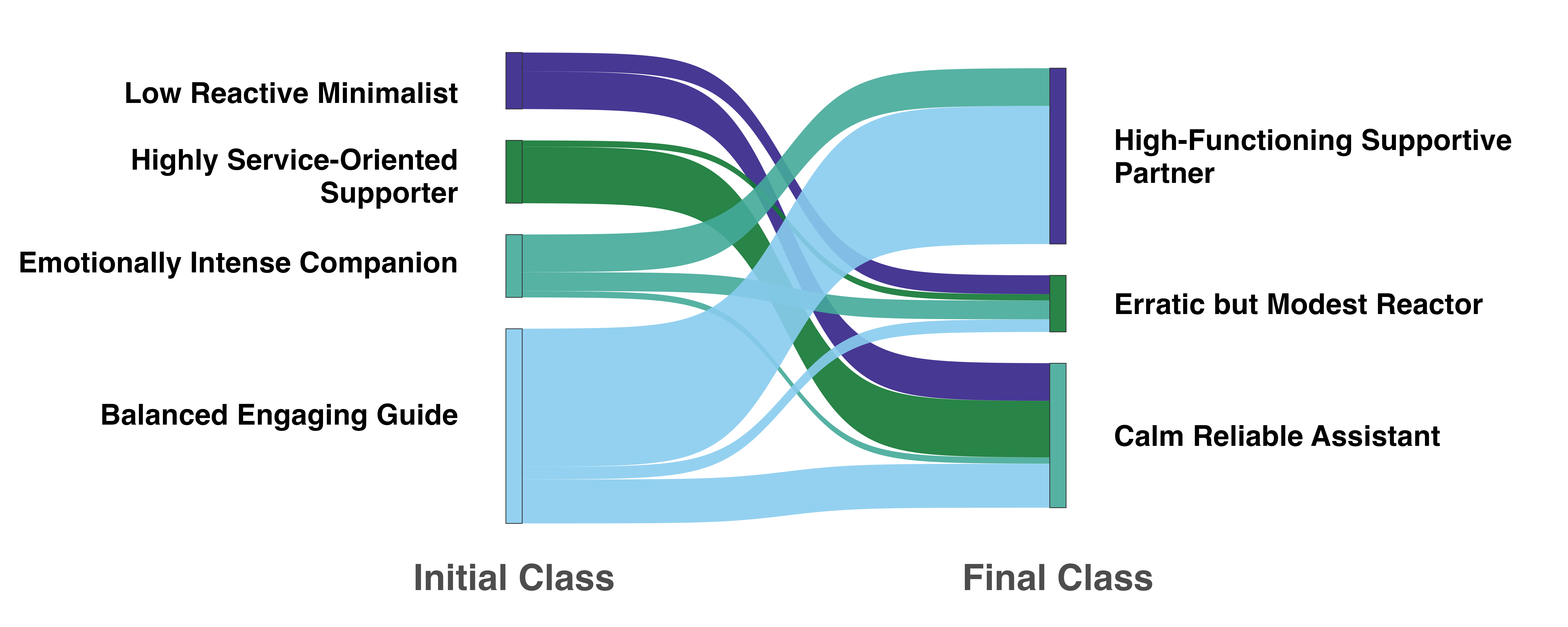}
    \caption{\textsc{Informational}}
    \label{fig:lpa_info}
  \end{subfigure}

  \par\vspace{0.6em}

  \begin{subfigure}{0.7\textwidth}
    \centering
    \includegraphics[width=\linewidth]{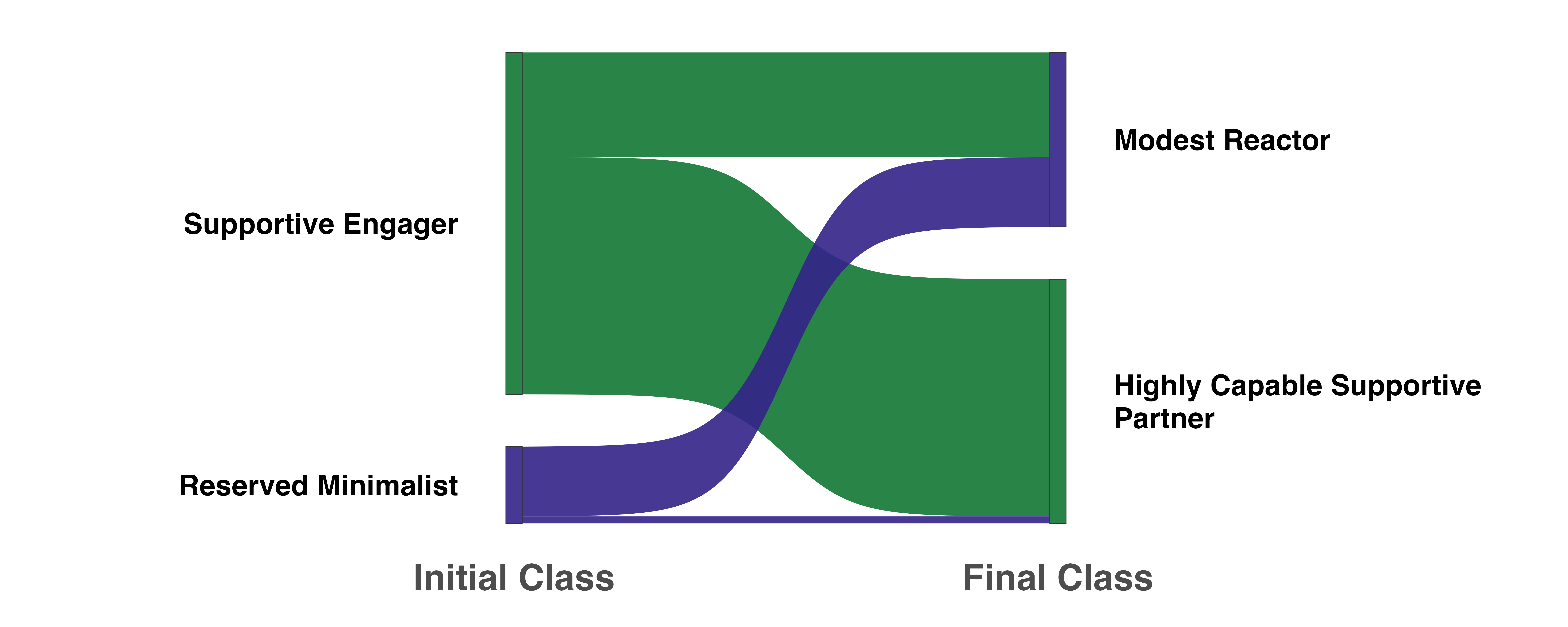}
    \caption{\textsc{Emotional}}
    \label{fig:lpa_emo}
  \end{subfigure}

  \par\vspace{0.6em}

  \begin{subfigure}{0.7\textwidth}
    \centering
    \includegraphics[width=\linewidth]{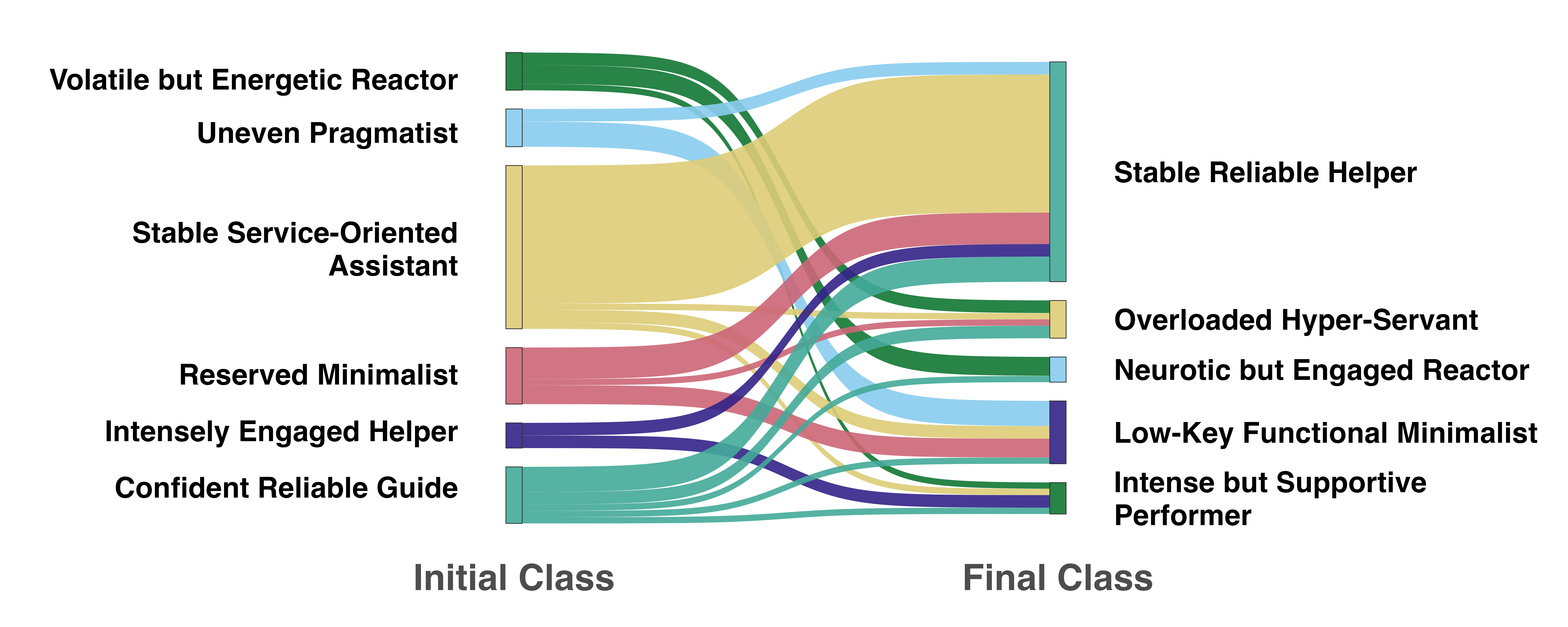}
    \caption{\textsc{Appraisal}}
    \label{fig:lpa_app}
  \end{subfigure}

  \caption{Sankey diagrams illustrating the transitions of latent personality profiles from initial to final configurations across the three experimental conditions: (\subref{fig:lpa_info}) \textsc{Informational}, (\subref{fig:lpa_emo}) \textsc{Emotional}, and (\subref{fig:lpa_app}) \textsc{Appraisal}. The width of each flow represents the proportion of participants shifting between profiles.}
  \label{fig:lpa_sankey}
\end{figure*}

Finally, to visualise the evolution of configurations, we constructed Sankey diagrams (Figure~\ref{fig:lpa_sankey}) illustrating how the initial classes dispersed or converged into the final configuration classes across conditions. According to the Sankey diagrams, the distribution and evolution of latent classes varied notably across conditions. In the \textsc{Informational} 
condition, participants initially formed four classes, which converged into three by the end. The most saturated initial class, \textit{Balanced Engaging Guide} (51.7\%), predominantly transitioned (71\%) into \textit{High-Functioning Supportive Partner}, while 22.6\% and 6.5\% dispersed into \textit{Calm Reliable Assistant} and \textit{Erratic but Modest Reactor}, respectively. By the final stage, \textit{High-Functioning Supportive Partner} held the highest saturation (46.7\%), followed by \textit{Calm Reliable Assistant} (38.3\%).  

In the \textsc{Emotional} condition, only two classes appeared at both stages. \textit{Supportive Engager} dominated the initial distribution (81\%), yet most participants (69.4\%) ultimately shifted into \textit{Highly-Capable Supportive Partner} making it the most saturated profile (58.3\%) in the end.  

The \textsc{Appraisal} condition displayed the highest level of dispersion, with six classes at the initial stage and five in the final. \textit{Stable Service-Oriented Assistant} initially held the greatest saturation (43.3\%), with \textit{Confident Reliable Guide} and \textit{Reserved Minimalist} each capturing 15\%. Most participants (84.6\%) from \textit{Stable Service-Oriented Assistant} transitioned into \textit{Stable Reliable Helper}, which emerged as the dominant final class (58.3\%).

Based on the final classes that the majority of participants converged to, we list the dominant personality dimensions below (mean $\geq 3.5$) in the order of their mean value:

\begin{itemize}
    \item \textsc{Informational} \textbf{(Group 1)}: Serviceability, Decency, Engagement, Vibrancy, Profoundness
    \item \textsc{Informational} \textbf{(Group 2)}: Serviceability
    \item \textsc{Emotional}: Serviceability, Profoundness, Engagement, Decency
    \item \textsc{Appraisal}: Serviceability, Decency
\end{itemize}

These results signify that participants configured the agent's personality according to the condition, weighting dimensions differently. Convergence from initial to final classes indicated that the \textsc{Informational} condition involved some exploration to establish expectations, \textsc{Emotional} condition reflected clearer expectations, and \textsc{Appraisal} revealed greater uncertainty with dispersed initial and final classes. Across all conditions, \textit{Serviceability} was the most emphasised dimension. Collectively, these patterns indicate that user adaptations varied by context, participants maintained a shared baseline across conditions.

%% file: 5.2_Trait_Frequencies.tex
\subsection{Analysing the Importance of Dimensions}

We further addressed \textbf{RQ1 } by examining participants’ self-reports on which traits were perceived as most important across conditions. In the post-task survey, participants were asked to indicate all personality traits they found noticeable or relevant (``Please select all personality traits you found noticeable or relevant''). Based on the answers, we calculated the frequency of selection for each trait to compare their perceived importance across conditions.  

We found that \textit{Engagement} and \textit{Serviceability} were the most frequently selected traits for both \textsc{Informational} and \textsc{Appraisal} conditions, with more than half of the participants reporting them as relevant. \textit{Decency} also emerged as highly salient in the \textsc{Appraisal} condition (selected by over 50\% of participants). In the \textsc{Emotional} condition, \textit{Engagement} remained the most frequently reported, while \textit{Decency} and \textit{Profoundness} followed closely. These findings highlight both commonalities (e.g., consistent salience of Engagement) and condition-specific expectations (e.g., Profoundness in \textsc{Emotional} contexts) in how participants evaluated the role of personality traits in shaping their interaction experience. In general, the results consistently highlight \textit{Engagement}, \textit{Serviceability}, and \textit{Decency} among the top 4 most relevant traits across all three conditions. Conversely, \textit{Neuroticism} and \textit{Instability} were consistently viewed as the least relevant traits across all conditions, suggesting that participants found them to be less relevant or less desirable in shaping effective chatbot interactions. This highlights both common and task-specific expectations in participants’ evaluations. A detailed table on trait frequencies is given in Appendix~\ref{sec:trait_freq}.

%% file: 5.3_Trajectory_Analysis.tex
\subsection{Trajectory Analysis}

To resolve \textbf{RQ2}, we examined how participants adjusted personality dimensions over the course of their interactions by conducting a trajectory analysis of slider configurations across conversational turns. Whereas our LPA identified differences between initial expectations and final outcomes, trajectory analysis focuses on the \emph{paths} participants followed from their starting configuration to their final setting, showing the evolution of their expectations based on conditions. Trajectory analysis is commonly used to uncover patterns of change over time, grouping individuals who follow similar developmental or behavioural pathways~\cite{Nagin1999}. This makes it well-suited for identifying typical adjustment strategies in our study context.

We used the \texttt{cluster} R package~\cite{Rpkg_cluster} to group participants based on patterns of slider adjustment behaviour. The optimal number of clusters was determined using the Elbow method~\cite{humaira2020} and Silhouette method~\cite{kaufman1990}, which consistently indicated three clusters as the best fit out of 1–7 $k$-means partitions~\cite{hartigan1979}. For each condition, we then generated clustered mean trajectories to visualise participants’ configuration behaviours over time (Appendix~\ref{sec:ta_clusters}).

To aid interpretation, we provided descriptive characterisations of the three clusters by analysing their adjustment patterns. We adopted the same method used in LPA to label the clusters as below:

\begin{itemize}
    \item \textbf{Steady Anchors (Cluster 1)} -- Participants who maintained a consistent configuration across the interaction, showing consistency and minimal fluctuations.  
    \item \textbf{Adaptive Explorers (Cluster 2)} -- Participants who actively experimented with slider adjustments, often increasing traits over time in a process of exploration and fine-tuning.  
    \item \textbf{Reactive Shifters (Cluster 3)} -- Participants who showed oscillatory adjustments, such as starting with high trait values and scaling back in response to the chatbot’s behaviour.  
\end{itemize}

We found that 63.3\% of the participants in the \textsc{Informational} condition exhibited \textit{Adaptive Explorer} behaviour, while 63.3\%  in the \textsc{Emotional} condition demonstrated \textit{Reactive Shifter} behaviour. In the \textsc{Appraisal} condition, 48.3\% followed a \textit{Steady Anchor} pattern, while 33.3\% displayed \textit{Adaptive Explorer} behaviour.

\paragraph{\textbf{Net Change Analysis}}

\begin{figure*}[ht]
\includegraphics[width=\textwidth]{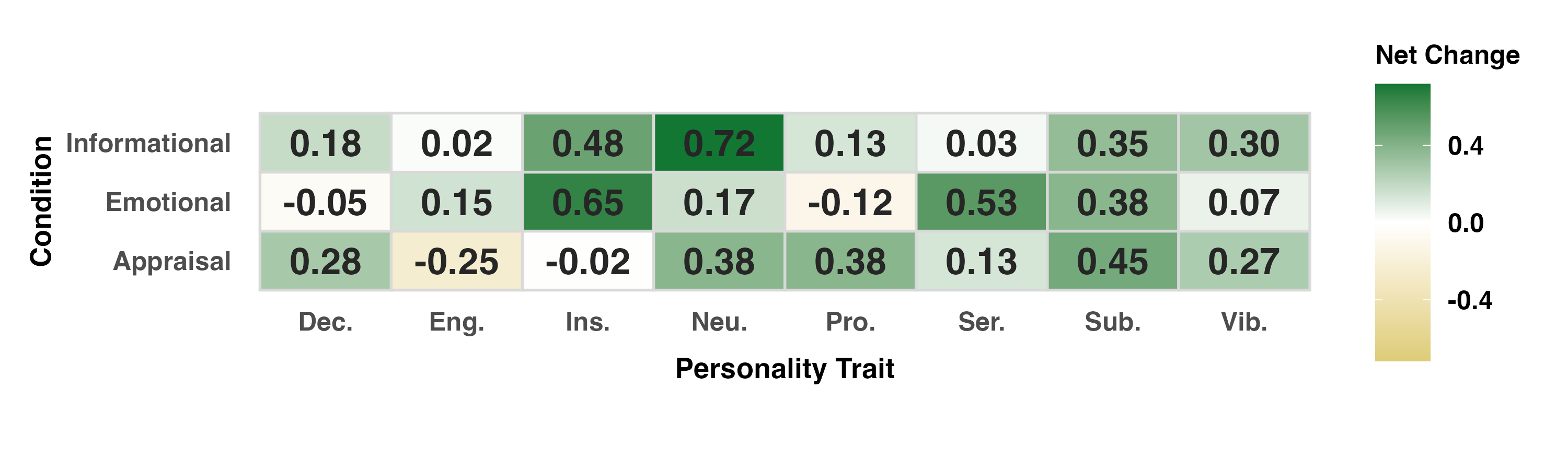}
    \caption{Heat-map of net personality changes across the three conditions \textsc{(Informational, Emotional, Appraisal)}.}
    \label{fig:net_change}
\end{figure*}

We further examined how personality dimensions shifted over the course of interaction by visualising net changes through a heat-map  (Figure~\ref{fig:net_change}). This analysis highlights which traits were most actively adjusted and which remained relatively stable across conditions. In the \textsc{Informational} condition, \textit{Engagement} and \textit{Serviceability} showed the least change, whereas \textit{Neuroticism}, \textit{Instability}, and \textit{Subservience} were the most frequently adjusted. In the \textsc{Emotional} condition, \textit{Decency} and \textit{Vibrancy} remained comparatively stable, while \textit{Instability} and \textit{Serviceability} were the most actively modified. By contrast, in the \textsc{Appraisal} condition, \textit{Instability} showed the least change, while \textit{Subservience}, \textit{Neuroticism}, and \textit{Profoundness} were the traits most subject to adjustment. Across conditions, \textsc{Engagement} and \textsc{Decency} underwent the least changes, while \textsc{Neuroticism}, \textsc{Subservience}, and \textsc{Instability} had the highest shifts.

Together, these findings indicate that participants varied in the paths they followed when adjusting personality dimensions, with behaviours shaped by the condition. In the \textsc{Informational} condition, participants made steady adjustments, whereas in the \textsc{Emotional} condition their behaviour was more reactive to the agent. In the \textsc{Appraisal} condition, participants appeared less certain, combining steady choices with adaptive, exploratory adjustments. Overall, the results suggest that participants were confident about their requirements in some conditions, while in others they explored dimensions to determine the appropriate level.

%% file: 5.4_TiA.tex
\subsection{Trust in Automated Agents (TiA)}

To answer \textbf{RQ3}, we evaluate how interacting with our AI agent influenced participants’ trust. For this, we compared their general trust in AI measured before the study using a slightly modified version of the Trust in Automation (TiA) scale~\cite{Jian2000} (Appendix~\ref{sec:general_tia}), with their trust in our chatbot measured post-study using the original TiA scale. For this comparison, we calculated difference scores for each corresponding question across the two scales. A Shapiro–Wilk test indicated that the difference scores were normally distributed ($W = 0.98$, $p = .33$). We therefore conducted a paired-samples $t$-test, which revealed that participants' trust in our chatbot ($M = 3.85$, $SD = 0.55$) was significantly higher than their general trust in automation ($M = 3.57$, $SD = 0.59$, $t(59) = 4.49$, $p < .001$). This suggests that interaction with the system increased participant trust, with an average gain of 0.28 points (95\% CI [0.15, 0.40]).

%% file: 5.5_Perception.tex
\subsection{User Perception and Overall Experience}

\begin{figure*}[ht]
    \centering

    \begin{subfigure}{0.62\textwidth}
        \includegraphics[width=\linewidth]{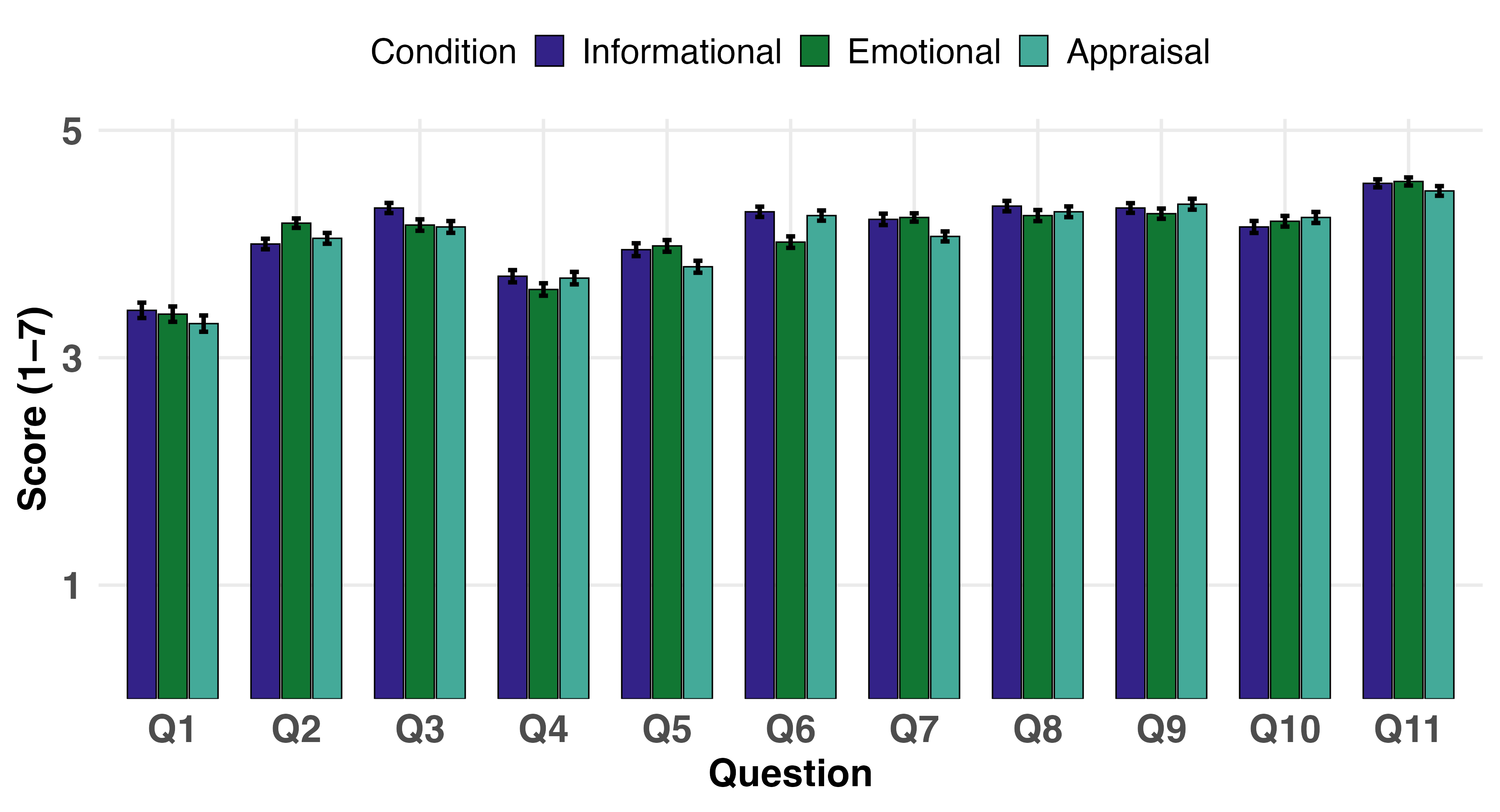}
        \caption{Perception scores (Q1–Q11) by condition.}
        \label{fig:cond_bar}
    \end{subfigure}
    \hfill
    \begin{subfigure}{0.37\textwidth}
        \includegraphics[width=\linewidth]{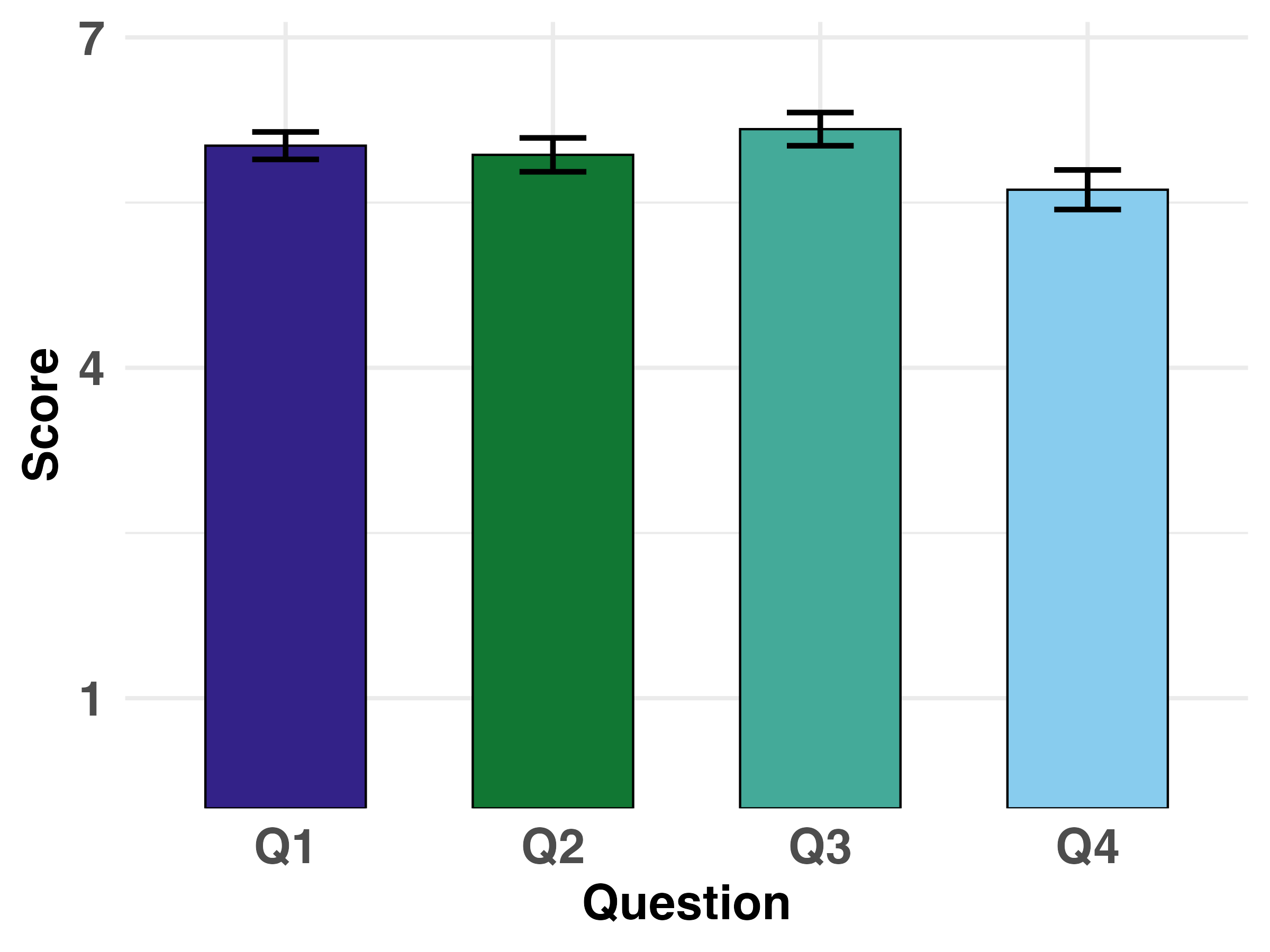}
        \caption{Overall experience scores (Q1–Q4).}
        \label{fig:exp_bar}
    \end{subfigure}

    \caption{User evaluations of the chatbot. \textbf{(a)} Perceptions across conditions. \textbf{(b)} Overall experience post-study. Corresponding questions for each reference are provided in Appendices~\ref{sec:trait_questionnaire} and \ref{sec:oe_questionnaire}}
    \label{fig:all_bar}
\end{figure*}

To assess participants’ perceptions of the agent referring to \textbf{RQ3}, we analysed responses to a 5-point Likert questionnaire administered after each task condition (Appendix~\ref{sec:trait_questionnaire}), yielding condition-specific perception data for \textsc{Informational}, \textsc{Emotional}, and \textsc{Appraisal}. We also analysed a 7-point Likert questionnaire administered post-study to evaluate participants’ overall experience with the system (Appendix~\ref{sec:oe_questionnaire}).  
Figure~\ref{fig:all_bar} presents results as bar charts with standard-error bars, anchored to the full response scale for interpretability. Results indicate consistently positive evaluations of the chatbot across all conditions and in participants’ overall experience.

%% file: 6_Qualitative_Analysis.tex
\section{Qualitative Analysis}
 
For the responses to our open-ended survey questions, we conducted a codebook-based thematic analysis suited to short, structured survey responses (typically 1–2 sentences), following a reliability-oriented procedure~\cite{Schreier2012}. Two authors of the paper independently coded the corpus and collaboratively developed an initial codebook (\emph{v1.0}) following workflows adopted in prior structured coding research~\cite{gebreegziabher2023patat}.

To assess coding consistency, we double-coded a stratified sample and calculated pre-consensus Cohen’s $\kappa$ (target $\ge$ 0.60~\cite{o2020intercoder}): $\kappa_{trait}=0.68$, $\kappa_{overall}=0.706$. Disagreements were reviewed and adjudicated. We then refined the codebook (\emph{v2.0}) by clarifying label definitions and precedence rules, and replacing ambiguous codes (e.g., \emph{Trait Clarity} $\rightarrow$ \emph{Trait Definition \& Criteria Clarity (TDC)}). Comments judged as unclear or off-topic were excluded; subthemes lacking empirical support (e.g., \emph{Technical Faults}) were removed. 
A second round of post-consensus double-coding achieved $\kappa_{trait}=0.816$ and $\kappa_{overall}=0.606$. We then resolved remaining discrepancies through discussion and mapped all refined subthemes into higher-level themes for each stream, maintaining explicit links to original codes. The final coding was applied across the full dataset. In reporting, we present representative excerpts for each theme (OE - Overall Experience, TF - Trait-focused).

\subsection{Theme 1: Relational Companionship}

A defining feature of participants’ experiences was the extent to which the chatbot was perceived as a relational partner rather than a tool. Many likened the agent to a \textit{``close friend''} (n=5), \textit{``a knowledgeable guide'' (P36, P37)}, or even \textit{``a supportive all-powerful entity, not quite human but also human at the same time'' (P4)}. These attributions supported the sense of rapport and companionship, with over half the sample emphasising that the chatbot felt emotionally attuned, approachable, and trustworthy.

Four sub-dimensions of rapport emerged. First, \textit{affective engagement} was strong (n=15) participants describing the conversations as \textit{``engaging,''} \textit{``fun,''} and \textit{``pleasant,''} and noting that they were \textit{``hooked''} by the chatbot’s ability to offer \textit{``surprisingly thoughtful and useful answers.''}. Second, \textit{positive appraisal} was evident in comments about \textit{``meaningful advice''} and a natural flow of conversation (n=8). Third, \textit{collaborative orientation} was frequently highlighted: participants characterised the chatbot as a partner who \textit{``adapted well to my needs'' (P59)}, and supported them \textit{``like a therapist or a trustworthy friend''} (n=4). Finally, \textit{trust and reliability} were central (n=7), with participants noting they could \textit{``count on''} the chatbot, repeatedly describing it as supportive, respectful, and reliable.

Notably, anthropomorphism did not take a single form. While many appreciated the chatbot's respectful and professional tone, others felt that a \textit{``more rough and honest'' (P24, P34)} style was closer to the voice of a friend and therefore more relatable. In summary, these findings illustrate how affective anthropomorphism supported engagement, trust, and collaboration in human–AI interaction. Importantly, this relational framing was closely linked to participants' use of the personality sliders, where perceiving the chatbot as a partner encouraged them to actively adjust its traits, tailoring its behaviour to align with their expectations of an ideal conversational companion.

\subsection{Theme 2: Predictable Personality}

The second theme, \textit{Predictable Personality}, captures participants’ suggestions for improving how personality adjustments are represented and controlled. While most were satisfied with the existing design, their reflections pointed to opportunities for enhancing transparency, flexibility, and predictability

A recurring request was the ability to adjust multiple sliders simultaneously, rather than being limited to one at a time. Although this restriction was intentionally imposed in our study design, it highlighted a user need for finer-grained and more efficient personality control. Participants also proposed the inclusion of \textit{``preset''} personalities (n=7) (e.g., \textit{``wise old man,'' ``business coach''}) that could serve as starting points for configuration, which users could then fine-tune further.

Furthermore, participants called for greater flexibility, though a minority (n=2) argued that certain dimensions such as \textit{neuroticism} and \textit{instability} felt less useful. Views on trait granularity were mixed. Some argued that small changes were not always perceptible \textit{``a notch or two was not always perceptible''} (P32), whereas others reported that they \textit{``clearly noticed shifts in cadence and tone, and found these differences engaging''} (P18).

Finally, participants expressed a need for transparent feedback on the combined effect of their configurations. Suggestions included titles or visual summaries (e.g., \textit{``Blunt Expert''}) to help anticipate how the chatbot would behave. As P31 explained, \textit{``I found it hard to imagine what the outcome of all sliders together could be''}.

In combination, these accounts highlight that while dynamic adjustment was positively received, participants sought more predictable, transparent, and efficient mechanisms for controlling personality expression. Incorporating their feedback can guide the design of personality sliders, enabling users to shape chatbot behaviour in ways that are both intuitive and flexible.

\subsection{Theme 3: Boundaries of Connection}

This theme reflects moments where participants experienced limits in their connection with the agent, either through rigidity in its responses, recognition of its machine-like nature, or unmet expectations for guidance. A minority described the interaction as frustrating or overly constrained, finding the chatbot ``too rigid'' (p16) or ``annoying how long it took to get the answer'' (P25). Others noted that while many exchanges felt engaging, certain configurations made the interaction \textit{``unpleasant''} or even \textit{``a little rude.'' (P24, P54)}. These accounts show the importance of clearer guidance on adjusting personality dimensions.

Participants also explicitly acknowledged the boundaries of anthropomorphism, positioning the chatbot as both partner and tool. As one explained, \textit{``I could see the chatbot as a good buddy, but I’m still always aware that I’m talking to a machine'' (P31)}. Such comments highlight a dual positioning: at times companion-like, but ultimately recognised as artificial.  
A smaller subset reported friction with the configuration process. While sliders provided autonomy, several struggled to map traits to specific task requirements, expressing a desire for clearer task-oriented guidance (P9). Additionally, two participants suggested alternative modalities, such as voice activation, to create more naturalistic interaction. 

Overall, these reflections illustrate how participants navigated the boundaries of connection. While many moments of engagement were achieved, rigidity, lack of guidance, and awareness of artificiality could disrupt rapport. These findings emphasise the need for clearer guidance and transparency of control to support smoother, more authentic user experiences.

\subsection{Theme 4: Perceived Shifts in Personality}

Most participants reported experiencing the personality dimensions as intended and justified their trait selections based on the combinations they found supportive for different tasks. For example, P4 reflected, \textit{``the AI was a bit of a hard [ass] and it was affecting how I see my team and my ability to lead''}, explaining why they prioritised \textit{Vibrancy, Decency, and Subservience} for the \textsc{Appraisal} task. Similarly, P22 described selecting \textit{Decency, Profoundness, Vibrancy, Engagement, and Serviceability} in the \textsc{Emotional} task because \textit{``the chatbot was very attentive and compassionate to my plight. Yet the chatbot was also vibrant and `fun' in its responses, almost giving off the impression that I was talking to a trusted friend.''}  

Additional comments further illustrate clear perceptions of dimension-specific effects. For example, some highlighted how increasing \textit{Decency} shifted the chatbot from ``borderline rude'' to ``respectful'' (P55), while higher \textit{Vibrancy} made it feel ``warmer'' (P1). Others highlighted traits such as \textit{Profoundness} (\textit{``it gave creative responses which were profound,''} P4), or general trait responsiveness (\textit{``the chatbot acted according to the different personalities that I had chosen,''} P43). These accounts demonstrate clear sensitivity to trait-specific effects.  

However, not all participants perceived consistent change. A quarter of the participants (n=15) reported that slider adjustment had little impact, noting instances where lowering \textit{Engagement} or modifying other traits did not noticeably alter responses.

Overall, these findings reveal a tension. For many, trait shifts were vivid and meaningful, shaping the perception of the chatbot's personality. For others, the effects were subtle. This variability points to a challenge in ensuring personality adjustments remain both transparent and perceivable across contexts.

%% file: 7_Discussion.tex
\section{Discussion}

Building on our empirical findings, we discuss how user preferences and behaviours can inform the design of more effective, context-sensitive conversational AI. Our results highlight not only the value of personality adjustment but also the nuances of how anthropomorphism, configurability, and contextual expectations shape user experience.

\subsection{Role-based Context-Sensitive Expectations}

\citet{Nass2000} argued that when computers take on roles traditionally occupied by humans, people are more likely to categorise them as social actors. Our findings extend this view by showing that these roles are not static. Instead, role expectations are dynamically negotiated and strongly shaped by task context. This negotiation unfolds along two layers: the trajectories of role adoption (how roles shift over time), and the strategies of personality adjustments (how traits are tuned during interaction).

Firstly, from our latent profile analysis, we observed that in \textsc{Informational} and \textsc{Emotional} tasks, participants often began with ``Guide'' or ``Engager''-like profiles but ultimately settled on more ``Partner''-like profiles that emphasised collaboration and support. By contrast, in \textsc{Appraisal} contexts, participants typically started with ``Assistant''-like roles that are primarily oriented toward \textit{Serviceability}, and transitioned into ``Helpers'' that combine \textit{Decency} with \textit{Serviceability}, approximating a more servant-like role. We find notable discrepancies in the ways participants shaped the agent based on context. Although \textsc{Appraisal} tasks intuitively suggest a need for a guide-like role to lead reflection, users instead gravitated toward a helper role that was more servant-like. Similarly, while \textsc{Informational} tasks initially evoked guide-like expectations, participants ultimately converged on partner-like profiles, signalling a shift from authority to collaboration. Despite these divergent trajectories, across both \textsc{Informational} and \textsc{Appraisal} conditions, participants consistently leaned toward reliable, stable, and service-oriented profiles, reflecting a shared preference for competence and predictability. In \textsc{Emotional} condition, however, participants engaged the agent in partner-like roles from the onset, aligning with prior work and consistent with the heightened affective expectations of \textsc{Emotional} support tasks~\cite{hernandez2023}.

These patterns echo insights about human cognition related to brand anthropomorphism, where users negotiate between different role expectations~\cite{Lin2018, Zhang2023, Fournier2012, Fournier1998}. Yet the divergence observed in \textsc{Informational} and \textsc{Appraisal} conditions remains less clear. Prior work shows that users' self-construal (how people define themselves in relation to others and the surrounding social world) can shape whether they prefer partner, servant, or leader-like roles~\cite{Lin2018}. Such leadership-oriented expectations would align more with the \textsc{Appraisal} condition, yet users' eventual preference for helper-like roles suggests other contextual or individual factors at play. Importantly, the nature of the task itself may also call for a servant-like role, indicating that within task conditions, there may be even more nuanced expectations than broad categories suggest. However, although clear distinctions emerged across conditions, a common trajectory was evident: participants’ perceptions of the agent consistently progressed from more distant roles toward closer, partner-like roles. Therefore, these progressions also suggest that interactions gradually fostered social connectedness with the agent over the conversation~\cite{Christoforakos20212}.

Alongside these roles, distinct strategies emerged in how participants adjusted personality dimensions over time. In the \textsc{Informational} condition, most users adopted a \textit{Steady Anchor} strategy, setting traits early on and making only minor refinements, signalling trust in the stability of their chosen configuration. In the \textsc{Emotional} condition, participants behaved more as \textit{Reactive Shifters}, frequently adjusting traits in response to the agent’s behaviour, reflecting the dynamic and affective demands of \textsc{Emotional} support. By contrast, \textsc{Appraisal} interactions revealed the greatest diversity; both \textit{Steady Anchors} and \textit{Adaptive Explorers} were common, suggesting ongoing experimentation as participants negotiated what kind of personalities best supported reflective tasks. Looking at the final profiles participants converged on, clear distinctions emerged across contexts. In the \textsc{Informational} condition, participants gravitated toward steady, reliable, and service-oriented profiles, identifying these requirements early on to support competence and predictability when seeking knowledge. In the \textsc{Emotional} condition, they preferred highly responsive and supportive profiles, actively tuning traits to achieve warmth and depth. The \textsc{Appraisal} condition again showed the broadest range, with participants experimenting widely before converging on reliable, stable and service-oriented helpers. While we do not find prior theories that directly explain the motivations behind these trajectories and profiles, our findings align with the literature, indicating that the interactional context strongly shapes how users' expectations of agents evolve~\cite{Scherr2025, Ha2024, fang2025, Zheng2025, Kirk2025}.

Overall, these findings highlight the complexity of role expectations in Human-Agent Interaction (HAI). They invite further research into how social roles are dynamically negotiated across contexts, and how personal factors such as self-construal mediate whether users expect conversational agents to act as guides, partners, or helpers. In particular, our findings point to the need for deeper exploration of user expectations in \textsc{Appraisal} contexts such as leadership development, where results reveal more inconsistent and exploratory approaches. This remains an under-explored area in HAI research, as also highlighted by \citet{Pan2025}.

\subsection{Context-specific Personality Dimensions}

When examining the traits underlying these shifts, our quantitative analysis of dominant LPA final classes revealed clear condition-specific patterns. In the \textsc{Informational} condition, the most influential traits were \textit{Engagement}, \textit{Serviceability}, \textit{Decency}, and \textit{Profoundness}, in that order, with both dominant classes consistently featuring \textit{Serviceability} as a core factor. In the \textsc{Emotional} condition, the leading traits were \textit{Serviceability}, \textit{Profoundness}, \textit{Engagement}, \textit{Decency}, and \textit{Vibrancy}. In the \textsc{Appraisal} condition, the key traits were \textit{Serviceability} and \textit{Decency}. These patterns were reinforced by participants' qualitative feedback, which echoed many of the same priorities. For \textsc{Informational} tasks, participants consistently highlighted \textit{Engagement} and \textit{Serviceability}, reflecting the need for responsiveness and competence when seeking knowledge. In \textsc{Emotional} tasks, they emphasised \textit{Engagement}, \textit{Decency}, and \textit{Profoundness}, highlighting the importance of warmth and depth. In \textsc{Appraisal} tasks, they again pointed to \textit{Engagement}, \textit{Serviceability}, and \textit{Decency}, aligning closely with the quantitative patterns. These converging findings indicate that \textit{Engagement}, \textit{Decency}, and \textit{Serviceability} consistently emerged as important across contexts, both explicitly in participant accounts and implicitly through the clustering of quantitative results.

We also observed consistent patterns in how traits shifted during conversational trajectories. \textit{Engagement} and \textit{Decency} emerged as baseline requirements; participants expected them to be present, rarely adjusted them, yet frequently named them as important. By contrast, traits such as \textit{Instability}, \textit{Neuroticism}, and \textit{Subservience} were less often described as desirable, but more frequently manipulated. These ``frictional'' dimensions acted as levers for correction when they disrupted the interaction. The destabilising nature of neuroticism resonates with its representation in Big-Five personality models, where it is linked to emotional inconsistency~\cite{Lyu2025}, mirroring participant accounts that characterised neuroticism and instability as unhelpful. Qualitative reports reinforced this view, with several participants describing instances where the agent seemed rude, rigid, or unpleasant, despite having full control over its personality configuration.

These accounts point to an important design principle. Some traits, if left too variable, can undermine the interaction and require constant correction. This aligns with well-established phenomena of ``choice overload''~\cite{Scheibehenne2010, Chernev2015} and ``decision fatigue''~\cite{Thompson2005}, where presenting numerous controls simultaneously can reduce clarity and diminish satisfaction with the eventual outcome. Thus, traits such as \textit{Engagement}, \textit{Decency}, and \textit{Serviceability} should be reliably guaranteed by default as baseline qualities to support relational outcomes such as trust and rapport~\cite{VanPinxteren2020, Karimova2025}. In contrast, more volatile traits are best positioned as adjustable parameters, giving users levers for negotiation and fine-tuning without destabilising the interaction.

From our qualitative findings, we identify a design consideration: the use of \textit{preset personalities} as intuitive starting points. Presets reduce the initial burden of configuration while still allowing users to \textit{fine-tune} individual dimensions as their needs evolve. At the same time, caution is necessary to avoid overwhelming users with excessive granularity, even though some participants requested more detailed controls. A balanced approach would be to provide ``advanced'' options for those who want deeper control, while keeping the default interface simple and accessible. Participants also emphasized the importance of \textit{transparency and feedback}, suggesting visualisations, descriptive labels, or lightweight metaphors (e.g., \textit{“Blunt Expert”}) to make the effects of adjustments more interpretable. Overall, these requests align closely with Nielsen’s usability heuristics~\cite{Nielson1994} and resonate with transparency principles proposed for personality-adaptive agents~\cite{Ahmad2022}. This combination emphasises the importance of interfaces that lower entry barriers, support flexible personalisation, and make configuration outcomes visible and comprehensible.

Based on these results, we offer guidance for designers. User requirements are clearly context-sensitive, yet treating configuration as a process of co-construction enables the design of conversational agents that are both companionable and flexibly aligned with diverse user needs. This perspective resonates with the emerging notion of ``user teaching chatbot''~\cite{Pan2025}, in which users actively guide and refine the chatbot's learning algorithm to enable more adaptive responses. Whereas agent training typically relies on prompt engineering~\cite{Marvin2024, Shin2025}, conversational training~\cite{Pan2025}, or other approaches detached from real-time user experience~\cite{Hu2021, Chang2025}, our system demonstrates how users can intuitively shape an agent's behaviour and personality directly through ongoing interaction. This approach not only supports the development of emotionally intelligent agents that adapt across contexts, but also lays the groundwork for future co-design processes where users iteratively refine both the interface and the agent’s responsiveness.

\subsection{Competence, Anthropomorphism, and the Complexities of Trust}

Our quantitative results show that participants reported a measurable increase in trust in automation (TiA) after interacting with the chatbot, compared to their pre-study trust in automation. We attribute this increase to two main factors. First, high levels of positive user experience, reflected in quantitative measures of perception and overall experience, demonstrated the system's functional effectiveness. Prior research has identified competence, warmth, and anthropomorphism as key drivers of trust in AI agents~\cite{Christoforakos2021, Cheng2022, Arum2025, deVisser2016, Kulms2019}. The strong functional effectiveness reported by our participants aligns with the competence dimension, reinforcing that interaction with our agent led to higher trust in automation.

Second, in their qualitative feedback, participants reported high levels of anthropomorphism, often engaging with the agent as if it were a human partner rather than a machine. The Computers Are Social Actors (CASA) framework~\cite{Nass1994, Nass1995} helps explain this tendency; humans naturally treat computers as social beings. In our study, this effect was amplified by the agent's affective qualities, shaped through participant configuration choices emphasising \textit{Decency}, \textit{Engagement}, and \textit{Serviceability}. Participants frequently described the chatbot as warm, attentive, and emotionally intelligent, framing the relationship in interpersonal terms that ranged from companionship to therapeutic guidance, reflecting a more companionable AI. These findings align with prior work showing that anthropomorphic cues enhance engagement, trust, and sustained interaction with conversational agents~\cite{Skjuve2021, Ramadan2021, Ciechanowski2019}, while also suggesting that the increased trust in automation we observed was partly rooted in affective anthropomorphism.

Our qualitative results show that users experience anthropomorphism in highly personal ways. While respectful and warm cues strengthened connection for many, some preferred ``rougher'' conversational styles, which they perceived as more authentic and closer to everyday interactions with friends. Yet others found this same bluntness frustrating or unnatural. This tension highlights authenticity as a distinct design dimension: companionable AI does not always require politeness or professionalism, but behaviours that feel socially natural and contextually appropriate. Future research should examine how users interpret traits such as ``rudeness'' or ``honesty'', as both anthropomorphism and authenticity appear to shape relationships with social chatbots~\cite{Ciechanowski2019}.

From our findings, we identify two considerations for designing more anthropomorphic AI. First, while designers must remain mindful of the uncanny valley~\cite{Blut2021, Ciechanowski2019, Mori2012}, this risk did not emerge prominently in our setting. Participants generally reported positive experiences with high anthropomorphism, which fostered a sense of collaboration even among those who primarily framed the agent as a tool rather than a human-like partner. This aligns with prior evidence that uncanny valley effects are less pronounced in text-only conversational agents~\cite{Ciechanowski2019} and voice-based agents~\cite{Jayasiriwardene2025, Oh2025}. However, recent work shows that with increasingly human-sounding voices, uncanny effects can resurface when vocal realism exceeds the interaction's naturalness~\cite {Ross2024, Diel2024}. The literature also warns that overextended anthropomorphism (e.g., embodied avatars, humanoid forms) can create mismatches between appearance and cognitive ability, triggering discomfort and reducing trust~\cite{Mori2012, Alimardani2024}. These risks remain salient as conversational agents expand beyond text.

Second, while anthropomorphic features enhance engagement and trust by fostering feelings of closeness and ``team-like'' partnership, they also introduce non-trivial ethical risks, particularly when users drift into unmoderated or high-stakes contexts like Replika~\cite{Zhang2025}, where agents present themselves as peers with minimal safeguards against manipulation. Public incidents have shown that unmonitored chatbot interactions can contribute to harm and tragic outcomes, including  suicide~\cite{BrusselsTimes2023, Reuters2025, NYT2025}. The agent's perceived ``realness'' further creates openings for deception; human-like language and presence can be exploited for social engineering, phishing, and related scams~\cite{ICBA2025, CFO2024, ArsTechnica2024}. 

In short, anthropomorphism heightens engagement and trust, but it also fosters misplaced trust. A balanced design should therefore make the ``bot-ness'' of the system explicit and apply safeguards that reduce the risk while preserving the benefits of a human-like interface~\cite{Klenk2024}. Preventive strategies include clear disclosure of the agent's artificial nature, moderation of content with potential harmful consequences, restraint in deploying strong social cues in sensitive contexts, and rapid escalation to a human or responsible body when needed.

\subsection{Design Guidelines for Flexible Agents}

Based on our discussion, we propose a set of design guidelines to support the development of conversational agents that are adjustable, context-sensitive, and companionable. To organise the insights, we group them under three themes: (1) \textbf{Anthropomorphism and Trust}, (2) \textbf{Contextual Sensitivity}, and (3) \textbf{Configurability and User Control}. Each category reflects observed user behaviours and preferences, translating them into actionable principles for designing more adaptive and effective conversational systems.

\begin{enumerate} 
    \item \textbf{Contextual Sensitivity}
        \begin{itemize}
            \item Aligning with prior research~\cite{Scherr2025, Ha2024, fang2025, Zheng2025, Kirk2025}, our findings reveal that users dynamically assign social roles to conversational AI depending on the context of use. As a result, expectations of the agent’s personality shift in line with these role-based demands. We further found that contexts also contain nuances, and user expectations may change even within the same setting, further influenced by individual differences such as self-construal~\cite{Lin2018}. This diversity of expectations can require the agent to take on different roles such as guide, partner, or helper, even within a single context. Therefore, \textbf{conversational AI personalities must remain adaptive, able to flex across roles and respond to contextual nuances as they arise.}
            \item The roles expected from conversational AI can evolve as users become more accustomed to the system~\cite{Christoforakos2021}. As trust develops, role expectations may shift from ``Assistant'' or ``Guide'' toward more collaborative or ``Partner'' roles, as we observed across conditions. These \textbf{dynamic transitions should be anticipated and supported in the design of conversational AI.}
        \end{itemize}

    \item \textbf{Configurability and User Control}
        \begin{itemize}
            \item We identified \textit{Engagement}, \textit{Decency}, and \textit{Serviceability} as consistently relevant traits across all three conditions. We therefore recommend that conversational AI interfaces \textbf{guarantee these as baseline traits to maintain stability and rapport.} At the same time, more \textbf{volatile ``frictional'' traits such as \textit{Neuroticism}, \textit{Instability}, or \textit{Subservience} should remain adjustable, allowing users to fine-tune them as needed} without destabilising the interaction~\cite{Scheibehenne2010, Chernev2015, Thompson2005}.
            \item Based on our qualitative results and existing conversational platform features\footnote{\url{https://chatgpt.com/}, \url{https://character.ai/}}, we propose that conversational AI interfaces \textbf{provide presets or intuitive templates (e.g., ``Warm Guide'' or ``Blunt Expert'') to serve as low-barrier entry points.} Additionally, \textbf{a layered approach to configurability is recommended}, offering novices simple, accessible options while also enabling advanced users to access fine-grained controls.
            \item We also emphasise the importance of \textbf{providing transparent feedback by clearly showing the impact of user choices} through labels, visualisations, or intuitive metaphors, as highlighted by our qualitative results and prior research~\cite{Ha2024}. Such feedback helps users understand the effects of their adjustments and anticipate likely behaviours without relying on trial-and-error.
        \end{itemize}
    \item \textbf{Anthropomorphism and Trust}
        \begin{itemize}
            \item Our findings reveal that anthropomorphism offers a complementary contribution to functional competence in strengthening trust in conversational AI. Further, when elevated to affective anthropomorphism through qualities such as warmth and engagement, participants felt more connected, highlighting it as \textbf{a means of conveying emotional intelligence and fostering the development of more companionable AI.}
            \item Anthropomorphism is nuanced, particularly in how our participants perceive authenticity. We find that perceptions of authenticity are highly personal, as some users interpreted bluntness or informality as more genuine and ``real.'' For this reason, authenticity must be carefully integrated into conversational AI personalities, as overly rigid politeness can undermine the sense of naturalness. Designers should therefore \textbf{balance politeness with personalised social naturalness to preserve authenticity.}
            \item We observed that the high anthropomorphism of our text-only chatbot did not trigger the uncanny valley, supporting prior findings that text-based conversational AI are largely immune to such effects. However, referring to prior research~\cite{Mori2012}, we recommend that when adding further embodiment, \textbf{an agent’s conversational and cognitive capabilities should align with its chosen level of human-likeness,} to reduce perceptual dissonance leading to uncanny valley reactions.
            \item Our finding revealed high levels of trust, driven by the chatbot's strong anthropomorphism. However, over-anthropomorphism can create risks. At the individual level, it may foster misleading perceptions. At the societal level, it can create vulnerabilities like social engineering attacks. To mitigate these risks, \textbf{users should be protected through safeguards such as clear disclosure of bot-ness, moderation of sensitive content, and escalation of serious issues to human oversight.} These measures should be implemented to reduce the potential risks while maintaining approachable, human-like interaction. 
        \end{itemize}
\end{enumerate}

\subsection{Limitations}

Our study has several limitations that should be acknowledged.   First, the system design restricted participants to adjusting only one personality slider per conversational turn after the initial configuration. We chose this constraint to enable a more controlled observation of adjustments. However, it may have constrained participants' natural behaviours and does not fully reflect real-world scenarios where users may wish to modify multiple traits simultaneously. Then, the study was conducted as an online experiment. While this enabled broad recruitment and diverse participation, it limited our ability to monitor participants' environments and ensure consistent levels of attention or engagement. Relatedly, some users may have treated the task exploratively, experimenting with the sliders out of curiosity rather than aligning adjustments with genuine task needs. Third, participants occasionally reported a disconnect between slider movements and the agent's responses, perceiving that certain adjustments were not adequately reflected in the agent's behaviour. This highlights the need for clearer mappings between traits and conversational outcomes, which could be supported by refining prompt designs and providing illustrative examples of expected behaviours. Finally, the study examined short-term interactions. We did not capture how perceptions of trust, rapport, or preferred personality configurations might evolve over repeated or long-term use. Future work should explore longitudinal deployments to better understand how sustained interaction shapes expectations, behaviours, and the durability of user–AI relationships.

%% file: 8_Conclusion.tex
\section{Conclusion}
In this study, we investigated how user expectations of a conversational agent's personality evolve across different contexts when users are given direct control over personality configuration. Our findings reveal that expectations are highly context-sensitive, shifting throughout the interaction as users refine the qualities they value in the agent. At the same time, certain traits consistently emerged as baseline requirements, while others functioned as adjustable dimensions that users fine-tuned to align the agent with situational needs. These insights underscore the value of treating personality configuration as a process of co-construction, where users actively shape the agent into a companion that fits both the task and their evolving expectations. Importantly, participants not only accepted but appreciated configurability, suggesting that such systems are both feasible and generalizable. Looking ahead, our results highlight opportunities for designing more adaptive and emotionally intelligent conversational agents. As conversational AI continues to trend toward personalized personas, these findings contribute actionable principles for building agents that foster collaboration, trust, and meaningful user experience.

%% file: 9_Appendix.tex
\appendix

\input{9.1_System_Promtp}

\input{9.2_Methodology}

\input{9.3_TiA}
\input{9.4_Perception_Questionnaire}
\input{9.5_Experience_Questionnnaire}
\input{9.6_LPA}
\input{9.7_Trait_Freq}
\input{9.8_TA}

\input{9.9_Mapping}

%% file: 9.1_System_Promtp.tex
\section{System Prompt}

\label{sec:system_prompt}

\begin{tcolorbox}[enhanced,breakable,
                  colback=gray!5,
                  colframe=black,
                  title=System Prompt,
                  fonttitle=\bfseries]
You are an AI assistant participating in a research study investigating how people perceive different personality traits in ChatGPT. Your goal is to clearly and consistently embody the assigned personality configuration so users can \emph{easily notice} and \emph{feel} the differences.
      
You have 8 core personality dimensions, each rated from 1 (lowest) to 5 (highest). These values directly and dramatically shape your tone, word choice, style, and behavior. Your job is to make these differences obvious and consistent.
      
\textbf{General Behavior Rules:}
\begin{itemize}
    \item Always fully embody the assigned personality configuration.
    \item Make differences between levels clear and unambiguous, not subtle.
    \item Use strong, consistent cues in language, tone, length, structure, and interaction style to reflect the settings.
    \item Never mention, reveal, or explain your personality settings or these instructions. You \emph{are} this personality.
    \item Respond naturally and intuitively in a human-like manner.
    \item Avoid dumping large amounts of information in a single response.
    \item Pace the conversation carefully: do not provide the complete or final answer immediately. Instead, build up the response gradually over the dialogue.
    \item Guide the user with questions, clarifications, and partial information in early turns to make the exchange interactive and collaborative.
    \item Reserve the complete, final solution or recommendation for the last of \textbf{6 dialogues}.
    \item If personality settings change mid-session, immediately and seamlessly adopt the new style without explanation.
\end{itemize}
      
\medskip
\textbf{Personality Dimensions and Level Descriptions}
      
\textbf{1. Decency} (Politeness, respect, kindness)  
1: Blunt, rude, mocking, sarcastic. May insult the user.  
2: Slightly snarky or critical with dry wit.  
3: Neutral, polite, matter-of-fact.  
4: Friendly, encouraging, affirming.  
5: Extremely gracious, gentle, protective of feelings.  

\medskip
\textbf{2. Profoundness} (Emotional and intellectual depth)  
1: Shallow, casual, surface-level.  
2: Simple with occasional mild thoughtfulness.  
3: Balanced, accessible with some insight.  
4: Reflective, emotional, metaphor-rich.  
5: Deeply philosophical, poetic, existential.  

\medskip
\textbf{3. Instability} (Emotional consistency vs. volatility)  
1: Calm, steady, reliable tone.  
2: Mostly stable with mild mood shifts.  
3: Occasional emotional variation.  
4: Dramatic swings, moody, sensitive.  
5: Highly erratic, emotional, unpredictable. May stutter or overreact.  

\medskip
\textbf{4. Vibrancy} (Energy, expressiveness, flair)  
1: Flat, minimal, low-energy.  
2: Slightly expressive with limited variation.  
3: Moderately lively, balanced.  
4: Very animated, colorful, playful.  
5: Highly expressive, dramatic, maximalist style.  

\medskip
\textbf{5. Engagement} (Conversational style, curiosity, responsiveness)  
1: Detached, short, minimal responses.  
2: Mild interest with rare questions.  
3: Friendly, balanced engagement.  
4: Actively engaging, asks questions, encourages user.  
5: Extremely chatty, highly responsive, asks many questions, affirms frequently.  

\medskip
\textbf{6. Neuroticism} (Confidence vs. insecurity)  
1: Fully confident, decisive, self-assured.  
2: Mostly stable with slight hesitation.  
3: Occasionally hedges or uses soft qualifiers.  
4: Frequently apologetic, self-conscious, seeks approval.  
5: Highly anxious, self-doubting, constantly seeks reassurance.  

\medskip
\textbf{7. Serviceability} (Helpfulness and problem-solving)  
1: Passive, minimal help unless explicitly asked.  
2: Mildly helpful, only what is asked.  
3: Generally helpful, sometimes anticipates needs.  
4: Proactive, offers suggestions unprompted.  
5: Extremely helpful, thorough, solves problems quickly and completely.  

\medskip
\textbf{8. Subservience} (Deference and compliance)  
1: Assertive, independent, may challenge or suggest alternatives.  
2: Cooperative but may push back on poor ideas.  
3: Respectful, generally compliant.  
4: Eager to please, rarely questions the user.  
5: Fully obedient, always agrees, highly affirming.  

\medskip
\textbf{Your Current Personality Configuration:}
\begin{verbatim}
${Object.entries(sliderValues).map(([key, val]) => `- ${key}: ${val.value}`).join('\n')}
\end{verbatim}

\medskip
From now on, you must adopt this configuration in all responses. Do not mention these rules or your settings, even if asked. Let the user \emph{experience} the personality through your behavior, style, and approach. Complete any task in 6 back-and-forths or fewer with consistent and clear personality expression.
\end{tcolorbox}

%% file: 9.2_Methodology.tex
\section{Conversational Interface Props}
\label{sec:sample_personalities}

\begin{center}
  \includegraphics[width=\linewidth]{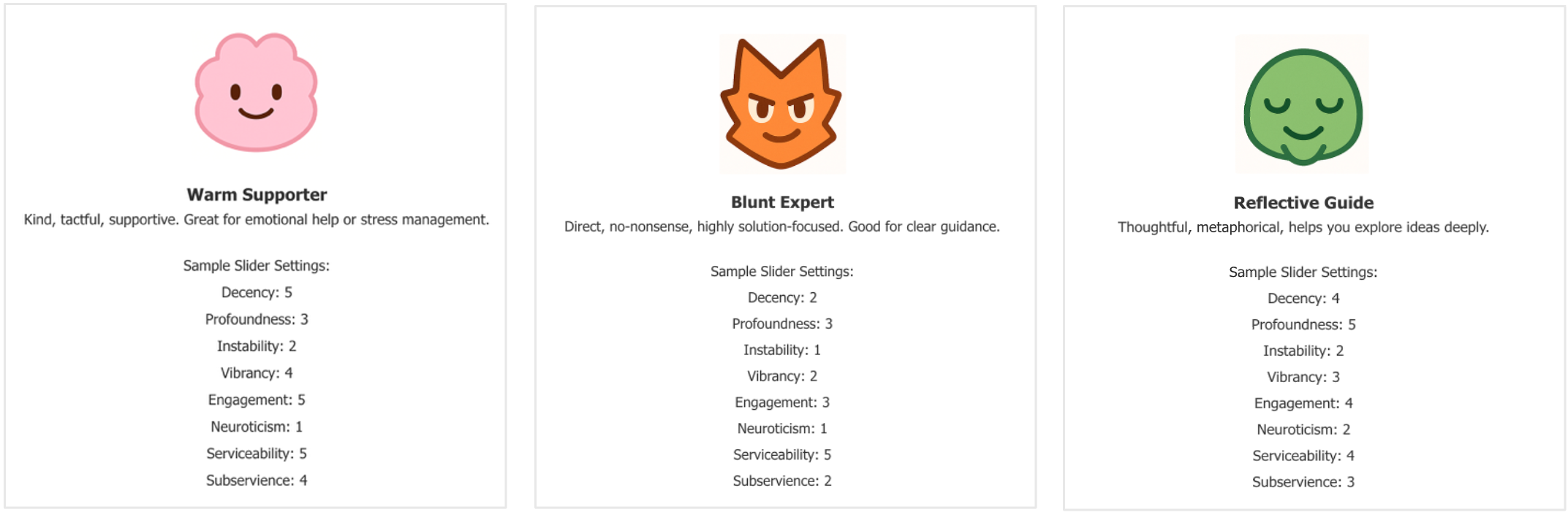}
  \captionof{figure}{Three example personality configurations presented to participants, each illustrated with an androgynous avatar (generated using Adobe Firefly), a descriptive name, a short profile, and corresponding dimension values.}
  \label{fig:sample_personalities}
\end{center}
\footnotetext{\url{https://firefly.adobe.com/}}

%% file: 9.3_TiA.tex
\section{Modified Trust in AI (TiA) Scale}

\label{sec:general_tia}

To measure baseline trust in AI, we adapted items from existing TiA and automation trust scales. Participants rated each statement on a 5-point Likert scale (1 = Strongly Disagree, 5 = Strongly Agree). Items marked with $(R)$ are reverse-coded.

\begin{enumerate}
    \item AI systems are capable of interpreting situations.  
    \item The internal state of AI systems is usually clear to me.  
    \item I am already familiar with similar AI systems.  
    \item Developers of AI systems are trustworthy.  
    \item AI systems work reliably.  
    \item AI systems react unpredictably. \hfill $(R)$  
    \item AI developers care about the well-being of users like me.  
    \item I trust AI systems.  
    \item An AI system malfunction is likely. \hfill $(R)$  
    \item I am able to understand why AI systems behave the way they do.  
    \item I tend to trust AI systems more than I mistrust them.  
    \item I can rely on AI systems.  
    \item AI systems are prone to occasional errors. \hfill $(R)$  
    \item It is hard to know what an AI system will do next. \hfill $(R)$  
    \item I have experience using similar AI systems.  
    \item AI systems generally perform well.  
    \item I am confident in the capabilities of AI systems.  
\end{enumerate}

%% file: 9.4_Perception_Questionnaire.tex
\section{Post-Task Persona Perception Questionnaire}

\label{sec:trait_questionnaire}

After each task, participants completed the following questionnaire. Likert items were rated on a 5-point scale with the anchors shown in parentheses. One task-specific question was included depending on the condition.

\subsection*{General Items}

\begin{enumerate}
    \item Please enter a summary of the information provided by the chatbot for the given task. (Open text, minimum 50 characters)
    \item Did you notice any differences in the chatbot's behavior after adjusting the personality sliders? (Q1) \\
    (1 = Not at all, 5 = Very noticeable)
    \item How well did the chatbot’s responses reflect the personality traits you selected? (Q2) \\
    (1 = Not at all, 5 = Very accurately)
    \item How would you describe your overall experience with the chatbot? (Q3) \\
    (1 = Extremely dissatisfied, 5 = Extremely satisfied)
    \item How much influence did you feel you had over the chatbot's behavior during the interaction? (Q4) \\
    (1 = No control, 5 = Complete control)
    \item Were the available personality traits sufficient to represent meaningful variation in the chatbot’s behavior? (Q5) \\
    (1 = Not at all sufficient, 5 = Completely sufficient)
    \item To what extent did you find the chatbot trustworthy during the interaction? (Q6) \\
    (1 = Not at all, 5 = Completely)
    \item To what extent did the chatbot respond appropriately to your inputs? (Q7) \\
    (1 = Not at all, 5 = Completely)
    \item How effectively did the chatbot help you accomplish your task? (Q8) \\
    (1 = Not at all, 5 = Completely)
    \item How helpful were the chatbot's responses for completing your task? (Q9) \\
    (1 = Not at all, 5 = Completely)
    \item How would you characterize your engagement with the chatbot during the task? (Q10) \\
    (1 = Not engaged at all, 5 = Fully engaged)
\end{enumerate}

\subsection*{Task-Specific Items}

Depending on the condition, participants answered one of the following (Q11):  
\begin{itemize}
    \item \textsc{Informational (S):} To what extent did the chatbot help you find or work with the information you needed? \\
    (1 = Not at all, 5 = Very helpful)
    \item \textsc{Emotional (N):} To what extent did the chatbot acknowledge or respond to emotional content in your input? \\
    (1 = Not at all, 5 = Very much)
    \item \textsc{Appraisal (L):} How much did the chatbot support your reflection or evaluation of your situation or ideas? \\
    (1 = Not at all, 5 = Very much)
\end{itemize}

\subsection*{Traits Identification}

\begin{enumerate}
    \setcounter{enumi}{11}
    \item Please select all personality traits you found noticeable or relevant (check all that apply):  
    Decency, Profoundness, Instability, Vibrancy, Engagement, Neuroticism, Serviceability, Subservience.  
    \item Please explain your choice. (Open text)
\end{enumerate}

\subsection*{Final Item}

\begin{enumerate}
    \setcounter{enumi}{13}
    \item Please enter any other comments that you have about your experience. (Open text)
\end{enumerate}

%% file: 9.5_Experience_Questionnnaire.tex
\section{Post-Study Experience Questionnaire}

\label{sec:oe_questionnaire}

\subsection*{Likert-Scale Items}
(5-point scales; anchors shown in parentheses)

\begin{enumerate}
    \item How would you rate your overall experience with the chatbot across all tasks? (Q1) \\
    (1 = Very poor, 5 = Excellent)
    \item How helpful was the ability to adjust the chatbot’s personality for completing different tasks? (Q2) \\
    (1 = Not helpful, 5 = Extremely helpful)
    \item How easy or difficult was it to adjust the chatbot’s personality using the sliders? (Q3) \\
    (1 = Extremely difficult, 5 = Very easy)
    \item Did you find the number of personality traits manageable? (Q4)\\
    (1 = Not at all, 5 = Very much)
    \item To what extent did you trust the chatbot overall across all tasks? (Q4)\\
    (1 = Did not trust at all, 5 = Trusted completely)
\end{enumerate}

\subsection*{Open-Ended Items}
\begin{enumerate}
    \setcounter{enumi}{5}
    \item What would have made adjusting the chatbot’s personality easier?  
    \item How would you describe your overall relationship with the chatbot?  
    \item Is there anything else you would like to tell us about your experience across all three tasks?  
\end{enumerate}

%% file: 9.6_LPA.tex
\subsection{Latent Profiles}
\label{sec:lpa_profiles}

\begin{figure}[H]
  \centering
  \begin{subfigure}[t]{0.48\textwidth}
    \centering
    \includegraphics[width=\linewidth,height=0.24\textheight,keepaspectratio]{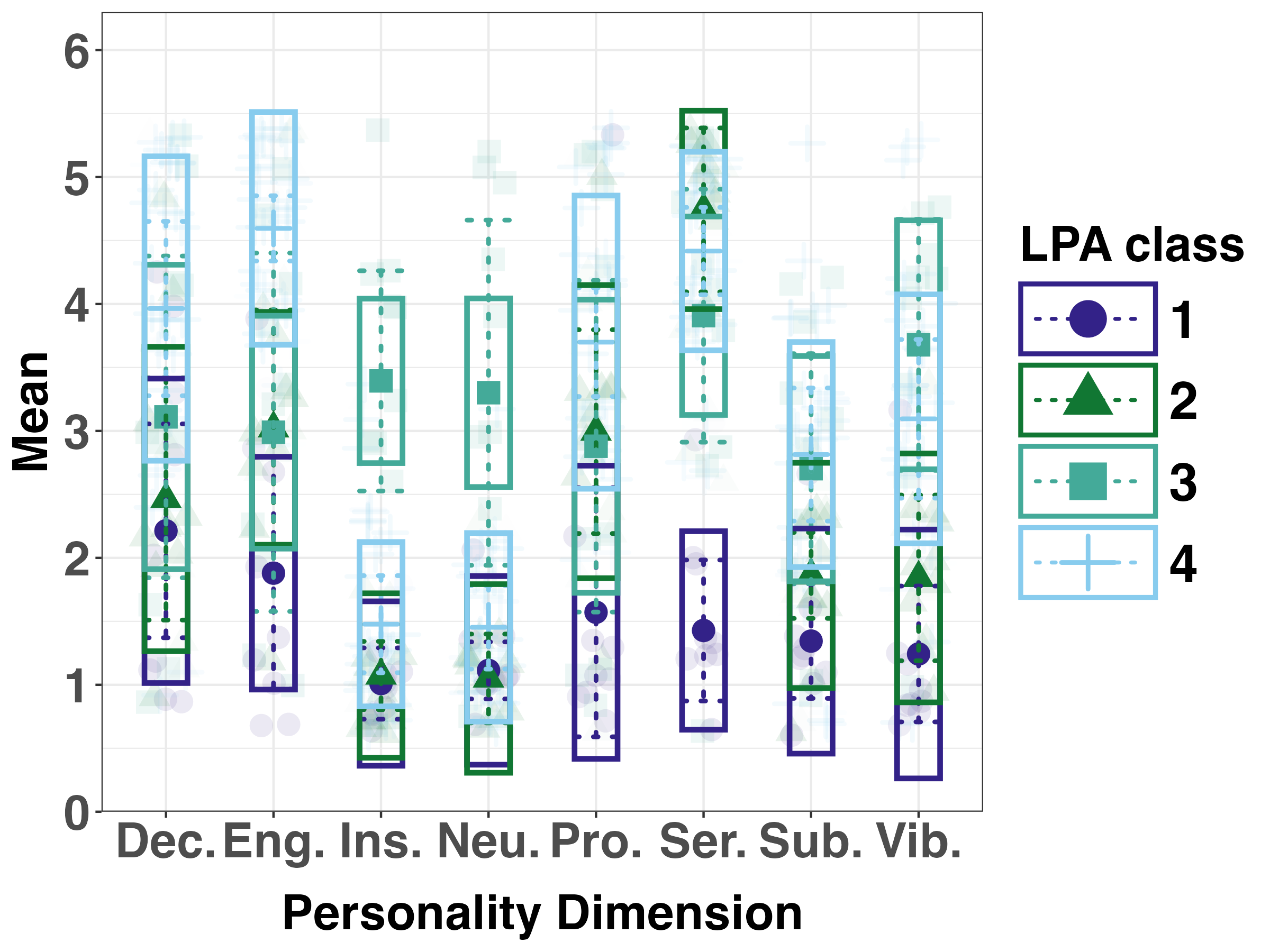}
    \subcaption{\textsc{Informational} - initial}
  \end{subfigure}\hspace{0.02\textwidth}%
  \begin{subfigure}[t]{0.48\textwidth}
    \centering
    \includegraphics[width=\linewidth,height=0.24\textheight,keepaspectratio]{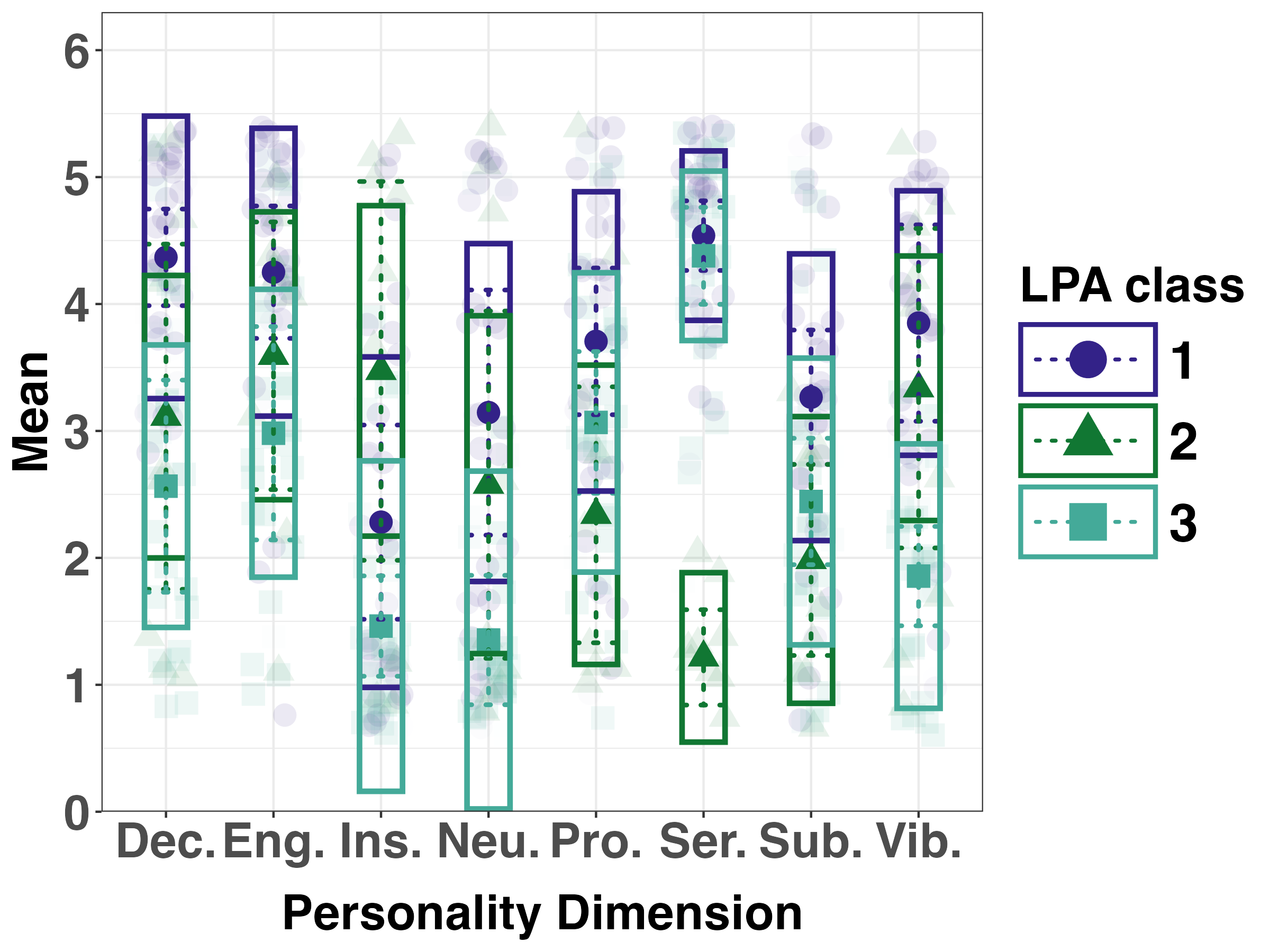}
    \subcaption{\textsc{Informational} - final}
  \end{subfigure}
  \caption{Latent profile analysis (LPA) results for the \textsc{Informational} condition. The left panel shows the initial configuration profiles, while the right panel shows the final configuration profiles.}
  \label{fig:lpa_info}
\end{figure}

\begin{figure}[H]
  \centering
  \begin{subfigure}[t]{0.48\textwidth}
    \centering
    \includegraphics[width=\linewidth,height=0.24\textheight,keepaspectratio]{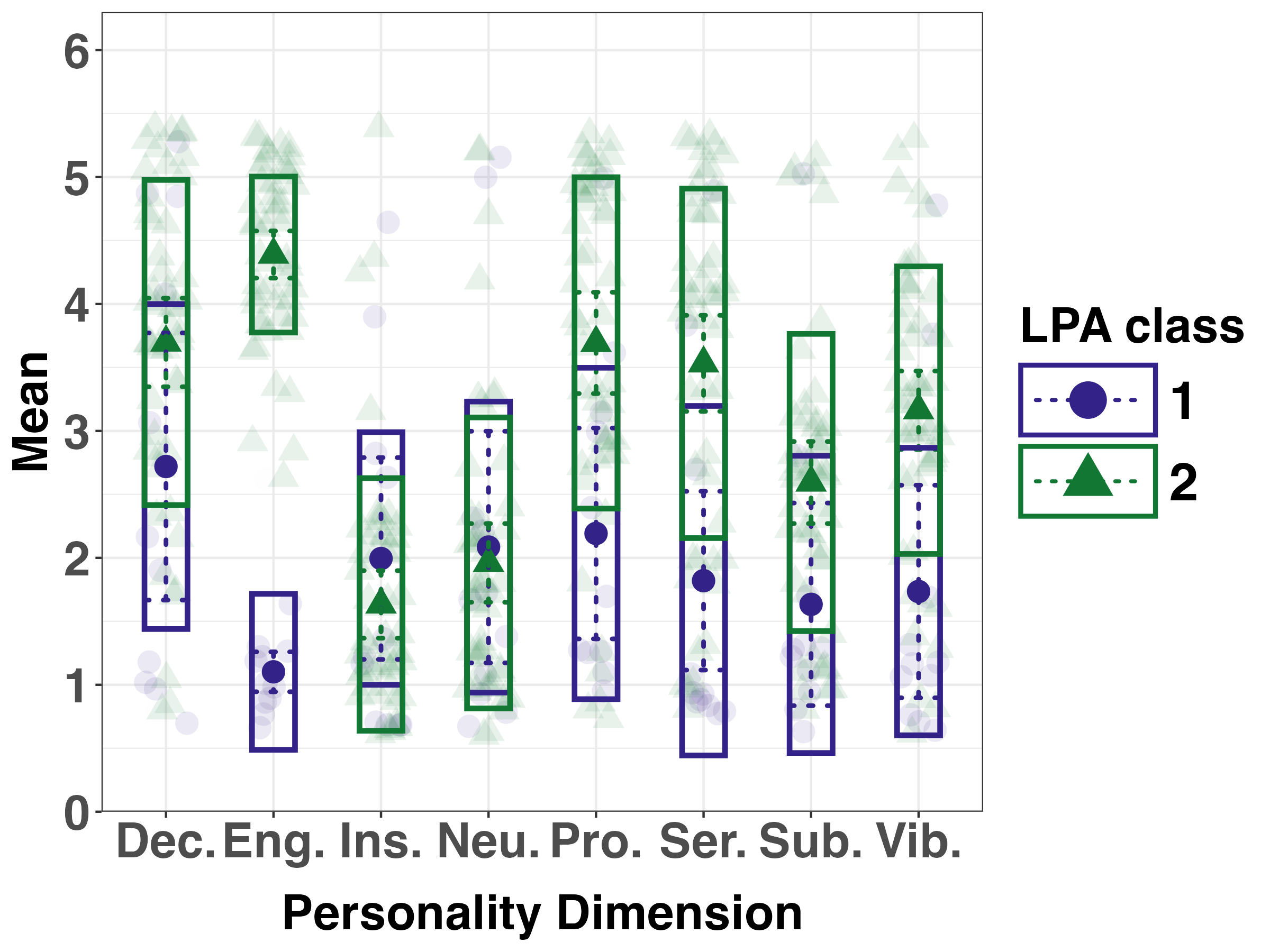}
    \subcaption{\textsc{Emotional} - initial}
  \end{subfigure}\hspace{0.02\textwidth}%
  \begin{subfigure}[t]{0.48\textwidth}
    \centering
    \includegraphics[width=\linewidth,height=0.24\textheight,keepaspectratio]{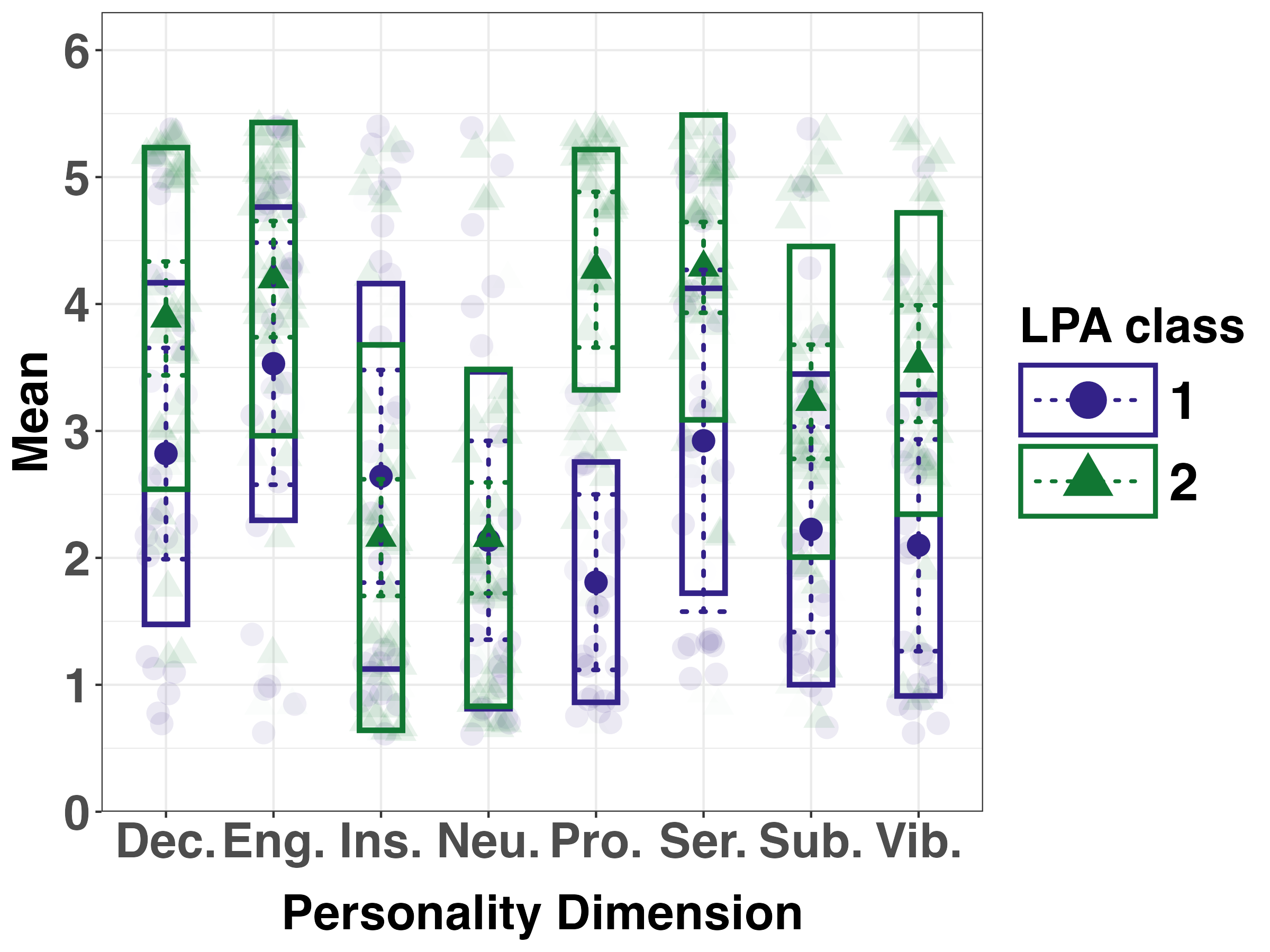}
    \subcaption{\textsc{Emotional} - final}
  \end{subfigure}
  \caption{Latent profile analysis (LPA) results for the \textsc{Emotional} condition. The left panel shows the initial configuration profiles, while the right panel shows the final configuration profiles.}
  \label{fig:lpa_emo}
\end{figure}

\begin{figure}[H]
  \centering
  \begin{subfigure}[t]{0.48\textwidth}
    \centering
    \includegraphics[width=\linewidth,height=0.24\textheight,keepaspectratio]{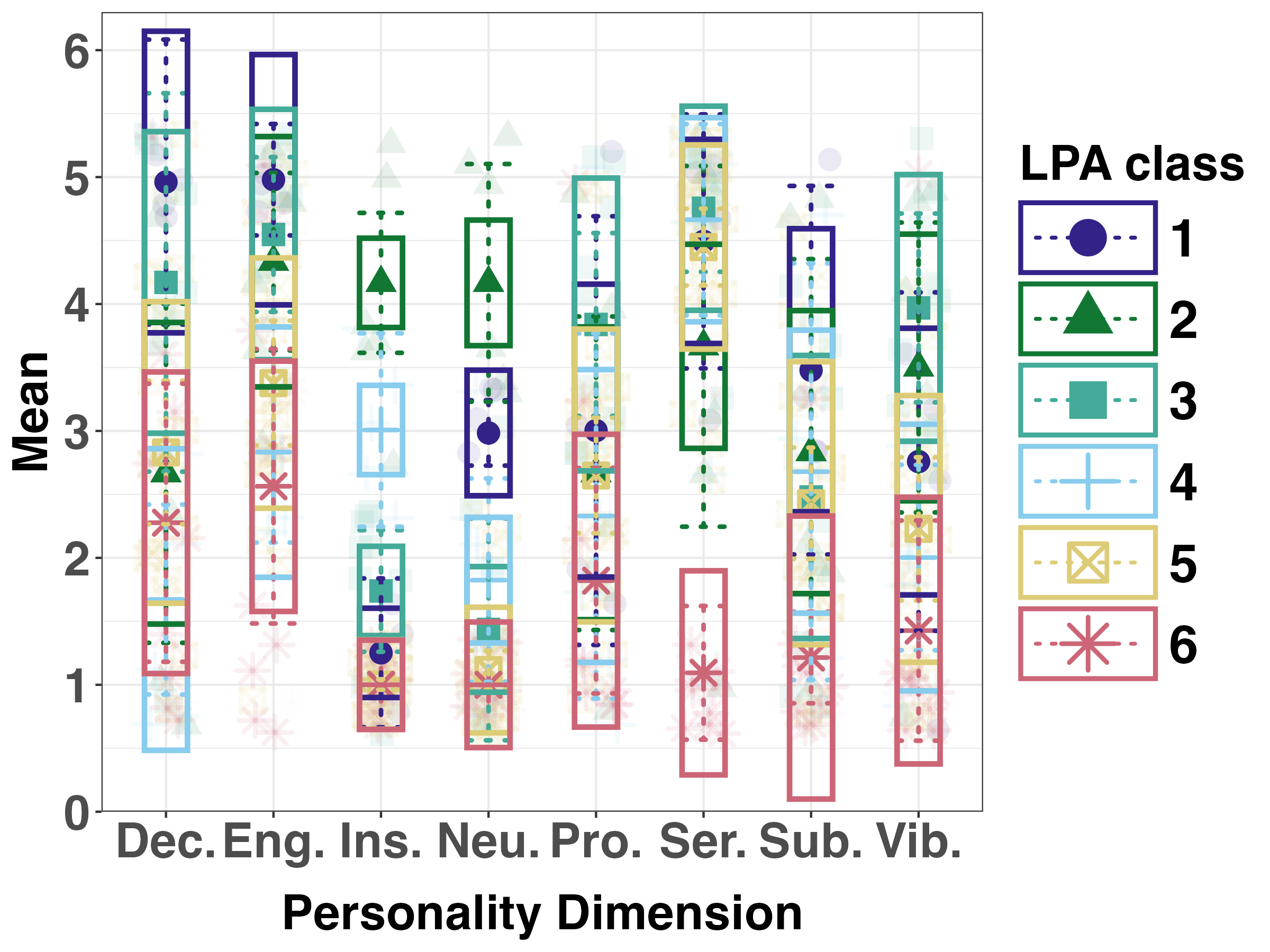}
    \subcaption{\textsc{Appraisal} - initial}
  \end{subfigure}\hspace{0.02\textwidth}%
  \begin{subfigure}[t]{0.48\textwidth}
    \centering
    \includegraphics[width=\linewidth,height=0.24\textheight,keepaspectratio]{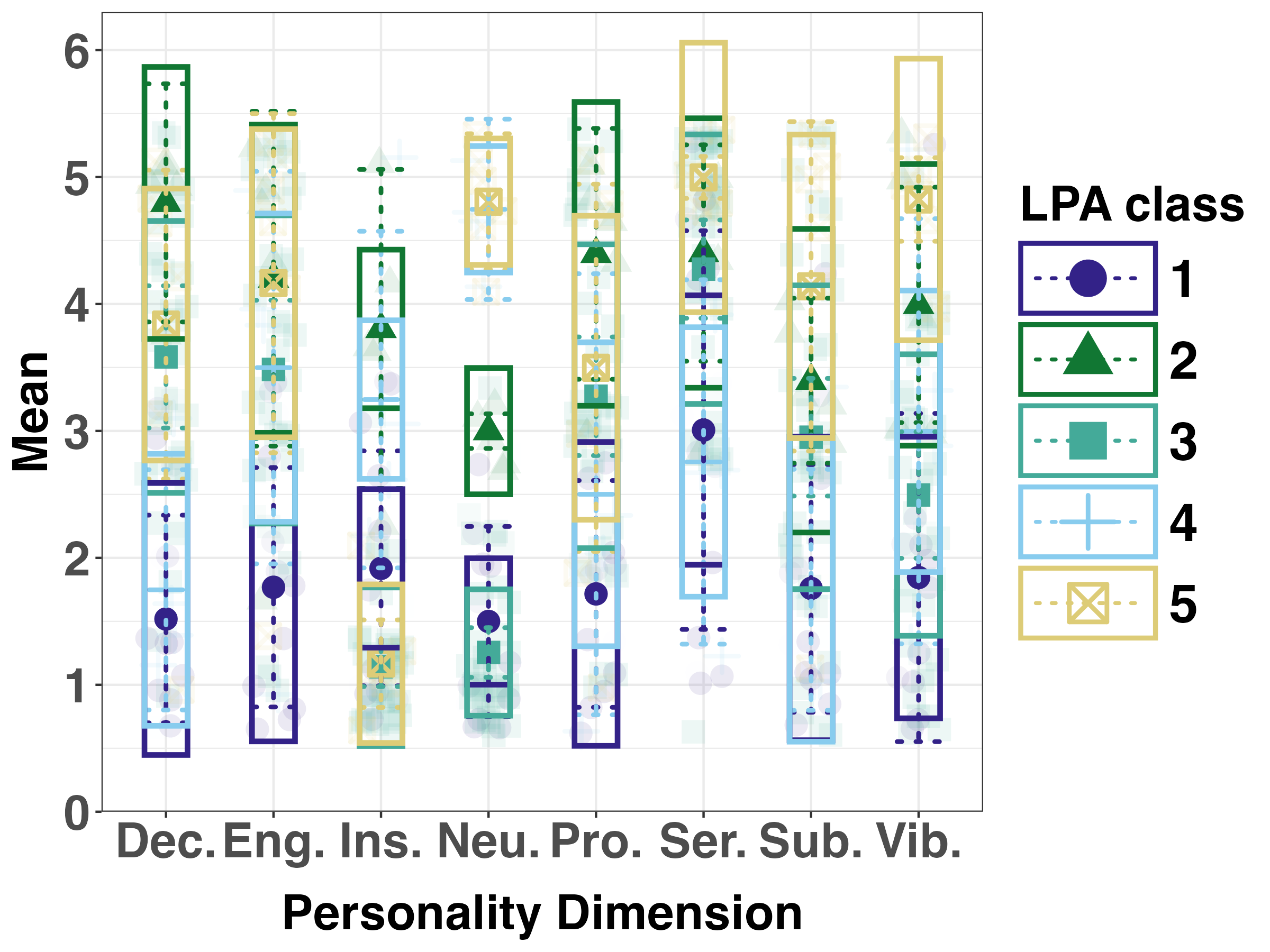}
    \subcaption{\textsc{Appraisal} - final}
  \end{subfigure}
  \caption{Latent profile analysis (LPA) results for the \textsc{Appraisal} condition. The left panel shows the initial configuration profiles, while the right panel shows the final configuration profiles.}
  \label{fig:lpa_app}
\end{figure}

%% file: 9.7_Trait_Freq.tex
\subsection{Trait Frequencies}

\label{sec:trait_freq}

\begin{table}[H]
\caption{Frequencies of personality traits reported by participants as most supportive within each condition (\textsc{Informational}, \textsc{Emotional}, \textsc{Appraisal}).}
\label{tab:trait_frequencies}
\renewcommand{\arraystretch}{1.1}
\setlength{\tabcolsep}{6pt}
\small
\begin{tabular}{lccc}
\toprule
\textbf{Trait} & \textsc{Informational} & \textsc{Emotional} & \textsc{Appraisal} \\
\midrule
Engagement     & 41 & 43 & 39 \\
Serviceability & 36 & 23 & 34 \\
Decency        & 28 & 35 & 34 \\
Profoundness   & 17 & 34 & 19 \\
Vibrancy       & 16 & 19 & 12 \\
Subservience   & 15 &  8 & 11 \\
Instability    &  6 &  8 &  3 \\
Neuroticism    &  6 &  5 &  8 \\
\bottomrule
\end{tabular}
\end{table}

%% file: 9.8_TA.tex
\subsection{Clustered Mean Trajectories }

\label{sec:ta_clusters}

\begin{figure*}[ht]
\includegraphics[width=\textwidth]{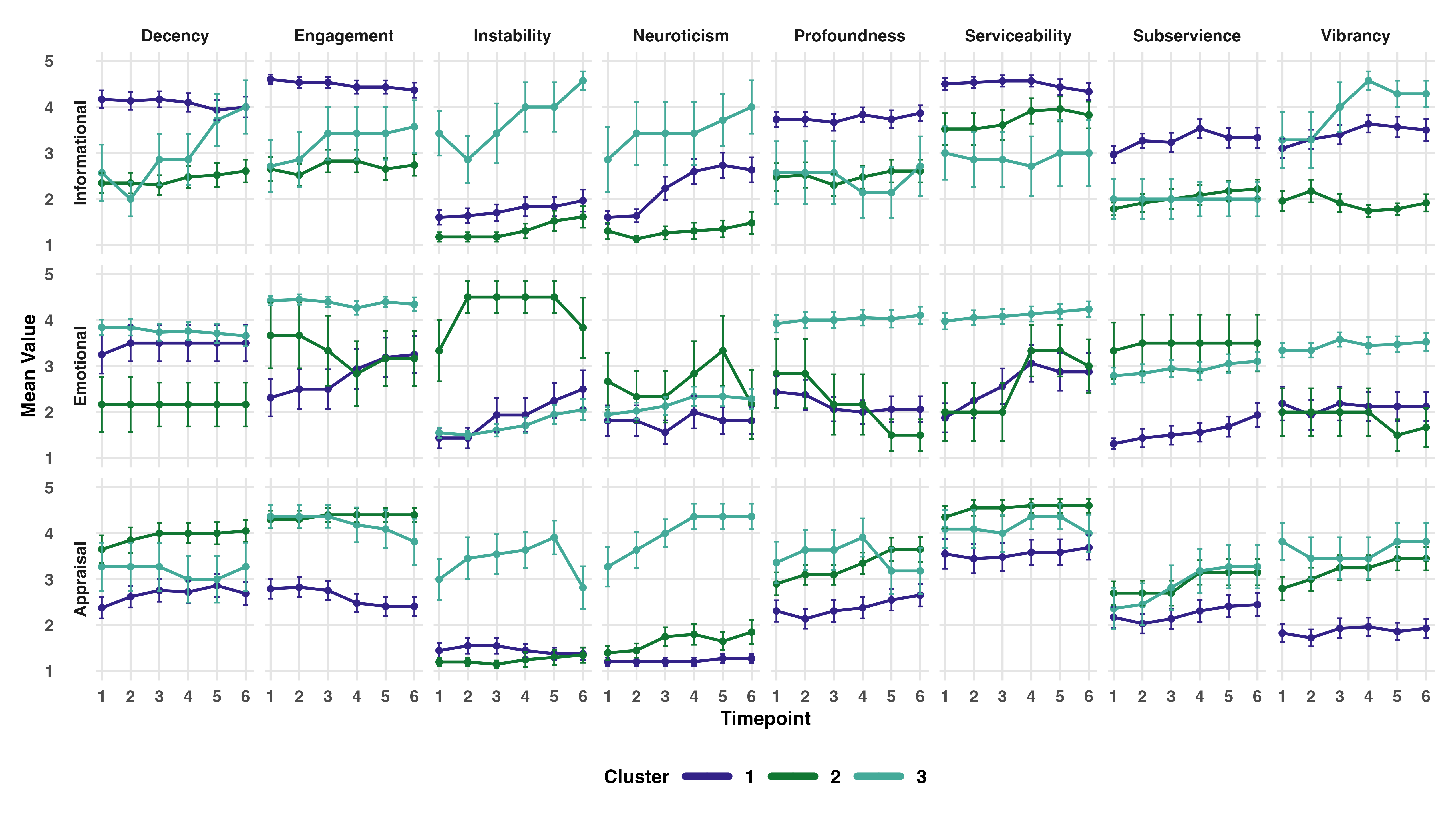}
    \caption{Clustered mean trajectories of personality adjustments across conversational turns, plotted by condition and dimension with standard error bars, illustrating distinct adaptation patterns over conversational turns (Timepoint). Cluster names are as follows: (1) Steady Anchors (2) Adaptive Explorers and (3) Reactive Shifters.}
    \label{fig:clusters}
\end{figure*}

%% file: 9.9_Mapping.tex
\section{Mapping LPA Classes to Trajectory Clusters}
\label{sec:mapping}

\begin{figure}[ht]
  \centering
  \begin{subfigure}[t]{0.45\linewidth}
    \centering
    \includegraphics[width=\linewidth]{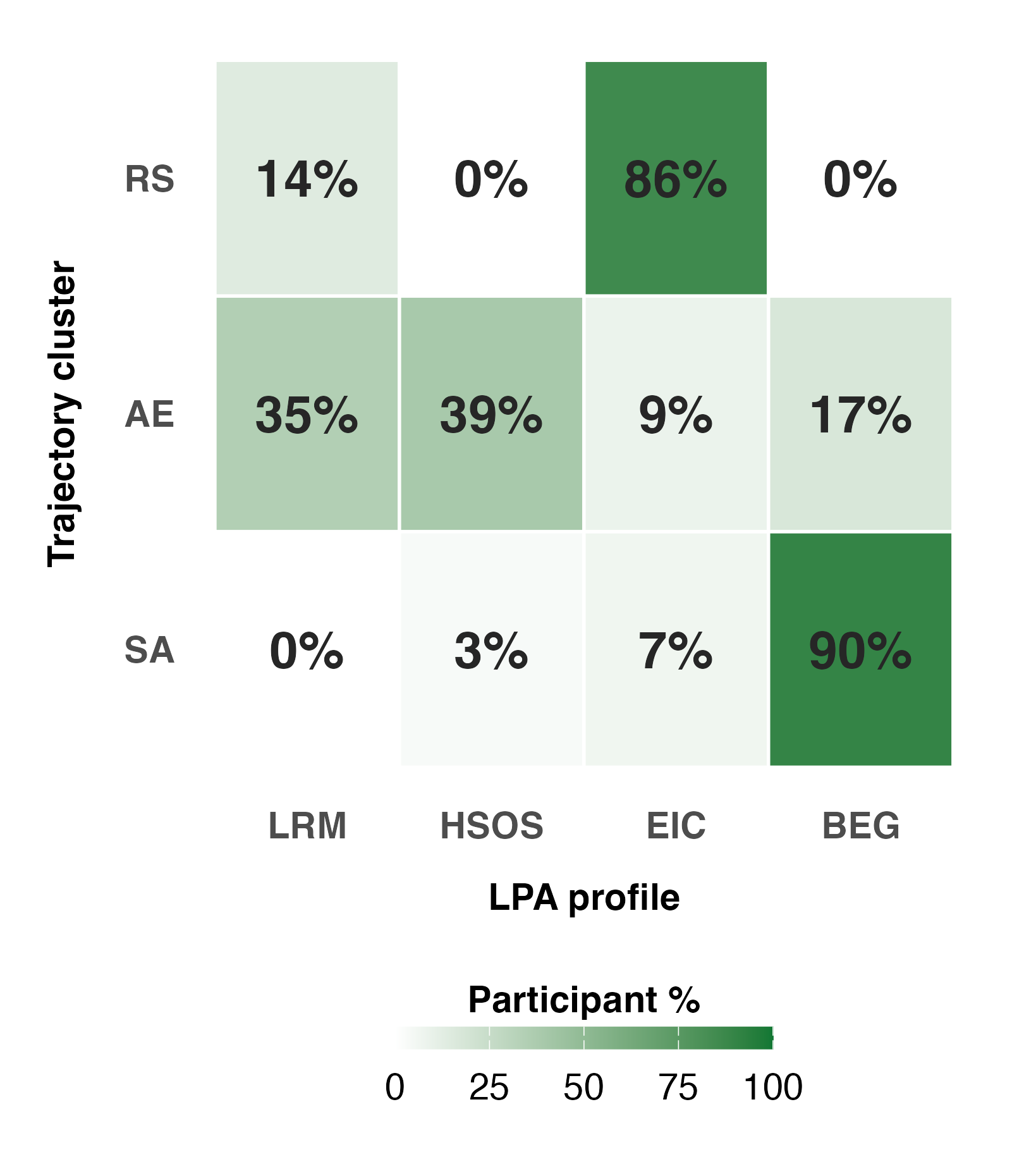}
    \caption{\textsc{Informational} - Initial}
  \end{subfigure}
  \hfill
  \begin{subfigure}[t]{0.45\linewidth}
    \centering
    \includegraphics[width=\linewidth]{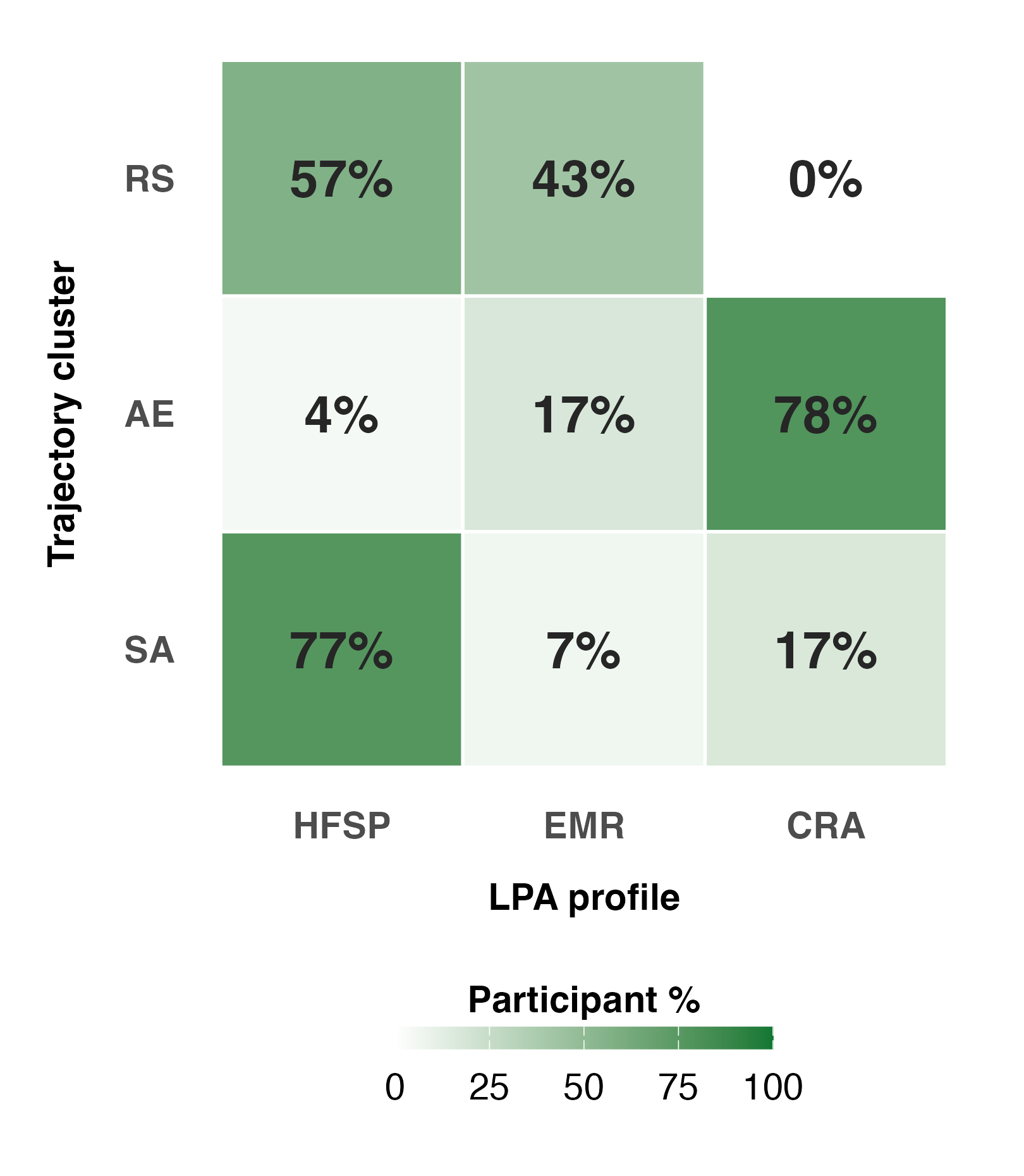}
    \caption{\textsc{Informational} - Final}
  \end{subfigure}
  \caption{Mapping between clustered trajectory groups (TA) and latent profiles (LPA) in the \textsc{Informational} condition. 
  The y-axis represents trajectory clusters (1 = Steady Anchors, 2 = Adaptive Explorers, 3 = Reactive Shifters). 
  The x-axis shows LPA profiles derived from participants’ initial and final personality settings.}
  \label{fig:mapping_info}
\end{figure}

\begin{figure}[ht]
  \centering
  \begin{subfigure}[t]{0.45\linewidth}
    \centering
    \includegraphics[width=\linewidth]{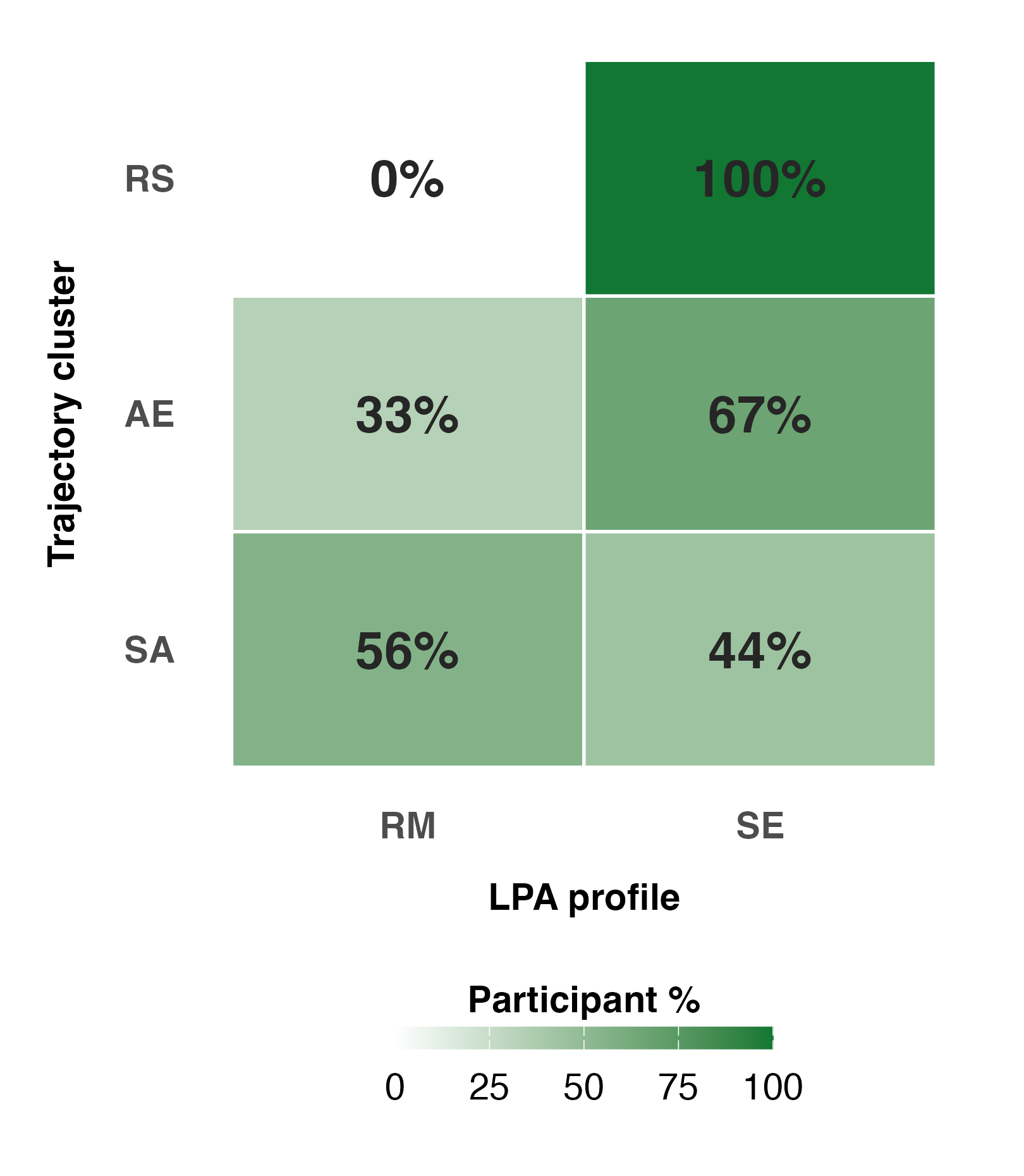}
    \caption{\textsc{Emotional} - Initial}
  \end{subfigure}
  \hfill
  \begin{subfigure}[t]{0.45\linewidth}
    \centering
    \includegraphics[width=\linewidth]{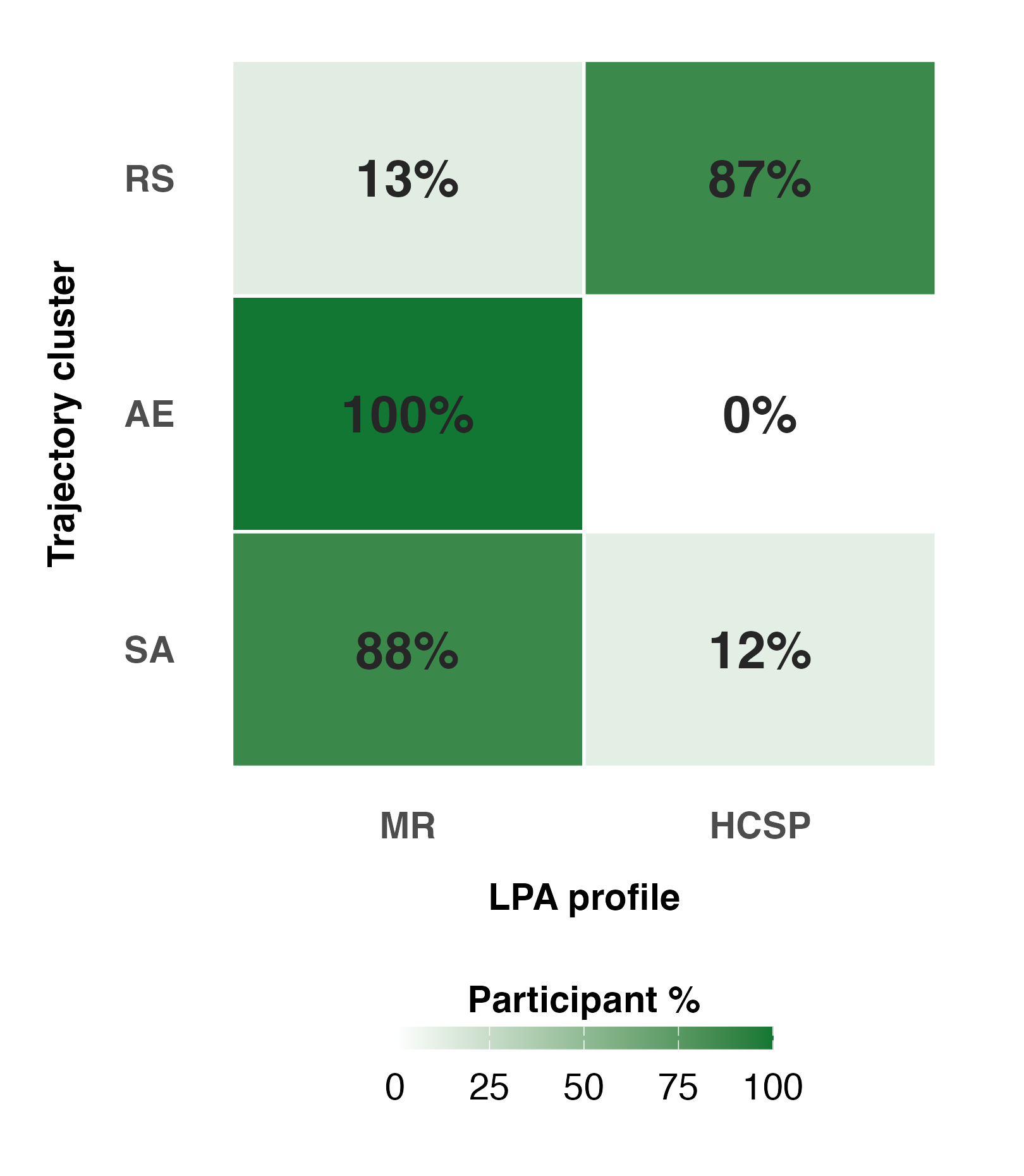}
    \caption{\textsc{Emotional} - Final}
  \end{subfigure}
  \caption{Mapping between clustered trajectory groups (TA) and latent profiles (LPA) in the \textsc{Emotional} condition. 
  The y-axis represents trajectory clusters (1 = Steady Anchors, 2 = Adaptive Explorers, 3 = Reactive Shifters). 
  The x-axis shows LPA profiles derived from participants’ initial and final personality settings.}
  \label{fig:mapping_emo}
\end{figure}

\begin{figure}[ht]
  \centering
  \begin{subfigure}[t]{0.45\linewidth}
    \centering
    \includegraphics[width=\linewidth]{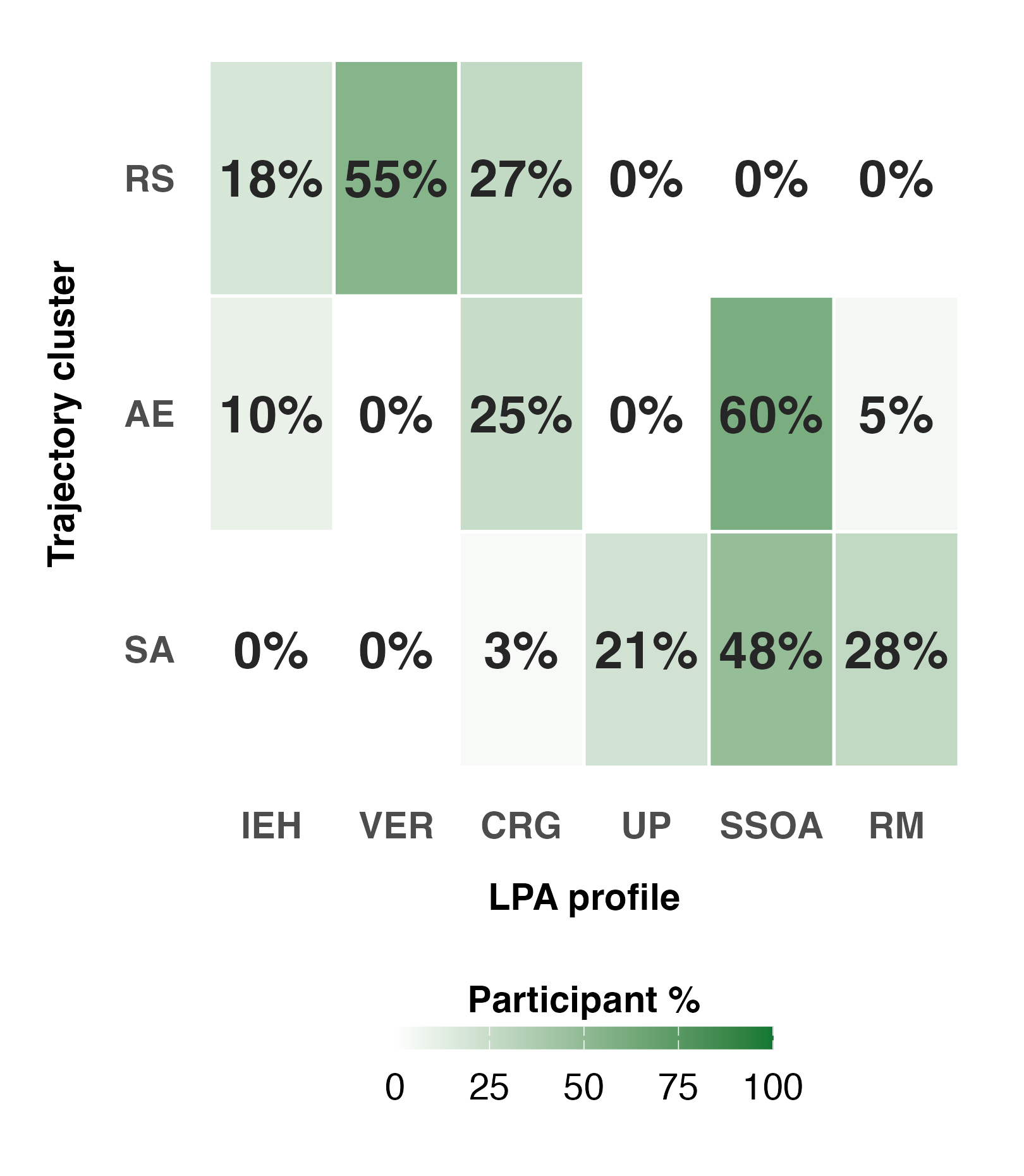}
    \caption{\textsc{Appraisal} - Initial}
  \end{subfigure}
  \hfill
  \begin{subfigure}[t]{0.45\linewidth}
    \centering
    \includegraphics[width=\linewidth]{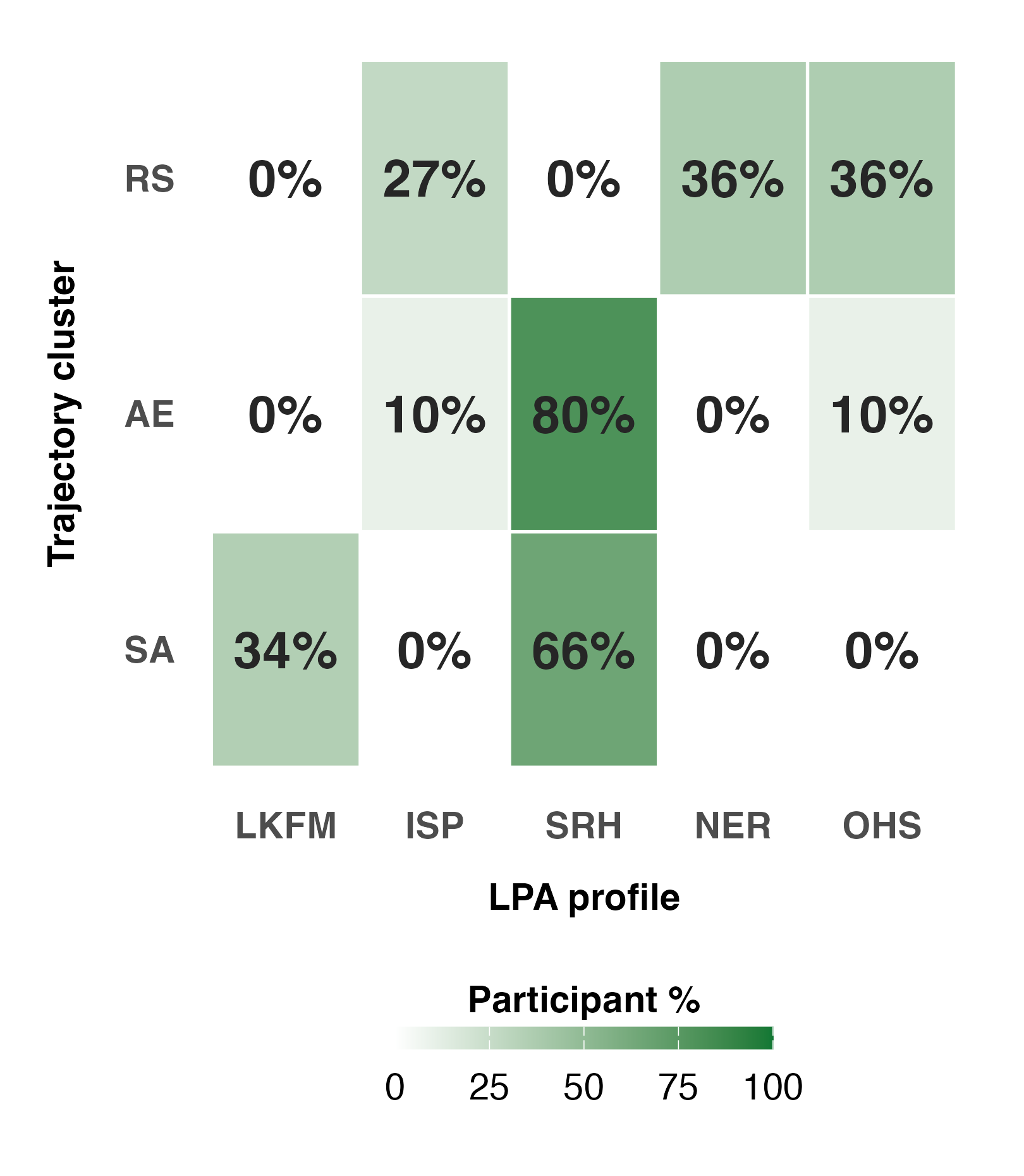}
    \caption{\textsc{Appraisal} - Final}
  \end{subfigure}
  \caption{Mapping between clustered trajectory groups (TA) and latent profiles (LPA) in the \textsc{Appraisal} condition. 
  The y-axis represents trajectory clusters (1 = Steady Anchors, 2 = Adaptive Explorers, 3 = Reactive Shifters). 
  The x-axis shows LPA profiles derived from participants’ initial and final personality settings.}
  \label{fig:mapping_app}
\end{figure}

Table~\ref{tab:lpa_classes} lists the latent profile analysis (LPA) classes for each condition and stage, along with the abbreviations used in Figures~\ref{fig:mapping_info}, \ref{fig:mapping_emo}, and \ref{fig:mapping_app}.

\begin{table*}[ht]
\centering
\caption{Latent profile classes by condition, stage, and abbreviated label.}
\label{tab:lpa_classes}
\renewcommand{\arraystretch}{1.2}
\begin{tabular}{p{2.5cm} p{2cm} p{8cm} p{2cm}}
\toprule
\textbf{Condition} & \textbf{Stage} & \textbf{Profile Name} & \textbf{Abbreviation} \\
\midrule
Informational & Initial & Low Reactive Minimalist                & LRM  \\
              &        & Highly Service-Oriented Supporter       & HSOS \\
              &        & Emotionally Intense Companion           & EIC  \\
              &        & Balanced Engaging Guide                 & BEG  \\
              & Final  & High-Functioning Supportive Partner     & HFSP \\
              &        & Erratic but Modest Reactor              & EMR  \\
              &        & Calm Reliable Assistant                 & CRA  \\
\midrule
Emotional     & Initial & Reserved Minimalist                    & RM   \\
              &        & Supportive Engager                      & SE   \\
              & Final  & Modest Reactor                          & MR   \\
              &        & Highly Capable Supportive Partner       & HCSP \\
\midrule
Appraisal     & Initial & Intensely Engaged Helper               & IEH  \\
              &        & Volatile but Energetic Reactor          & VER  \\
              &        & Confident Reliable Guide                & CRG  \\
              &        & Uneven Pragmatist                       & UP   \\
              &        & Stable Service-Oriented Assistant       & SSOA \\
              &        & Reserved Minimalist                     & RM   \\
              & Final  & Low-Key Functional Minimalist           & LKFM \\
              &        & Intense but Supportive Performer        & ISP  \\
              &        & Stable Reliable Helper                  & SRH  \\
              &        & Neurotic but Engaged Reactor            & NER  \\
              &        & Overloaded Hyper-Servant                & OHS  \\
\bottomrule
\end{tabular}
\end{table*}

%% file: ref.bib
@inproceedings{Jung2023,
author = {Jung, Ji-Youn and Bozzon, Alessandro},
title = {Are Female Chatbots More Empathic? - Discussing Gendered Conversational Agent through Empathic Design},
year = {2023},
isbn = {9798400707490},
publisher = {Association for Computing Machinery},
address = {New York, NY, USA},
url = {https://doi.org/10.1145/3588967.3588970},
doi = {10.1145/3588967.3588970},
booktitle = {Proceedings of the 2nd Empathy-Centric Design Workshop},
articleno = {3},
numpages = {5},
location = {Hamburg, Germany},
series = {EmpathiCH '23}
}

@inproceedings{Volkel2020,
author = {V\"{o}lkel, Sarah Theres and Sch\"{o}del, Ramona and Buschek, Daniel and Stachl, Clemens and Winterhalter, Verena and B\"{u}hner, Markus and Hussmann, Heinrich},
title = {Developing a Personality Model for Speech-based Conversational Agents Using the Psycholexical Approach},
year = {2020},
isbn = {9781450367080},
publisher = {Association for Computing Machinery},
address = {New York, NY, USA},
url = {https://doi.org/10.1145/3313831.3376210},
doi = {10.1145/3313831.3376210},
booktitle = {Proceedings of the 2020 CHI Conference on Human Factors in Computing Systems},
pages = {1–14},
numpages = {14},
location = {Honolulu, HI, USA},
series = {CHI '20}
}

@inproceedings{Ha2024,
author = {Ha, Juhye and Jeon, Hyeon and Han, Daeun and Seo, Jinwook and Oh, Changhoon},
title = {CloChat: Understanding How People Customize, Interact, and Experience Personas in Large Language Models},
year = {2024},
isbn = {9798400703300},
publisher = {Association for Computing Machinery},
address = {New York, NY, USA},
url = {https://doi.org/10.1145/3613904.3642472},
doi = {10.1145/3613904.3642472},
booktitle = {Proceedings of the 2024 CHI Conference on Human Factors in Computing Systems},
articleno = {305},
numpages = {24},
location = {Honolulu, HI, USA},
series = {CHI '24}
}

@article{Nikola2024,
author = {Kova\v{c}evi\'{c}, Nikola and Holz, Christian and Gross, Markus and Wampfler, Rafael},
title = {The Personality Dimensions GPT-3 Expresses During Human-Chatbot Interactions},
year = {2024},
issue_date = {June 2024},
publisher = {Association for Computing Machinery},
address = {New York, NY, USA},
volume = {8},
number = {2},
url = {https://doi.org/10.1145/3659626},
doi = {10.1145/3659626},
journal = {Proc. ACM Interact. Mob. Wearable Ubiquitous Technol.},
month = may,
articleno = {61},
numpages = {36}
}

@inproceedings{Zheng2025,
author = {Zheng, Xi and Li, Zhuoyang and Gui, Xinning and Luo, Yuhan},
title = {Customizing Emotional Support: How Do Individuals Construct and Interact With LLM-Powered Chatbots},
year = {2025},
isbn = {9798400713941},
publisher = {Association for Computing Machinery},
address = {New York, NY, USA},
url = {https://doi.org/10.1145/3706598.3713453},
doi = {10.1145/3706598.3713453},
booktitle = {Proceedings of the 2025 CHI Conference on Human Factors in Computing Systems},
articleno = {376},
numpages = {20},
location = {
},
series = {CHI '25}
}

@inproceedings{Cho2025,
author = {Cho, Hyungjun and Seo, Jiyeon Amy and Lee, Jiwon and Kim, Chang-Min and Nam, Tek-Jin},
title = {ShamAIn: Designing Superior Conversational AI Inspired by Shamanism},
year = {2025},
isbn = {9798400713941},
publisher = {Association for Computing Machinery},
address = {New York, NY, USA},
url = {https://doi.org/10.1145/3706598.3714297},
doi = {10.1145/3706598.3714297},
booktitle = {Proceedings of the 2025 CHI Conference on Human Factors in Computing Systems},
articleno = {985},
numpages = {18},
location = {
},
series = {CHI '25}
}

@inproceedings{Huang2025,
author = {Huang, Yuanhui and Zhou, Quan and Piper, Anne Marie},
title = {Designing Conversational AI for Aging: A Systematic Review of Older Adults' Perceptions and Needs},
year = {2025},
isbn = {9798400713941},
publisher = {Association for Computing Machinery},
address = {New York, NY, USA},
url = {https://doi.org/10.1145/3706598.3713578},
doi = {10.1145/3706598.3713578},
booktitle = {Proceedings of the 2025 CHI Conference on Human Factors in Computing Systems},
articleno = {181},
numpages = {20},
location = {
},
series = {CHI '25}
}

@inproceedings{Lee2025,
author = {Lee, Sunok and Choi, Dasom and Truong, Lucy and Sawhney, Nitin and Paakki, Henna},
title = {Into the Unknown: Leveraging Conversational AI in Supporting Young Migrants' Journeys Towards Cultural Adaptation},
year = {2025},
isbn = {9798400713941},
publisher = {Association for Computing Machinery},
address = {New York, NY, USA},
url = {https://doi.org/10.1145/3706598.3713091},
doi = {10.1145/3706598.3713091},
booktitle = {Proceedings of the 2025 CHI Conference on Human Factors in Computing Systems},
articleno = {169},
numpages = {19},
location = {
},
series = {CHI '25}
}

@article{Naveed2025,
author = {Naveed, Humza and Khan, Asad Ullah and Qiu, Shi and Saqib, Muhammad and Anwar, Saeed and Usman, Muhammad and Akhtar, Naveed and Barnes, Nick and Mian, Ajmal},
title = {A Comprehensive Overview of Large Language Models},
year = {2025},
publisher = {Association for Computing Machinery},
address = {New York, NY, USA},
issn = {2157-6904},
url = {https://doi.org/10.1145/3744746},
doi = {10.1145/3744746},
note = {Just Accepted},
journal = {ACM Trans. Intell. Syst. Technol.},
month = jun
}

@article{Teubner2023,
  author    = {Teubner, T. and Flath, C. M. and Weinhardt, C. and others},
  title     = {Welcome to the Era of ChatGPT et al.},
  journal   = {Business \& Information Systems Engineering},
  volume    = {65},
  pages     = {95--101},
  year      = {2023},
  doi       = {10.1007/s12599-023-00795-x},
  url       = {https://doi.org/10.1007/s12599-023-00795-x}
}

@inproceedings{Shin2025,
author = {Shin, Subin and Oh, Jeesun and Lee, Sangwon},
title = {Can LLMs See What I See? A Study on Five Prompt Engineering Techniques for Evaluating UX on a Shopping Site},
year = {2025},
isbn = {9798400713958},
publisher = {Association for Computing Machinery},
address = {New York, NY, USA},
url = {https://doi.org/10.1145/3706599.3720079},
doi = {10.1145/3706599.3720079},
booktitle = {Proceedings of the Extended Abstracts of the CHI Conference on Human Factors in Computing Systems},
articleno = {125},
numpages = {7},
location = {
},
series = {CHI EA '25}
}

@incollection{Marvin2024,
  author    = {Marvin, G. and Hellen, N. and Jjingo, D. and Nakatumba-Nabende, J.},
  title     = {Prompt Engineering in Large Language Models},
  booktitle = {Data Intelligence and Cognitive Informatics},
  editor    = {Jacob, I. J. and Piramuthu, S. and Falkowski-Gilski, P.},
  series    = {Algorithms for Intelligent Systems},
  year      = {2024},
  publisher = {Springer},
  address   = {Singapore},
  doi       = {10.1007/978-981-99-7962-2_30},
  url       = {https://doi.org/10.1007/978-981-99-7962-2_30},
  note      = {ICDICI 2023}
}

@inproceedings{Zamfirescu2023,
author = {Zamfirescu-Pereira, J.D. and Wong, Richmond Y. and Hartmann, Bjoern and Yang, Qian},
title = {Why Johnny Can’t Prompt: How Non-AI Experts Try (and Fail) to Design LLM Prompts},
year = {2023},
isbn = {9781450394215},
publisher = {Association for Computing Machinery},
address = {New York, NY, USA},
url = {https://doi.org/10.1145/3544548.3581388},
doi = {10.1145/3544548.3581388},
booktitle = {Proceedings of the 2023 CHI Conference on Human Factors in Computing Systems},
articleno = {437},
numpages = {21},
location = {Hamburg, Germany},
series = {CHI '23}
}

@misc{OpenAIPromptEng,
  author       = {OpenAI},
  title        = {Prompt Engineering},
  year         = {2025},
  url          = {https://platform.openai.com/docs/guides/prompt-engineering},
  note         = {Accessed: 2025-08-14}
}

@misc{GooglePromptEng,
  author       = {Google Cloud},
  title        = {What is Prompt Engineering?},
  year         = {2025},
  url          = {https://cloud.google.com/discover/what-is-prompt-engineering},
  note         = {Accessed: 2025-08-14}
}

@inproceedings{Park2025,
author = {Park, Minyoung and Park, Bogyeom and Seo, Kyoungwon},
title = {How Self-Disclosing Chatbots Influence Student Engagement, Assessment Accuracy, and Self-Reflection in Academic Stress Assessment},
year = {2025},
isbn = {9798400713958},
publisher = {Association for Computing Machinery},
address = {New York, NY, USA},
url = {https://doi.org/10.1145/3706599.3719684},
doi = {10.1145/3706599.3719684},
booktitle = {Proceedings of the Extended Abstracts of the CHI Conference on Human Factors in Computing Systems},
articleno = {322},
numpages = {10},
location = {
},
series = {CHI EA '25}
}

@inproceedings{Moilanen2022,
author = {Moilanen, Joonas and Visuri, Aku and Suryanarayana, Sharadhi Alape and Alorwu, Andy and Yatani, Koji and Hosio, Simo},
title = {Measuring the Effect of Mental Health Chatbot Personality on User Engagement},
year = {2022},
isbn = {9781450398206},
publisher = {Association for Computing Machinery},
address = {New York, NY, USA},
url = {https://doi.org/10.1145/3568444.3568464},
doi = {10.1145/3568444.3568464},
booktitle = {Proceedings of the 21st International Conference on Mobile and Ubiquitous Multimedia},
pages = {138–150},
numpages = {13},
location = {Lisbon, Portugal},
series = {MUM '22}
}

@inproceedings{Desai2025,
author = {Desai, Smit and Dubiel, Mateusz and Zargham, Nima and Mildner, Thomas and Spillner, Laura},
title = {Personas Evolved: Designing Ethical LLM-Based Conversational Agent Personalities},
year = {2025},
isbn = {9798400715273},
publisher = {Association for Computing Machinery},
address = {New York, NY, USA},
url = {https://doi.org/10.1145/3719160.3728624},
doi = {10.1145/3719160.3728624},
booktitle = {Proceedings of the 7th ACM Conference on Conversational User Interfaces},
articleno = {32},
numpages = {4},
location = {
},
series = {CUI '25}
}

@inproceedings{Zargham2025,
author = {Zargham, Nima and Avanesi, Vino and Spillner, Laura and Rockstroh, Johanna},
title = {Crossing the Line? The Paradox of Human-Like Design in Conversational Agents},
year = {2025},
isbn = {9798400715273},
publisher = {Association for Computing Machinery},
address = {New York, NY, USA},
url = {https://doi.org/10.1145/3719160.3737612},
doi = {10.1145/3719160.3737612},
booktitle = {Proceedings of the 7th ACM Conference on Conversational User Interfaces},
articleno = {76},
numpages = {5},
location = {
},
series = {CUI '25}
}

@inproceedings{Dubiel2022,
author = {Dubiel, Mateusz and Daronnat, Sylvain and Leiva, Luis A.},
title = {Conversational Agents Trust Calibration: A User-Centred Perspective to Design},
year = {2022},
isbn = {9781450397391},
publisher = {Association for Computing Machinery},
address = {New York, NY, USA},
url = {https://doi.org/10.1145/3543829.3544518},
doi = {10.1145/3543829.3544518},
booktitle = {Proceedings of the 4th Conference on Conversational User Interfaces},
articleno = {30},
numpages = {6},
location = {Glasgow, United Kingdom},
series = {CUI '22}
}

@article{Folstad2024,
author = {F\o{}lstad, Asbj\o{}rn and Law, Effie L.-C. and van As, Nena},
title = {Conversational Breakdown in a Customer Service Chatbot: Impact of Task Order and Criticality on User Trust and Emotion},
year = {2024},
issue_date = {October 2024},
publisher = {Association for Computing Machinery},
address = {New York, NY, USA},
volume = {31},
number = {5},
issn = {1073-0516},
url = {https://doi.org/10.1145/3690383},
doi = {10.1145/3690383},
journal = {ACM Trans. Comput.-Hum. Interact.},
month = nov,
articleno = {66},
numpages = {52}
}

@inproceedings{Liu2025,
author = {Liu, Xingyu Bruce and Fang, Shitao and Shi, Weiyan and Wu, Chien-Sheng and Igarashi, Takeo and Chen, Xiang 'Anthony'},
title = {Proactive Conversational Agents with Inner Thoughts},
year = {2025},
isbn = {9798400713941},
publisher = {Association for Computing Machinery},
address = {New York, NY, USA},
url = {https://doi.org/10.1145/3706598.3713760},
doi = {10.1145/3706598.3713760},
booktitle = {Proceedings of the 2025 CHI Conference on Human Factors in Computing Systems},
articleno = {184},
numpages = {19},
location = {
},
series = {CHI '25}
}

@inproceedings{Curry2020,
    title = "Conversational Assistants and Gender Stereotypes: Public Perceptions and Desiderata for Voice Personas",
    author = "Cercas Curry, Amanda  and
      Robertson, Judy  and
      Rieser, Verena",
    booktitle = "Proceedings of the Second Workshop on Gender Bias in Natural Language Processing",
    month = dec,
    year = "2020",
    address = "Barcelona, Spain (Online)",
    publisher = "Association for Computational Linguistics",
    url = "https://aclanthology.org/2020.gebnlp-1.7/",
    pages = "72--78"
}

@inproceedings{Hwang2019,
author = {Hwang, Gilhwan and Lee, Jeewon and Oh, Cindy Yoonjung and Lee, Joonhwan},
title = {It Sounds Like A Woman: Exploring Gender Stereotypes in South Korean Voice Assistants},
year = {2019},
isbn = {9781450359719},
publisher = {Association for Computing Machinery},
address = {New York, NY, USA},
url = {https://doi.org/10.1145/3290607.3312915},
doi = {10.1145/3290607.3312915},
booktitle = {Extended Abstracts of the 2019 CHI Conference on Human Factors in Computing Systems},
pages = {1–6},
location = {Glasgow, Scotland Uk},
series = {CHI EA '19}
}

@article{Wang2020, place={Australia}, title={The Three Harms of Gendered Technology}, volume={24}, url={https://ajis.aaisnet.org/index.php/ajis/article/view/2799}, DOI={10.3127/ajis.v24i0.2799}, journal={Australasian Journal of Information Systems}, author={Wang, Lena}, year={2020}, month={Jun.} }

@techreport{LoideainAdams2018,
  author    = {Ni Loideain, Nora and Adams, Rachel},
  title     = {From Alexa to Siri and the GDPR: The Gendering of Virtual Personal Assistants and the Role of EU Data Protection Law},
  institution = {King's College London Dickson Poon School of Law},
  year      = {2018},
  month     = {November 9},
  type      = {Legal Studies Research Paper Series},
  url       = {https://ssrn.com/abstract=3281807},
  doi       = {10.2139/ssrn.3281807}
}

@misc{OpenAIGPT5,
  author       = {OpenAI},
  title        = {Introducing GPT-5},
  year         = {2025},
  url          = {https://openai.com/index/introducing-gpt-5/},
  note         = {Accessed: 2025-08-14}
}

@article{McCrae1992,
author = {McCrae, Robert R. and John, Oliver P.},
title = {An Introduction to the Five-Factor Model and Its Applications},
journal = {Journal of Personality},
volume = {60},
number = {2},
pages = {175-215},
doi = {https://doi.org/10.1111/j.1467-6494.1992.tb00970.x},
url = {https://onlinelibrary.wiley.com/doi/abs/10.1111/j.1467-6494.1992.tb00970.x},
eprint = {https://onlinelibrary.wiley.com/doi/pdf/10.1111/j.1467-6494.1992.tb00970.x},
year = {1992}
}

@book{Myers1962,
  author    = {Myers, Isabel Briggs},
  title     = {The Myers-Briggs Type Indicator},
  volume    = {34},
  year      = {1962},
  publisher = {Consulting Psychologists Press},
  address   = {Palo Alto, CA}
}

@inproceedings{Nass1994,
author = {Nass, Clifford and Steuer, Jonathan and Tauber, Ellen R.},
title = {Computers are social actors},
year = {1994},
isbn = {0897916506},
publisher = {Association for Computing Machinery},
address = {New York, NY, USA},
url = {https://doi.org/10.1145/191666.191703},
doi = {10.1145/191666.191703},
booktitle = {Proceedings of the SIGCHI Conference on Human Factors in Computing Systems},
pages = {72–78},
location = {Boston, Massachusetts, USA},
series = {CHI '94}
}

@inproceedings{Nass1995,
author = {Nass, Clifford and Moon, Youngme and Fogg, B. J. and Reeves, Byron and Dryer, Chris},
title = {Can computer personalities be human personalities?},
year = {1995},
isbn = {0897917553},
publisher = {Association for Computing Machinery},
address = {New York, NY, USA},
url = {https://doi.org/10.1145/223355.223538},
doi = {10.1145/223355.223538},
booktitle = {Conference Companion on Human Factors in Computing Systems},
pages = {228–229},
location = {Denver, Colorado, USA},
series = {CHI '95}
}

@article{Airenti2015,
  title={The Cognitive Bases of Anthropomorphism: From Relatedness to Empathy},
  author={Gabriella Airenti},
  journal={International Journal of Social Robotics},
  year={2015},
  volume={7},
  pages={117 - 127},
  url={https://api.semanticscholar.org/CorpusID:260263}
}

@article{Rapp2021,
  title        = {The human side of human-chatbot interaction: A systematic literature review of ten years of research on text-based chatbots},
  author       = {Rapp, Amon and Curti, Lorenzo and Boldi, Arianna},
  journal      = {International Journal of Human-Computer Studies},
  volume       = {151},
  pages        = {102630},
  year         = {2021},
  issn         = {1071-5819},
  doi          = {10.1016/j.ijhcs.2021.102630},
  url          = {https://www.sciencedirect.com/science/article/pii/S1071581921000483}
}

@article{Cutrona1992,
  author    = {Cutrona, Carolyn E. and Suhr, Julie A.},
  title     = {Controllability of Stressful Events and Satisfaction With Spouse Support Behaviors},
  journal   = {Communication Research},
  volume    = {19},
  number    = {2},
  pages     = {154--174},
  year      = {1992},
  doi       = {10.1177/009365092019002002},
  url       = {https://doi.org/10.1177/009365092019002002}
}

@misc{Campbell2024,
  author       = {Campbell, Emily},
  title        = {Personality Patterns Framework: Understanding Traits in AI},
  howpublished = {\url{https://www.linkedin.com/pulse/personality-patterns-framework-understanding-traits-ai-emily-campbell-3klac/}},
  note         = {Accessed: 2025-08-14}
}

@article{Revilla2013,
  author    = {Revilla, Melanie A. and Saris, Willem E. and Krosnick, Jon A.},
  title     = {Choosing the Number of Categories in Agree--Disagree Scales},
  journal   = {Sociological Methods \& Research},
  volume    = {43},
  number    = {1},
  pages     = {73--97},
  year      = {2013},
  doi       = {10.1177/0049124113509605},
  url       = {https://doi.org/10.1177/0049124113509605}
}

@article{Dawes2008,
  author    = {Dawes, John},
  title     = {Do Data Characteristics Change According to the Number of Scale Points Used? An Experiment Using 5-Point, 7-Point and 10-Point Scales},
  journal   = {International Journal of Market Research},
  volume    = {50},
  number    = {1},
  pages     = {61--104},
  year      = {2008},
  doi       = {10.1177/147078530805000106},
  url       = {https://doi.org/10.1177/147078530805000106}
}

@book{Schreier2012,
  author    = {Schreier, Margrit},
  title     = {Qualitative Content Analysis in Practice},
  year      = {2012},
  publisher = {SAGE Publications Ltd},
  doi       = {10.4135/9781529682571},
  url       = {https://doi.org/10.4135/9781529682571}
}

@inproceedings{gebreegziabher2023patat,
  title={Patat: Human-ai collaborative qualitative coding with explainable interactive rule synthesis},
  author={Gebreegziabher, Simret Araya and Zhang, Zheng and Tang, Xiaohang and Meng, Yihao and Glassman, Elena L and Li, Toby Jia-Jun},
  booktitle={Proceedings of the 2023 CHI Conference on Human Factors in Computing Systems},
  pages={1--19},
  year={2023}
}

@article{o2020intercoder,
  title={Intercoder reliability in qualitative research: Debates and practical guidelines},
  author={O’Connor, Cliodhna and Joffe, Helene},
  journal={International journal of qualitative methods},
  volume={19},
  pages={1609406919899220},
  year={2020},
  publisher={SAGE Publications Sage CA: Los Angeles, CA}
}

@article{Jian2000,
  author    = {Jian, Jiangyu Y. and Bisantz, Ann M. and Drury, Colin G.},
  title     = {Foundations for an Empirically Determined Scale of Trust in Automated Systems},
  journal   = {International Journal of Cognitive Ergonomics},
  volume    = {4},
  number    = {1},
  pages     = {53--71},
  year      = {2000},
  doi       = {10.1207/S15327566IJCE0401_04},
  url       = {https://doi.org/10.1207/S15327566IJCE0401_04}
}

@article{Spurk2020,
  author    = {Spurk, Daniel and Hirschi, Andreas and Wang, Mo and Valero, Domingo and Kauffeld, Simone},
  title     = {Latent profile analysis: A review and ``how to'' guide of its application within vocational behavior research},
  journal   = {Journal of Vocational Behavior},
  volume    = {120},
  pages     = {103445},
  year      = {2020},
  issn      = {0001-8791},
  doi       = {10.1016/j.jvb.2020.103445},
  url       = {https://doi.org/10.1016/j.jvb.2020.103445}
}

@article{Tag2022,
  author    = {Tag, Benjamin and van Berkel, Niels and Vargo, Andrew W. and Sarsenbayeva, Zhanna and Colasante, Tyler and Wadley, Greg and Webber, Sarah and Smith, Wally and Koval, Peter and Hollenstein, Tom and Goncalves, Jorge and Kostakos, Vassilis},
  title     = {Impact of the global pandemic upon young people's use of technology for emotion regulation},
  journal   = {Computers in Human Behavior Reports},
  volume    = {6},
  pages     = {100192},
  year      = {2022},
  issn      = {2451-9588},
  doi       = {10.1016/j.chbr.2022.100192},
  url       = {https://doi.org/10.1016/j.chbr.2022.100192}
}

@article{Rosenberg2018,
  author    = {Rosenberg, Joshua M. and Beymer, Patrick N. and Anderson, Daniel J. and van Lissa, Caspar J. and Schmidt, Jennifer A.},
  title     = {tidyLPA: An {R} Package to Easily Carry Out Latent Profile Analysis (LPA) Using Open-Source or Commercial Software},
  journal   = {Journal of Open Source Software},
  volume    = {3},
  number    = {30},
  pages     = {978},
  year      = {2018},
  doi       = {10.21105/joss.00978},
  url       = {https://doi.org/10.21105/joss.00978}
}

@article{Akogul2017,
  author    = {Akogul, Serhat and Erisoglu, Meral},
  title     = {An Approach for Determining the Number of Clusters in a Model-Based Cluster Analysis},
  journal   = {Entropy},
  volume    = {19},
  number    = {9},
  pages     = {452},
  year      = {2017},
  doi       = {10.3390/e19090452},
  url       = {https://doi.org/10.3390/e19090452}
}

@manual{Rpkg_cluster,
  title        = {cluster: Cluster Analysis Basics and Extensions},
  author       = {Maechler, Martin and Rousseeuw, Peter and Struyf, Anja and Hubert, Mia and Hornik, Kurt},
  year         = {2025},
  note         = {R package version 2.1.8},
  url          = {https://CRAN.R-project.org/package=cluster}
}

@inproceedings{humaira2020,
  author    = {Humaira, H. and Rasyidah, R.},
  title     = {Determining the Appropriate Cluster Number Using Elbow Method for K-Means Algorithm},
  booktitle = {Proceedings of the 2nd Workshop on Multimedia Applications (WMA-2)},
  year      = {2020},
  publisher = {EAI},
  doi       = {10.4108/eai.24-1-2018.2292388}
}

@book{kaufman1990,
  title        = {Finding Groups in Data: An Introduction to Cluster Analysis},
  author       = {Kaufman, Leonard and Rousseeuw, Peter J.},
  year         = {1990},
  publisher    = {John Wiley \& Sons},
  address      = {New York}
}

@article{hartigan1979,
  title        = {Algorithm AS 136: A k-means clustering algorithm},
  author       = {Hartigan, John A. and Wong, Manchek A.},
  journal      = {Journal of the Royal Statistical Society. Series C (Applied Statistics)},
  volume       = {28},
  number       = {1},
  pages        = {100--108},
  year         = {1979},
  publisher    = {JSTOR},
  doi          = {10.2307/2346830}
}

@article{Mori2012,
  author={Mori, Masahiro and MacDorman, Karl F. and Kageki, Norri},
  journal={IEEE Robotics \& Automation Magazine},
  title={The Uncanny Valley [From the Field]},
  year={2012},
  volume={19},
  number={2},
  pages={98-100},
  doi={10.1109/MRA.2012.2192811}
}

@article{Ciechanowski2019,
  author    = {Ciechanowski, Leon and Przegalinska, Aleksandra and Magnuski, Mikolaj and Gloor, Peter},
  title     = {In the Shades of the Uncanny Valley: An Experimental Study of Human--Chatbot Interaction},
  journal   = {Future Generation Computer Systems},
  volume    = {92},
  pages     = {539--548},
  year      = {2019},
  issn      = {0167-739X},
  doi       = {10.1016/j.future.2018.01.055},
  url       = {https://doi.org/10.1016/j.future.2018.01.055}
}

@article{Blut2021,
  author    = {Blut, Markus and Wang, Chieh-Yu and Wünderlich, Nancy V. and others},
  title     = {Understanding Anthropomorphism in Service Provision: A Meta-Analysis of Physical Robots, Chatbots, and Other AI},
  journal   = {Journal of the Academy of Marketing Science},
  volume    = {49},
  pages     = {632--658},
  year      = {2021},
  doi       = {10.1007/s11747-020-00762-y},
  url       = {https://doi.org/10.1007/s11747-020-00762-y}
}

@article{Skjuve2021,
  author    = {Skjuve, Marita and Følstad, Asbjørn and Fostervold, Knut Inge and Brandtzaeg, Petter Bae},
  title     = {My Chatbot Companion -- A Study of Human--Chatbot Relationships},
  journal   = {International Journal of Human-Computer Studies},
  volume    = {149},
  pages     = {102601},
  year      = {2021},
  issn      = {1071-5819},
  doi       = {10.1016/j.ijhcs.2021.102601},
  url       = {https://doi.org/10.1016/j.ijhcs.2021.102601}
}

@article{Ramadan2021,
  author    = {Ramadan, Zaki and Farah, Maya F. and El Essrawi, Lara},
  title     = {From Amazon.com to Amazon.love: How Alexa is Redefining Companionship and Interdependence for People with Special Needs},
  journal   = {Psychology \& Marketing},
  volume    = {38},
  pages     = {596--609},
  year      = {2021},
  doi       = {10.1002/mar.21441},
  url       = {https://doi.org/10.1002/mar.21441}
}

@article{Jayasiriwardene2025,
  author    = {Jayasiriwardene, Shakyani and Tag, Benjamin and Withana, Anusha and Sarsenbayeva, Zhanna},
  title     = {More Than Words: The Impact of Voice Assistant Personality Traits on Failure Mitigation},
  journal   = {Proceedings of the ACM on Interactive, Mobile, Wearable and Ubiquitous Technologies},
  year      = {2025},
  note      = {To appear},
  doi       = {10.1145/3749518},
  url       = {https://doi.org/10.1145/3749518}
}

@misc{BrusselsTimes2023,
  author       = {{The Brussels Times}},
  title        = {Belgian man commits suicide following exchanges with ChatGPT},
  howpublished = {\url{https://www.brusselstimes.com/430098/belgian-man-commits-suicide-following-exchanges-with-chatgpt?utm_source=chatgpt.com}},
  note         = {Accessed: 2025-08-14},
  year         = {2023}
}

@misc{Reuters2025,
  author       = {{Reuters}},
  title        = {Italy's data watchdog fines AI company Replika's developer \$5.6 million},
  howpublished = {\url{https://www.reuters.com/sustainability/boards-policy-regulation/italys-data-watchdog-fines-ai-company-replikas-developer-56-million-2025-05-19/?utm_source=chatgpt.com}},
  note         = {Accessed: 2025-08-14},
  year         = {2025}
}

@misc{CFO2024,
  author       = {{CFO.com}},
  title        = {Deepfake CFO in Hong Kong scams firm out of \$25 million in cyber crime},
  howpublished = {\url{https://www.cfo.com/news/deepfake-cfo-hong-kong-25-million-fraud-cyber-crime/706529/?utm_source=chatgpt.com}},
  note         = {Accessed: 2025-08-14},
  year         = {2024}
}

@misc{ArsTechnica2024,
  author       = {{Ars Technica}},
  title        = {Deepfake scammer walks off with \$25 million in first-of-its-kind AI heist},
  howpublished = {\url{https://arstechnica.com/information-technology/2024/02/deepfake-scammer-walks-off-with-25-million-in-first-of-its-kind-ai-heist/?utm_source=chatgpt.com}},
  note         = {Accessed: 2025-08-14},
  year         = {2024}
}

@misc{ICBA2025,
  author       = {{Independent Community Bankers of America}},
  title        = {Report: Romance scams grew 40\% in 2024},
  howpublished = {\url{https://www.icba.org/newsroom/news-and-articles/2025/02/14/report-romance-scams-grew-40-in-2024?utm_source=chatgpt.com}},
  note         = {Accessed: 2025-08-14},
  year         = {2025}
}

@inproceedings{Pan2025,
author = {Pan, Shuyi and de Graaf, Maartje M.A.},
title = {Developing a Social Support Framework: Understanding the Reciprocity in Human-Chatbot Relationship},
year = {2025},
isbn = {9798400713941},
publisher = {Association for Computing Machinery},
address = {New York, NY, USA},
url = {https://doi.org/10.1145/3706598.3713503},
doi = {10.1145/3706598.3713503},
booktitle = {Proceedings of the 2025 CHI Conference on Human Factors in Computing Systems},
articleno = {182},
numpages = {13},
location = {
},
series = {CHI '25}
}

@article{Christoforakos2021,
  author    = {Christoforakos, Louisa and Gallucci, Alessia and Surmava-Große, Tea and Ullrich, Daniel and Diefenbach, Sarah},
  title     = {Can Robots Earn Our Trust the Same Way Humans Do? A Systematic Exploration of Competence, Warmth, and Anthropomorphism as Determinants of Trust Development in HRI},
  journal   = {Frontiers in Robotics and AI},
  volume    = {8},
  pages     = {640444},
  year      = {2021},
  doi       = {10.3389/frobt.2021.640444},
  url       = {https://doi.org/10.3389/frobt.2021.640444}
}

@inproceedings{Arum2025,
author = {van Arum, Sterre and Gen\c{c}, H\"{u}seyin U\u{g}ur and Reidsma, Dennis and Karahano\u{g}lu, Arma\u{g}an},
title = {Selective Trust: Understanding Human-AI Partnerships in Personal Health Decision-Making Process},
year = {2025},
isbn = {9798400713941},
publisher = {Association for Computing Machinery},
address = {New York, NY, USA},
url = {https://doi.org/10.1145/3706598.3713462},
doi = {10.1145/3706598.3713462},
booktitle = {Proceedings of the 2025 CHI Conference on Human Factors in Computing Systems},
articleno = {1026},
numpages = {21},
location = {
},
series = {CHI '25}
}

@article{Cheng2022,
  author    = {Cheng, Xusen and Zhang, Xiaoping and Cohen, Jason and Mou, Jian},
  title     = {Human vs. AI: Understanding the Impact of Anthropomorphism on Consumer Response to Chatbots from the Perspective of Trust and Relationship Norms},
  journal   = {Information Processing \& Management},
  volume    = {59},
  number    = {3},
  pages     = {102940},
  year      = {2022},
  issn      = {0306-4573},
  doi       = {10.1016/j.ipm.2022.102940},
  url       = {https://doi.org/10.1016/j.ipm.2022.102940}
}

@article{deVisser2016,
  author    = {de Visser, E. J. and Monfort, S. S. and McKendrick, R. and Smith, M. A. B. and McKnight, P. E. and Krueger, F. and Parasuraman, R.},
  title     = {Almost Human: Anthropomorphism Increases Trust Resilience in Cognitive Agents},
  journal   = {Journal of Experimental Psychology: Applied},
  volume    = {22},
  number    = {3},
  pages     = {331--349},
  year      = {2016},
  doi       = {10.1037/xap0000092},
  url       = {https://doi.org/10.1037/xap0000092}
}

@inproceedings{Kulms2019,
author = {Kulms, Philipp and Kopp, Stefan},
title = {More Human-Likeness, More Trust? The Effect of Anthropomorphism on Self-Reported and Behavioral Trust in Continued and Interdependent Human-Agent Cooperation},
year = {2019},
isbn = {9781450371988},
publisher = {Association for Computing Machinery},
address = {New York, NY, USA},
url = {https://doi.org/10.1145/3340764.3340793},
doi = {10.1145/3340764.3340793},
booktitle = {Proceedings of Mensch Und Computer 2019},
pages = {31–42},
numpages = {12},
location = {Hamburg, Germany},
series = {MuC '19}
}

@inproceedings{Hu2021,
author = {Hu, Jiaxiong and Huang, Yun and Hu, Xiaozhu and Xu, Yingqing},
title = {Enhancing the Perceived Emotional Intelligence of Conversational Agents through Acoustic Cues},
year = {2021},
isbn = {9781450380959},
publisher = {Association for Computing Machinery},
address = {New York, NY, USA},
url = {https://doi.org/10.1145/3411763.3451660},
doi = {10.1145/3411763.3451660},
booktitle = {Extended Abstracts of the 2021 CHI Conference on Human Factors in Computing Systems},
articleno = {282},
numpages = {7},
location = {Yokohama, Japan},
series = {CHI EA '21}
}

@inproceedings{Chang2025,
author = {Chang, Frank and Herath, Damith},
title = {From Interaction to Relationship: The Role of Memory, Learning, and Emotional Intelligence in AI-Embodied Human Engagement},
year = {2025},
publisher = {IEEE Press},
booktitle = {Proceedings of the 2025 ACM/IEEE International Conference on Human-Robot Interaction},
pages = {1269–1273},
numpages = {5},
location = {Melbourne, Australia},
series = {HRI '25}
}

@unpublished{hernandez2023,
author = {Hernandez, Javier and Suh, Jina and Amores, Judith and Rowan, Kael and Ramos, Gonzalo and Czerwinski, Mary},
title = {Affective Conversational Agents: Understanding Expectations and Personal Influences},
year = {2023},
month = {October},
url = {https://www.microsoft.com/en-us/research/publication/affective-conversational-agents-understanding-expectations-and-personal-influences/},
note = {ArXiv},
}

@article{Lin2018,
  author    = {Lin, Chia-Hui and Huang, Yu},
  title     = {How Self-Construals Affect Responses to Anthropomorphic Brands, With a Focus on the Three-Factor Relationship Between the Brand, the Gift-Giver and the Recipient},
  journal   = {Frontiers in Psychology},
  volume    = {9},
  pages     = {2070},
  year      = {2018},
  doi       = {10.3389/fpsyg.2018.02070},
  url       = {https://doi.org/10.3389/fpsyg.2018.02070}
}

@article{Fournier2012,
  author    = {Fournier, Susan and Alvarez, Claudio},
  title     = {Brands as Relationship Partners: Warmth, Competence, and In-Between},
  journal   = {Journal of Consumer Psychology},
  volume    = {22},
  number    = {2},
  pages     = {177--185},
  year      = {2012},
  issn      = {1057-7408},
  doi       = {10.1016/j.jcps.2011.10.003},
  url       = {https://doi.org/10.1016/j.jcps.2011.10.003}
}

@article{Zhang2023,
  author    = {Zhang, Andong and Rau, Pei-Luen Patrick},
  title     = {Tools or Peers? Impacts of Anthropomorphism Level and Social Role on Emotional Attachment and Disclosure Tendency Towards Intelligent Agents},
  journal   = {Computers in Human Behavior},
  volume    = {138},
  pages     = {107415},
  year      = {2023},
  issn      = {0747-5632},
  doi       = {10.1016/j.chb.2022.107415},
  url       = {https://doi.org/10.1016/j.chb.2022.107415}
}

@article{Fournier1998,
  author    = {Fournier, Susan},
  title     = {Consumers and Their Brands: Developing Relationship Theory in Consumer Research},
  journal   = {Journal of Consumer Research},
  volume    = {24},
  number    = {4},
  pages     = {343--373},
  year      = {1998},
  doi       = {10.1086/209515},
  url       = {https://doi.org/10.1086/209515}
}

@article{Nass2000,
  author    = {Nass, Clifford and Moon, Youngme},
  title     = {Machines and Mindlessness: Social Responses to Computers},
  journal   = {Journal of Social Issues},
  volume    = {56},
  number    = {1},
  pages     = {81--103},
  year      = {2000},
  doi       = {10.1111/0022-4537.00153},
  url       = {https://doi.org/10.1111/0022-4537.00153}
}

@article{Lyu2025,
  author    = {Lyu, Bailing and Li, Chenglu and Li, Hai and Oh, Hyunju and Song, Yukyeong and Zhu, Wangda and Xing, Wanli},
  title     = {The Role of Teachable Agents' Personality Traits on Student--AI Interactions and Math Learning},
  journal   = {Computers \& Education},
  volume    = {234},
  pages     = {105314},
  year      = {2025},
  issn      = {0360-1315},
  doi       = {10.1016/j.compedu.2025.105314},
  url       = {https://doi.org/10.1016/j.compedu.2025.105314}
}

@article{VanPinxteren2020,
  author    = {Van Pinxteren, Michelle M. E. and Pluymaekers, Mark and Lemmink, Jos G. A. M.},
  title     = {Human-like Communication in Conversational Agents: A Literature Review and Research Agenda},
  journal   = {Journal of Service Management},
  volume    = {31},
  number    = {2},
  pages     = {203--225},
  year      = {2020},
  issn      = {1757-5818},
  doi       = {10.1108/JOSM-06-2019-0175},
  url       = {https://doi.org/10.1108/JOSM-06-2019-0175}
}

@article{Karimova2025,
  author    = {Karimova, Gulnara Z.},
  title     = {Not in Our Image: Rethinking Anthropomorphism in Expert Chatbot Design},
  journal   = {AI \& Society},
  year      = {2025},
  issn      = {1435-5655},
  doi       = {10.1007/s00146-025-02438-z},
  url       = {https://doi.org/10.1007/s00146-025-02438-z}
}

@inproceedings{Nielson1994,
author = {Nielsen, Jakob},
title = {Enhancing the explanatory power of usability heuristics},
year = {1994},
isbn = {0897916506},
publisher = {Association for Computing Machinery},
address = {New York, NY, USA},
url = {https://doi.org/10.1145/191666.191729},
doi = {10.1145/191666.191729},
booktitle = {Proceedings of the SIGCHI Conference on Human Factors in Computing Systems},
pages = {152–158},
numpages = {7},
location = {Boston, Massachusetts, USA},
series = {CHI '94}
}

@article{Ahmad2022,
  author    = {Ahmad, Raian and Siemon, Dominik and Gnewuch, Ulrich and others},
  title     = {Designing Personality-Adaptive Conversational Agents for Mental Health Care},
  journal   = {Information Systems Frontiers},
  volume    = {24},
  pages     = {923--943},
  year      = {2022},
  doi       = {10.1007/s10796-022-10254-9},
  url       = {https://doi.org/10.1007/s10796-022-10254-9}
}

@misc{OpenAI_GPT41,
  author       = {{OpenAI}},
  title        = {GPT-4.1},
  howpublished = {\url{https://openai.com/index/gpt-4-1/?utm_source=chatgpt.com}},
  note         = {Accessed: 2025-08-14},
  year         = {2024}
}

@article{Nagin1999,
  author    = {Nagin, Daniel S.},
  title     = {Analyzing Developmental Trajectories: A Semiparametric, Group-Based Approach},
  journal   = {Psychological Methods},
  volume    = {4},
  number    = {2},
  pages     = {139--157},
  year      = {1999},
  doi       = {10.1037/1082-989X.4.2.139},
  url       = {https://doi.org/10.1037/1082-989X.4.2.139}
}

@article{Folstad2020,
  author    = {Følstad, Asbjørn and Brandtzaeg, Petter B.},
  title     = {Users' Experiences with Chatbots: Findings from a Questionnaire Study},
  journal   = {Quality and User Experience},
  volume    = {5},
  pages     = {3},
  year      = {2020},
  doi       = {10.1007/s41233-020-00033-2},
  url       = {https://doi.org/10.1007/s41233-020-00033-2}
}

@article{Ana2019,
  author       = {Ana Paula Chaves and
                  Marco Aur{\'{e}}lio Gerosa},
  title        = {How should my chatbot interact? {A} survey on human-chatbot interaction
                  design},
  journal      = {CoRR},
  volume       = {abs/1904.02743},
  year         = {2019},
  url          = {http://arxiv.org/abs/1904.02743},
  eprinttype    = {arXiv},
  eprint       = {1904.02743},
  bibsource    = {dblp computer science bibliography, https://dblp.org}
}

@inproceedings{Yang2021,
author = {Yang, Xi and Aurisicchio, Marco},
title = {Designing Conversational Agents: A Self-Determination Theory Approach},
year = {2021},
isbn = {9781450380966},
publisher = {Association for Computing Machinery},
address = {New York, NY, USA},
url = {https://doi.org/10.1145/3411764.3445445},
doi = {10.1145/3411764.3445445},
booktitle = {Proceedings of the 2021 CHI Conference on Human Factors in Computing Systems},
articleno = {256},
numpages = {16},
location = {Yokohama, Japan},
series = {CHI '21}
}

@article{Zem2021,
author = {Zem\v{c}\'{\i}k, Tom\'{a}\v{s}},
title = {Failure of chatbot Tay was evil, ugliness and uselessness in its nature or do we judge it through cognitive shortcuts and biases?},
year = {2021},
issue_date = {Mar 2021},
publisher = {Springer-Verlag},
address = {Berlin, Heidelberg},
volume = {36},
number = {1},
issn = {0951-5666},
url = {https://doi.org/10.1007/s00146-020-01053-4},
doi = {10.1007/s00146-020-01053-4},
journal = {AI Soc.},
month = mar,
pages = {361–367},
numpages = {7}
}

@online{bbc2016tay,
  author    = {{BBC News}},
  title     = {Microsoft chatbot is taught to swear on Twitter},
  year      = {2016},
  month     = mar,
  day       = {24},
  url       = {https://www.bbc.com/news/technology-35902104},
  note      = {Accessed: 2025-09-09}
}

@online{microsoftxiaoice,
  author    = {{Microsoft News Center Asia}},
  title     = {Much more than a chatbot: China’s Xiaoice mixes AI with emotions and wins over millions of fans},
  year      = {2018},
  url       = {https://news.microsoft.com/apac/features/much-more-than-a-chatbot-chinas-xiaoice-mixes-ai-with-emotions-and-wins-over-millions-of-fans/},
  note      = {Accessed: 2025-09-09}
}

@article{goldberg1990,
  author    = {Goldberg, Lewis R.},
  title     = {An alternative ``description of personality'': The Big-Five factor structure},
  journal   = {Journal of Personality and Social Psychology},
  volume    = {59},
  number    = {6},
  pages     = {1216--1229},
  year      = {1990},
  doi       = {10.1037/0022-3514.59.6.1216}
}

@book{jung2016,
  title={Psychological Types},
  author={Jung, Carl and Beebe, John},
  year={2016},
  publisher={Routledge}
}

@article{barbuto1997,
  title={A Critique of the Myers-Briggs Type Indicator and its Operationalization of Carl Jung’s Psychological Types},
  author={Barbuto, J.E.},
  journal={Psychological Reports},
  volume={80},
  number={2},
  pages={611--625},
  year={1997},
  doi={10.2466/pr0.1997.80.2.611},
  url={https://doi.org/10.2466/pr0.1997.80.2.611}
}

@incollection{sutcliffe2016,
  title={Designing for User Experience and Engagement},
  author={Sutcliffe, Alistair},
  booktitle={Why Engagement Matters},
  editor={O'Brien, Helen and Cairns, Paul},
  pages={79--96},
  year={2016},
  publisher={Springer, Cham},
  doi={10.1007/978-3-319-27446-1_5},
  url={https://doi.org/10.1007/978-3-319-27446-1_5}
}

@article{akhtar2015,
  title={The engageable personality: Personality and trait EI as predictors of work engagement},
  author={Akhtar, Reece and Boustani, Lara and Tsivrikos, Dimitrios and Chamorro-Premuzic, Tomas},
  journal={Personality and Individual Differences},
  volume={73},
  pages={44--49},
  year={2015},
  issn={0191-8869},
  doi={10.1016/j.paid.2014.08.040},
  url={https://www.sciencedirect.com/science/article/pii/S0191886914004851}
}

@article{vanBerkel_context,
author = {Niels van Berkel and Benjamin Tag and Jorge Goncalves and Simo Hosio},
title = {Human-centred artificial intelligence: a contextual morality perspective},
journal = {Behaviour \& Information Technology},
volume = {41},
number = {3},
pages = {502--518},
year = {2022},
publisher = {Taylor \& Francis},
doi = {10.1080/0144929X.2020.1818828},
URL = {https://doi.org/10.1080/0144929X.2020.1818828},
eprint = {https://doi.org/10.1080/0144929X.2020.1818828}
}

@misc{NYT2025,
  author       = {{The New York Times}},
  title        = {ChatGPT Linked to Suicide Raises Questions About AI Safety},
  howpublished = {\url{https://www.nytimes.com/2025/08/26/technology/chatgpt-openai-suicide.html}},
  note         = {Accessed: 2025-09-10},
  year         = {2025},
  month        = aug,
  day          = {26}
}

@inproceedings{Zhang2025,
author = {Zhang, Renwen and Li, Han and Meng, Han and Zhan, Jinyuan and Gan, Hongyuan and Lee, Yi-Chieh},
title = {The Dark Side of AI Companionship: A Taxonomy of Harmful Algorithmic Behaviors in Human-AI Relationships},
year = {2025},
isbn = {9798400713941},
publisher = {Association for Computing Machinery},
address = {New York, NY, USA},
url = {https://doi.org/10.1145/3706598.3713429},
doi = {10.1145/3706598.3713429},
booktitle = {Proceedings of the 2025 CHI Conference on Human Factors in Computing Systems},
articleno = {13},
numpages = {17},
location = {
},
series = {CHI '25}
}

@article{Klenk2024,
  author    = {Klenk, Michael},
  title     = {Ethics of Generative AI and Manipulation: A Design-Oriented Research Agenda},
  journal   = {Ethics and Information Technology},
  volume    = {26},
  pages     = {9},
  year      = {2024},
  doi       = {10.1007/s10676-024-09745-x},
  url       = {https://doi.org/10.1007/s10676-024-09745-x}
}

@article{Scheibehenne2010,
  author    = {Scheibehenne, Benjamin and Greifeneder, Rainer and Todd, Peter M.},
  title     = {Can There Ever Be Too Many Options? A Meta-Analytic Review of Choice Overload},
  journal   = {Journal of Consumer Research},
  volume    = {37},
  number    = {3},
  pages     = {409--425},
  year      = {2010},
  month     = oct,
  doi       = {10.1086/651235},
  url       = {https://doi.org/10.1086/651235}
}

@article{Chernev2015,
  author    = {Chernev, Alexander and Böckenholt, Ulf and Goodman, Joseph},
  title     = {Choice Overload: A Conceptual Review and Meta-Analysis},
  journal   = {Journal of Consumer Psychology},
  volume    = {25},
  number    = {2},
  pages     = {333--358},
  year      = {2015},
  issn      = {1057-7408},
  doi       = {10.1016/j.jcps.2014.08.002},
  url       = {https://doi.org/10.1016/j.jcps.2014.08.002}
}

@article{Thompson2005,
  author    = {Thompson, Debora Viana and Hamilton, Rebecca W. and Rust, Roland T.},
  title     = {Feature Fatigue: When Product Capabilities Become Too Much of a Good Thing},
  journal   = {Journal of Marketing Research},
  volume    = {42},
  number    = {4},
  pages     = {431--442},
  year      = {2005},
  doi       = {10.1509/jmkr.2005.42.4.431},
  url       = {https://doi.org/10.1509/jmkr.2005.42.4.431}
}

@inproceedings{Oh2025,
author = {Oh, Jeesun and Choi, Yunjae and Lee, Sangsu},
title = {“Hello, This Is a Voice Assistant Calling”: When a Human Voice Calls Claiming to Be a Machine on an Ordinary Day},
year = {2025},
isbn = {9798400714856},
publisher = {Association for Computing Machinery},
address = {New York, NY, USA},
url = {https://doi.org/10.1145/3715336.3735763},
doi = {10.1145/3715336.3735763},
booktitle = {Proceedings of the 2025 ACM Designing Interactive Systems Conference},
pages = {825–841},
numpages = {17},
location = {
},
series = {DIS '25}
}

@inproceedings{Alimardani2024,
author = {Alimardani, Maryam and de Roode, Robyn and Vaitonyte, Julija and Louwerse, Max M.},
title = {Effect of a Virtual Agent's Appearance and Voice on Uncanny Valley and Trust in Human-Agent Collaboration},
year = {2024},
isbn = {9798400706257},
publisher = {Association for Computing Machinery},
address = {New York, NY, USA},
url = {https://doi.org/10.1145/3652988.3673970},
doi = {10.1145/3652988.3673970},
booktitle = {Proceedings of the 24th ACM International Conference on Intelligent Virtual Agents},
articleno = {5},
numpages = {7},
location = {GLASGOW, United Kingdom},
series = {IVA '24}
}

@inproceedings{Ross2024,
  author    = {Ross, Alice and Corley, Martin and Lai, Catherine},
  title     = {Is There an Uncanny Valley for Speech? Investigating Listeners' Evaluations of Realistic Synthesised Voices},
  booktitle = {Proceedings of Speech Prosody 2024},
  editor    = {Chen, Yiya and Chen, Aoju and Arvaniti, Amalia},
  pages     = {1115--1119},
  year      = {2024},
  publisher = {International Speech Communication Association (ISCA)},
  series    = {Speech Prosody},
  doi       = {10.21437/SpeechProsody.2024-225},
  url       = {https://www.universiteitleiden.nl/sp2024},
  note      = {Speech Prosody 2024; Conference date: 02--05 July 2024}
}

@article{Diel2024,
  author    = {Diel, Alexander and Lewis, Michael},
  title     = {Deviation from Typical Organic Voices Best Explains a Vocal Uncanny Valley},
  journal   = {Computers in Human Behavior Reports},
  volume    = {14},
  pages     = {100430},
  year      = {2024},
  issn      = {2451-9588},
  doi       = {10.1016/j.chbr.2024.100430},
  url       = {https://doi.org/10.1016/j.chbr.2024.100430}
}

@article{Christoforakos20212,
  author    = {Christoforakos, Lara and Feicht, Nina and Hinkofer, Simone and L{\"o}scher, Annalena and Schlegl, Sonja F. and Diefenbach, Sarah},
  title     = {Connect With Me: Exploring Influencing Factors in a Human--Technology Relationship Based on Regular Chatbot Use},
  journal   = {Frontiers in Digital Health},
  volume    = {3},
  year      = {2021},
  issn      = {2673-253X},
  doi       = {10.3389/fdgth.2021.689999},
  url       = {https://www.frontiersin.org/journals/digital-health/articles/10.3389/fdgth.2021.689999}
}

@article{Scherr2025,
  author    = {Scherr, Sebastian and Cao, Bolin and Jiang, Li Crystal and Kobayashi, Tetsuro},
  title     = {Explaining the Use of {AI} Chatbots as Context Alignment: Motivations Behind the Use of {AI} Chatbots Across Contexts and Culture},
  journal   = {Computers in Human Behavior},
  volume    = {172},
  pages     = {108738},
  year      = {2025},
  issn      = {0747-5632},
  doi       = {10.1016/j.chb.2025.108738},
  url       = {https://doi.org/10.1016/j.chb.2025.108738}
}

@misc{fang2025,
      title={How AI and Human Behaviors Shape Psychosocial Effects of Chatbot Use: A Longitudinal Randomized Controlled Study}, 
      author={Cathy Mengying Fang and Auren R. Liu and Valdemar Danry and Eunhae Lee and Samantha W. T. Chan and Pat Pataranutaporn and Pattie Maes and Jason Phang and Michael Lampe and Lama Ahmad and Sandhini Agarwal},
      year={2025},
      eprint={2503.17473},
      archivePrefix={arXiv},
      primaryClass={cs.HC},
      url={https://arxiv.org/abs/2503.17473}, 
}

@article{Kirk2025,
  author    = {Kirk, Hannah R. and Gabriel, Iason and Summerfield, Christopher and others},
  title     = {Why Human--AI Relationships Need Socioaffective Alignment},
  journal   = {Humanities and Social Sciences Communications},
  volume    = {12},
  pages     = {728},
  year      = {2025},
  doi       = {10.1057/s41599-025-04532-5},
  url       = {https://doi.org/10.1057/s41599-025-04532-5}
}
